\documentclass[prx,aps,letterpaper,groupedaddress,twocolumn,superscriptaddress,floatfix,nofootinbib]{revtex4-2}

\usepackage{mathptmx}
\usepackage[utf8]{inputenc}
\usepackage[T1]{fontenc}

\usepackage{amssymb,amsmath,amsfonts}
\usepackage{mathtools}
\usepackage{mathrsfs}
\usepackage{graphicx}
\usepackage{color}
\usepackage{bm}
\usepackage{xspace,soul}
\usepackage{siunitx}
\usepackage{dsfont}
\usepackage{braket}
\usepackage{booktabs,multirow}
\graphicspath{{figures/}}

\usepackage{comment}
\usepackage{float}
\usepackage[vmargin=1in, hmargin=0.9in]{geometry}
\usepackage{times}
\usepackage[normalem]{ulem}
\usepackage[breaklinks=true,colorlinks=true,urlcolor=blue,linkcolor=blue,citecolor=red]{hyperref}
\usepackage{xr}

\DeclarePairedDelimiter{\abs}{\lvert}{\rvert}
\newcommand{\ketbra}[1]{ | #1 \rangle\!\langle #1 |}
\newcommand{\Tr} {\operatorname{Tr}}
\newcommand{\Ident} {\mathds 1}
\newcommand{\id}{\mbox{$1 \hspace{-1.0mm} {\bf I}$}}
\newcommand{\dket}[1]{\mbox{$\left|\left.#1\right\rangle\!\right\rangle$}}
\newcommand{\dbra}[1]{\mbox{$\left\langle\!\left\langle #1\right.\right|$}}
\newcommand{\dbradket}[2]{\mbox{$\langle\!\langle #1|#2\rangle\!\rangle$}}
\newcommand{\dketdbra}[2]{\mbox{$|#1\rangle\!\rangle\langle\!\langle #2|$}}

\newcommand{\bLozenge}{\mathbin{\blacklozenge}}

\makeatletter
\@ifundefined{textcolor}{}
{\definecolor{BLACK}{gray}{0}
 \definecolor{WHITE}{gray}{1}
 \definecolor{RED}{rgb}{1,0,0}
 \definecolor{GREEN}{rgb}{0,.4,0}
 \definecolor{BLUE}{rgb}{0,0,1}
 \definecolor{CYAN}{cmyk}{1,0,0,0}
 \definecolor{MAGENTA}{cmyk}{0,1,0,0}
 \definecolor{YELLOW}{cmyk}{0,.3,1,0}}
 \definecolor{light-gray}{gray}{0.90}
\makeatother

\usepackage[dvipsnames, table]{xcolor}

\newcommand{\RS}[1]{\textcolor{blue}{[RS]: #1}}

\begin{document}

\title{Characterizing quantum instruments:\\ from non-demolition measurements to quantum error correction} 

\author{Roman Stricker}
\affiliation{Institut f\"ur Experimentalphysik, Universit\"at Innsbruck, Technikerstr.  25, A-6020 Innsbruck, Austria}
\author{Davide Vodola}
\affiliation{Dipartimento di Fisica e Astronomia dell'Universit\`a di Bologna}
\affiliation{INFN, Sezione di Bologna, I-40127 Bologna, Italy}
\author{Alexander Erhard}
\affiliation{Institut f\"ur Experimentalphysik, Universit\"at Innsbruck, Technikerstr.  25, A-6020 Innsbruck, Austria}
\author{Lukas Postler}
\affiliation{Institut f\"ur Experimentalphysik, Universit\"at Innsbruck, Technikerstr.  25, A-6020 Innsbruck, Austria}
\author{Michael Meth}
\affiliation{Institut f\"ur Experimentalphysik, Universit\"at Innsbruck, Technikerstr.  25, A-6020 Innsbruck, Austria}
\author{Martin Ringbauer}
\affiliation{Institut f\"ur Experimentalphysik, Universit\"at Innsbruck, Technikerstr.  25, A-6020 Innsbruck, Austria}
\author{Philipp Schindler}
\affiliation{Institut f\"ur Experimentalphysik, Universit\"at Innsbruck, Technikerstr.  25, A-6020 Innsbruck, Austria}
\author{Rainer Blatt}
\affiliation{Institut f\"ur Experimentalphysik, Universit\"at Innsbruck, Technikerstr.  25, A-6020 Innsbruck, Austria}
\affiliation{Institut  f\"ur  Quantenoptik  und  Quanteninformation, \"Osterreichische  Akademie  der  Wissenschaften, Otto-Hittmair-Platz  1, A-6020 Innsbruck, Austria}
\author{Markus M\"uller}
\affiliation{Institute for Quantum Information, RWTH Aachen University, D-52056 Aachen, Germany}
\affiliation{Peter Gr\"unberg Institute, Theoretical Nanoelectronics, Forschungszentrum J\"ulich, D-52425 J\"ulich, Germany}
\author{Thomas Monz}
\affiliation{Institut f\"ur Experimentalphysik, Universit\"at Innsbruck, Technikerstr.  25, A-6020 Innsbruck, Austria}
\affiliation{Alpine Quantum Technologies GmbH, 6020 Innsbruck, Austria}

\begin{abstract}
In quantum information processing quantum operations are often processed alongside measurements which result in classical data. Due to the information gain of classical measurement outputs non-unitary dynamical processes can take place on the system, for which common quantum channel descriptions fail to describe the time evolution. Quantum measurements are correctly treated by means of so-called quantum instruments capturing both classical outputs and post-measurement quantum states. Here we present a general recipe to characterize quantum instruments alongside its experimental implementation and analysis. Thereby, the full dynamics of a quantum instrument can be captured, exhibiting details of the quantum dynamics that would be overlooked with common tomography techniques. For illustration, we apply our characterization technique to a quantum instrument used for the detection of qubit loss and leakage, which was recently implemented as a building block in a quantum error correction (QEC) experiment \cite{Stricker2020}. Our analysis reveals unexpected and in-depth information about the failure modes of the implementation of the quantum instrument. We then numerically study the implications of these experimental failure modes on QEC performance, when the instrument is employed as a building block in QEC protocols on a logical qubit. Our results highlight the importance of careful characterization and modelling of failure modes in quantum instruments, as compared to simplistic hardware-agnostic phenomenological noise models, which fail to predict the undesired behavior of faulty quantum instruments. The presented methods and results are directly applicable to generic quantum instruments.
\end{abstract}

\maketitle

The most general operation an experimenter can perform on a quantum system is a quantum instrument~\cite{Davies1970, Ozawa1984}. It includes both quantum and classical inputs as well as outputs and thereby generalizes quantum measurements, state preparation, and operations~\cite{Dressel2013}. Quantum instruments are commonly used to describe scenarios where one needs to keep track of a classical input or output of a quantum operation, occurring e.g.~in the description of quantum networks~\cite{Chiribella2009}, quantum causality~\cite{Oreshkov2012}, measurement uncertainty trade-offs~\cite{Buscemi2006, Knips2018}, and weak measurements~\cite{Lloyd2000, Sun2010, Kim2012}. 

In the context of quantum information processing, we are  increasingly confronted with situations where quantum computations are not simple linear evolutions, but contain elements such as in-sequence measurements and feed-forward. A prime example of this are quantum error correction (QEC) codes~\cite{Gottesman1998, Schindler1059, sun_tracking_2014, kelly_repetetive_2015, cramer_repeated_2016, Jelezko2016, negnevitsky_repeated_2018} and quantum non-demolition (QND) measurements~\cite{Unnikrishnan2015, Hume2007, Sayrin_real-time_2011, Hatridge2013, Blok_backaction_2014, Stricker2020, Rudinger2021}, where it is important to keep track of the measurement outcome in each cycle. More subtly, imperfections in realizations of quantum systems are caused by coupling to other quantum systems \cite{Breuer2006, Ringbauer2015, Rotter2015, barreiro_open-system_2011, schindler_quantum_2013, Mueller_2011, Sieberer_2016}. As a consequence, the operations that are performed on what is considered a qubit usually feature a small non-unitary component due to coupling to and ignorance of other relevant degrees of freedom. Such small deviations often go unnoticed when enforcing a unitary description onto the system~\cite{Ringbauer2015}. Device-independent~\cite{Wagner2020} and self-testing~\cite{Miklin2019,Mohan2019} protocols have been developed to assess the performance of positive operator-valued measures (POVMs)~\cite{Gomez2016,Smania2020} and quantum instruments. However, these methods do not give full information on the dynamics that is required in the context of high-precision quantum computation~\cite{helsen2020RB, Wineland2008, Gaebler2016, blume-kohout_gst_2017, erhard_characterizing_2019, MCKay2019, Ballance2020} and QEC.

Here we describe a characterization method for quantum instruments that will be particularly useful to characterize building blocks of quantum information processors. We identify quantum instruments where conventional quantum process tomography fails and introduce tomography procedures that are suitable to completely reconstruct such quantum instruments. As an example, we analyse in detail a recently implemented QND measurement for detection of qubit loss and leakage \cite{Stricker2020}, which are processes that can drastically deteriorate the performance of QEC codes, if these loss-mechanisms go unnoticed \cite{Fowler2013,Ghosh2013, Stace2018}.
Based on an experimental quantum instrument reconstruction, we derive a full instrument description for a faulty QND loss detection unit, and study its effect on a QEC cycle on a low-distance near-term logical qubit. Our results highlight the importance of developing microscopic, experimentally informed noise models of faulty quantum instruments over widely used generic hardware-agnostic noise models. The methods introduced are widely applicable to other quantum instruments including the fields of quantum information and quantum metrology~\cite{Roos2006}.

\section{Introduction to quantum instruments}
\label{Sec:QInstruments}
\begin{figure}[ht]
\includegraphics[width=0.45\textwidth]{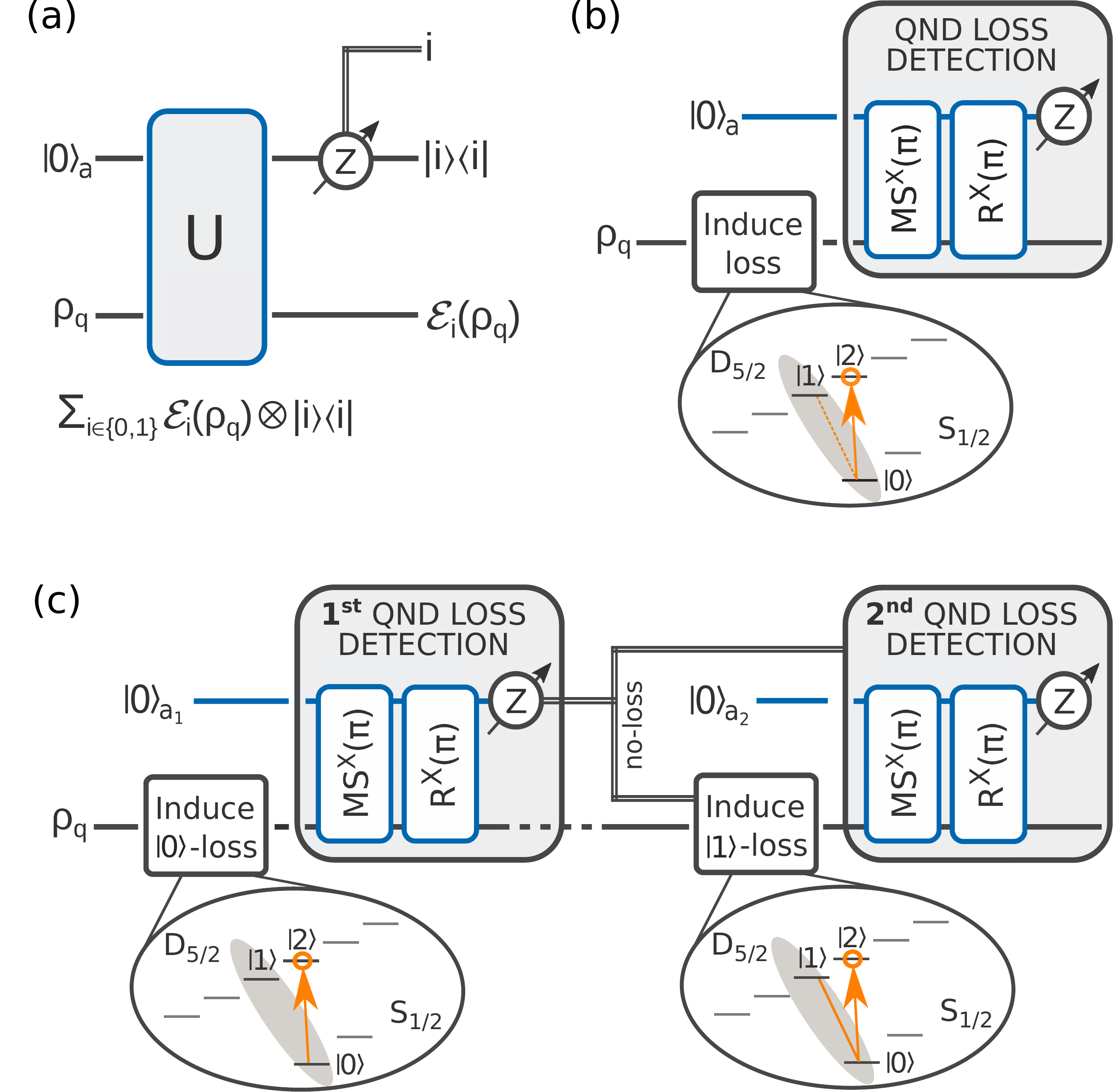}
\caption{\textbf{Quantum Instruments}. \textbf{(a)} The operation $U$ realizes a generic quantum instrument on the system in the initial state $\rho_q$ and writes the index $i$ of the applied operation into the state of an ancilla initialized in the state $\ket{0}_a$. The ancilla is finally projected onto the computational basis to read out the classical index. \textbf{(b)} A QND qubit loss detection unit as our application example for a quantum instrument. The system qubit is encoded in the computational subspace $\lbrace\ket{0}_q,\ket{1}_q\rbrace$ and is affected by loss to a third level $\ket{2}_q$. For details see text. \textbf{(c)} A quantum erasure-channel implemented by first inducing partial loss from $\ket{0}_q$ followed by its detection using the gadget from (b). Conditional on the qubit not being lost, the same partial loss is induced from $\ket{1}_q$ and subsequently detected.}
\label{Fig:QInstrument}
\end{figure}

Formally, a quantum instrument $\mathcal{I}$ is a set of trace non-increasing completely-positive (CP) maps $\{\mathcal{E}_j\}_{j \in I}$, labelled by an index $j\in I$, such that their sum is trace-preserving, $\Tr\left(\sum_j \mathcal{E}_j(\rho)\right) = \Tr(\rho)$, for every state $\rho$, see Fig.~\ref{Fig:QInstrument}a. For example, when $\mathcal{I}$ describes a quantum measurement, then $j\in I$ labels the measurement outcomes and $\mathcal{E}_j$ transforms the input state $\rho$ to the eigenstate corresponding to outcome $j$. The quantum instrument $\mathcal{I}: \mathcal{H}_1 \mapsto \mathcal{H}_2 \otimes \mathds{C}^{\abs{I}}$ thus maps the input Hilbert space $\mathcal{H}_1$ to an output Hilbert space $\mathcal{H}_2$ of potentially different size, and a classical space $\mathds{C}^{\abs{I}}$. The index of the applied operation can be represented by a set of orthogonal projectors $\ketbra{j} \in \mathds{C}^{\abs{I}}$. In practice, one might realize such a measurement by coupling the system to an ancilla, such that the ancilla state encodes the classical index, see Fig.~\ref{Fig:QInstrument}a. In the following we will focus on the simplest case with two possible measurement outcomes ($\abs{I}=2$), but all results can be straightforwardly extended to the general case.

\section{Tomography of Quantum Instruments}
\label{Sec:Tomography}
For qubit systems, complete information on their quantum evolution can be gained by quantum process tomography~\cite{Chuang1997}. However, when the evolution is described by a quantum instrument, the constituent maps are, in general, not individually trace-preserving. For example if leakage from the qubit level is present the tomographic measurements do not probe the full Hilbert space. In this case standard reconstruction techniques such as maximum likelihood estimation~\cite{Hradil1997, James2001} will not be able to describe the quantum dynamics faithfully, because they force the reconstructed map to be trace preserving. To approach this problem, we rely on a relaxed tomography algorithm that does not enforce trace preservation~\cite{Ringbauer2015, Maciel2015, Bongioanni2010}.

In order to reconstruct the quantum channel $\mathcal{E}$, we make use of the Choi-Jamiolkowsky isomorphism~\cite{choi1975} to relate $\mathcal{E}$ to an (unnormalized) quantum state $\Lambda_{\mathcal{E}}$, the \emph{Choi operator}. The correspondence between $\Lambda_{\mathcal{E}}$ and $\mathcal{E}$ is given by
\begin{equation*}
    \mathcal{E}(\rho) = \Tr_{1}[(\rho^T\otimes\id_{\mathcal X_2})\Lambda_{\mathcal{E}}] .
\end{equation*}
The Choi operator $\Lambda_{\mathcal{E}}$ with respect to the basis $\{\ket{k}\}_{k=0}^{d-1}$ can be explicitly constructed as
\begin{equation*}
    \Lambda_{\mathcal{E}} = \sum_{k,l}^{d-1}\ket{k}\!\bra{l} \otimes \mathcal{E}(\ket{k}\!\bra{l}) , 
\end{equation*}
where $d$ is the dimension of the Hilbert space. Following the notation of Ref.~\cite{Ringbauer2015}, the probability $p_{i,j}$ for observing the outcome state $\rho_j$ after preparing the state $\rho_i$ and subjecting it to the non-trace-preserving channel described by the Choi operator $\Lambda$ is given by
\begin{align}
p_{ij}
	&= \Tr\big[\rho_j^\dagger \Tr_{1}[(\rho_i^T\otimes\id_{\mathcal X_2})\Lambda]\big]	\nonumber\\
	&= \Tr\big[(\rho_i^T\otimes\rho_j^\dagger)\Lambda\big] .
\label{eq:Probs_tr}
\end{align}
Defining the projector $\Pi_{ij} \equiv \rho_i^\ast\otimes\rho_j$ and the column vector $\dket{\Lambda}=\sum_{i,j}^{d-1} \Lambda_{i,j}\ket{j}\otimes\ket{i}$, obtained by stacking the columns of $\Lambda$ (similarly for $\dket{\Pi_{ij}}$), we can identify the trace in Eq.~\eqref{eq:Probs_tr} with an inner product of the vectorized operators:
\begin{equation}
    p_{ij} = \dbradket{\Pi_{ij}}{\Lambda} .
    \label{eq:Probs}
\end{equation}
We now define the vector of observed frequencies $\vec{f}$, and the quadratic form $S$, as follows:
\begin{align}
\vec{f} &= \sum_{i,j} f_{ij}\ket{i,j}	\nonumber\\
S &= \sum_{i,j} \ket{i,j}\dbra{\Pi_{ij}} \nonumber
\end{align}

The most direct way to reconstruct the non-trace-preserving Choi operator $\Lambda$, is by inverting the above relation, a technique known as \emph{linear inversion}, 
\begin{equation}
\hat \Lambda = \mbox{argmin}_{\Lambda} \| S\dket{\Lambda}-\ket{p} \|_2 ,
\label{eq:least-sq}
\end{equation}
where $\|\cdot\|_2$ denotes the vector 2-norm, and the estimator $\hat \Lambda$ is analytically given by:
\begin{equation*}
\hat \Lambda = \sum_{i,j} p_{ij} \left(\sum_{l,m}\dketdbra{\Pi_{lm}}{\Pi_{lm}}\right)^{-1}\dket{\Pi_{ij}} .
\label{eq:LEestimator}
\end{equation*}
Unfortunately, linear inversion can produce unphysical results, especially in situations where the true (Choi)-state is close to pure~\cite{Hradil1997}. To avoid these problems, we can use modified maximum likelihood estimation by constraining the estimator to be positive semi-definite, i.e. a physical state:
\begin{align}
&\text{minimize}\quad \| WS\dket{\Lambda}-W\ket{p} \|_2
\nonumber\\
&\text{subject to: } \quad\Lambda \ge 0 .
\label{eq:MLEopt}
\end{align}
Here $W =  \sum_{i,j} \sqrt{\frac{N_j}{p_j(1-p_j)}}\ketbra{i,j}$ is a weight matrix, taking into account the multinomial distribution of observed frequencies. Note that in contrast to standard maximum likelihood quantum process tomography, we do not enforce the map to be trace preserving, i.e.\ $\Tr[\Lambda] = d$.

\section{Example: qubit loss detection}
\label{Sec:QND}
We now study in detail the example of a quantum instrument devised for a QND detection of qubit loss or leakage. Using a trapped-ion quantum processor we implement the qubit-loss detection circuit in Fig.~\ref{Fig:QInstrument}b, based on our previous work in Ref.~\cite{Stricker2020}. Our quantum processor uses trapped $^{40}$Ca$^{+}$ ions~\cite{toolbox}, where the qubits are encoded in two (meta-)stable electronic states $S_{1/2}(m=-1/2) \equiv \ket{0}$ and $D_{5/2}(m=-1/2) \equiv \ket{1}$. A universal set of quantum gate operations is realized by single-qubit rotations by an angle $\theta$ around the x- or y-axis of the Bloch sphere,  $\text{R}^\sigma(\theta) = \exp(-i{\theta}\sum_j {\sigma_j}/{2})$ with the Pauli operators $\sigma = X$ or $Y$, together with two-qubit M\o{}lmer-S\o{}renson entangling gate operations $\text{MS}_{i,j}(\theta) = \exp(-i{\theta}\sum_{i,j} X_i X_j /2)$~\cite{MSgate}; see Supplementary Material for more details~\cite{supp_mat}.

\begin{figure}[ht]
\includegraphics[width=0.43\textwidth]{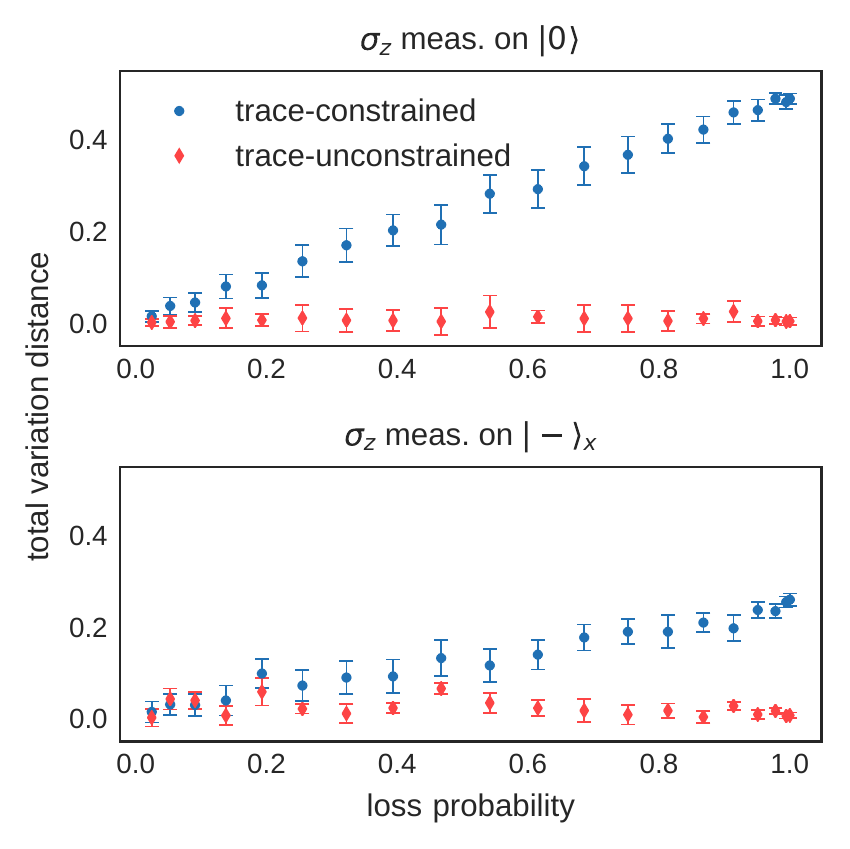}
\caption{\textbf{Comparison of trace-constrained and trace-unconstrained tomography for the non-unitary map $\mathcal{E}_0$ from Fig.~\ref{Fig:QInstrument}b.} We compute the total variation distance between directly measured frequencies and those predicted from the reconstructed Choi operators. Standard MLE increasingly fails to capture the underlying dynamics for higher loss probabilities. Error bars correspond to one standard deviation of statistical uncertainty due to quantum projection noise.}
\label{Fig:TomographyTechnique}
\end{figure} 
The dominant loss mechanism in a trapped-ion quantum processor is leakage from the qubit subspace $\lbrace\ket{0},\ket{1}\rbrace$ to other electronic states, which can occur due to radiative decay from  meta-stable electronic qubit states~\cite{Kreuter2004}, in Raman transitions ~\cite{Hayes2020}, or due to imperfections in spectroscopic decoupling pulses~\cite{Nigg2014} when additional electronic states outside the computational subspace are used deliberately. Hence, loss can be induced in a controlled fashion by transferring part of the population from either computational basis state to an auxiliary level $D_{5/2}(m=+1/2) \equiv \ket{2}$ referred to as loss-transition $R_\text{loss}(\phi)$ denoting a full coherent transfer in case of $\phi = \pi$. We then apply the QND unit to map the information about a loss of the system qubit (q) onto an ancilla qubit (a), which is subsequently read out. In the language of quantum instruments, this means that one of two possible maps (`loss' or `no loss') has been applied to the system, with the classical index of the applied map stored qubit states $\ket{0}_a$ and $\ket{1}_a$ of the ancilla. Similar QND loss detection protocols have been devised using various other physical platform ~\cite{Niemietz2021, Varbanov2020, Vala2005}.

Notably, for both ancilla outcomes the system qubit is subject to a map that is completely-positive (CP), but in general not trace-preserving (TP). This non-unitarity of the individual maps leads to several counter-intuitive effects: For example, in the present case, the evolution of the system qubit differs from the identity map, even in the case where no loss is detected, if loss occurs asymmetrically, i.e.~from only one of the computational basis states. More precisely, for loss restricted to occur from $\ket{0}$ the system qubit follows (up to normalization) a non-unitary evolution given by $\rho_q \mapsto\mathcal{E}_0\rho_q\mathcal{E}^\dagger_0$ with 
\begin{equation}
\label{eq:no-loss-map}
\mathcal{E}_0 = \ket{1}_q\!\bra{1} + \cos (\phi/2) \ket{0}_q\!\bra{0}
\end{equation}
considering the coherent loss operation $R_\text{loss}(\phi)$. This is a consequence of the information gain in the no-loss case, given by the ancilla measurement~\cite{Stricker2020}. In either case, the reconstruction becomes challenging, since standard reconstruction techniques for quantum process tomography enforce the reconstructed processes to be completely-positive and trace-preserving (CPTP), thereby suppressing the deviations from this condition characteristic for quantum instruments. This becomes evident in Fig.~\ref{Fig:TomographyTechnique}, where we compare the accuracy of quantum process reconstructions of the `no-loss' dynamics obtained via standard maximum likelihood technique (MLE) referred to as trace-constrained approach in contrast to the trace-unconstrained approach of Eq.~\eqref{eq:MLEopt}. As a figure of merit we use the total variation distance between the measured frequencies and the measurement outcomes that are predicted from the reconstructed Choi operators. This highlights how the trace-constrained approach can fail to capture the dynamics, an error that might go unnoticed for maps that are close to trace preserving.

\section{Results}
\label{Sec:Data}
We now discuss features associated in experiments with QND measurements that can only be captured using a full description as a quantum instrument. We start by characterizing the instrument acting on a two-level system (qubit), followed by a complete characterization in a higher dimensional Hilbert space that captures the entire dynamics of the QND measurement. 

\begin{figure*}[ht]
\includegraphics[width=0.9\textwidth]{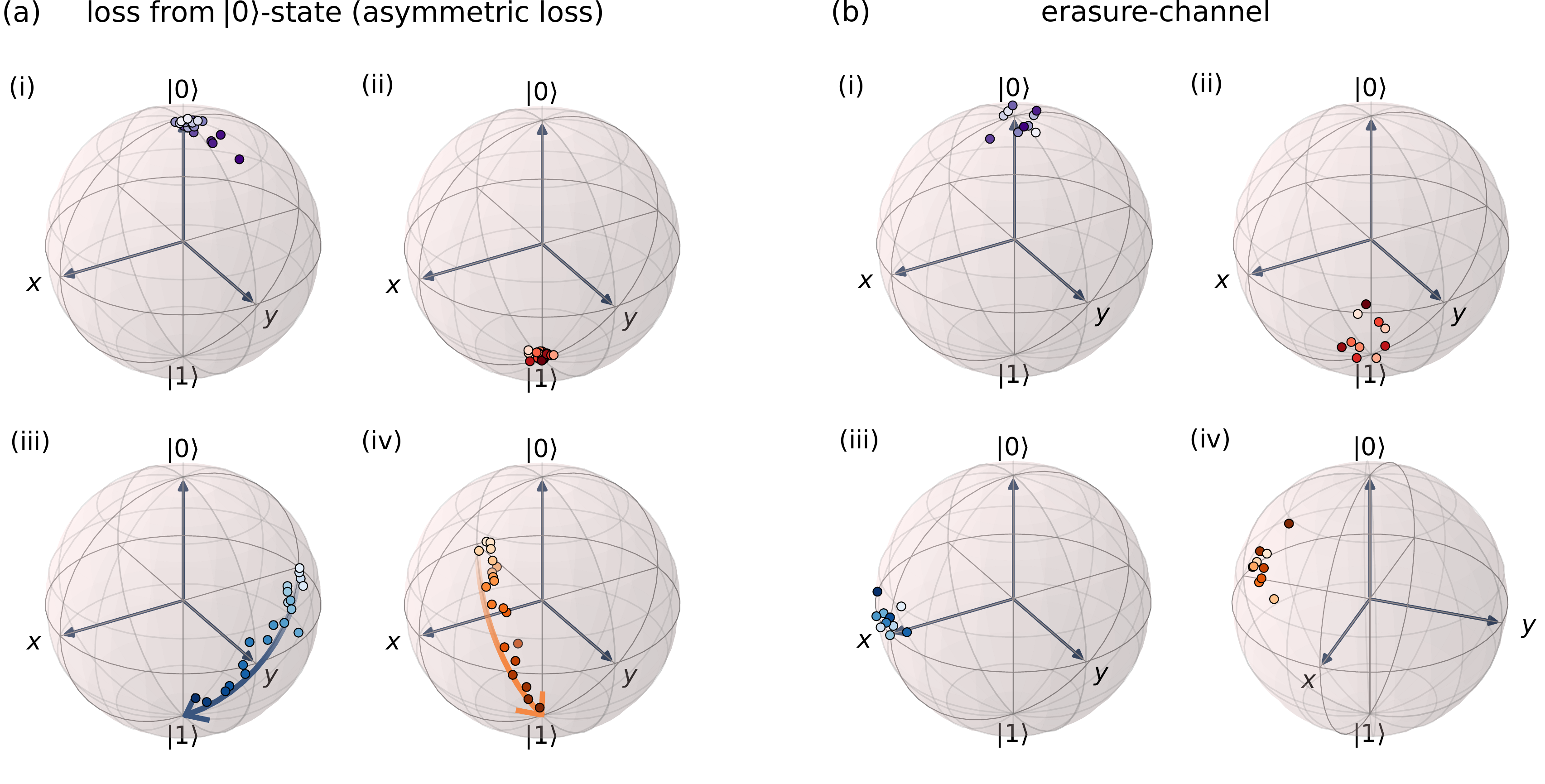}
\caption{\textbf{Bloch-vectors after undergoing the QND-detection in the no-loss case for different loss channels}.\textbf{(a)} State vector evolution for asymmetric loss from $\ket{0}_q$ is captured by the color gradient, ranging from $\SI{0}{\%}$ loss (bright points) to \SI{100}{\%} (dark points) for various input states \textbf{(i)} $\ket{0}_q$, \textbf{(ii)} $\Ket{1}_q$, \textbf{(iii)} $\Ket{-}_{X,q}$ and \textbf{(iv)} $\Ket{-}_{Y,q}$. Notably, the Bloch-vectors remain close to the surface of the sphere independent of the loss probability, see Supplementary Material~\cite{supp_mat}. The initial superposition states $\ket{-}_{X,q}$ and $\ket{-}_{Y,q}$ are found transitioning to the basis state not affected by the loss. \textbf{(b)} The erasure-channel is realized by consecutively inducing the same partial loss from $\ket{0}_q$ followed by $\ket{1}_q$ and post-selecting to both no-loss cases, i.e. both ancilla's $\ket{0}_a$ outcome. Our results support the theory derivation of a map $\propto\rho$ leaving the initial states up to noise unaltered; see Supplementary Material~\cite{supp_mat}.}
\label{Fig:SingleQND_Bloch}
\end{figure*}

We implement the circuit in Fig.~\ref{Fig:QInstrument}b on a two-ion string studying several input states $\lbrace\ket{0}_q$, $\ket{-}_{X,q}=1/\sqrt{2}(\ket{0}_q-\ket{1}_q)$, $\ket{-}_{Y,q}=1/\sqrt{2}(\Ket{0}_q-i\ket{1}_q)$, $\ket{1}_q\rbrace$ on the system qubit for a range of loss probabilities. We apply quantum state tomography for the runs that signal no-loss events, effectively applying the ``no-loss'' map $\mathcal{E}_0$ as given by Eq.~(\ref{eq:no-loss-map}). We focus on the no-loss outcome $\mathcal{E}_0$ given that in a realistic scenario the system qubit would remain intact, as opposed to the loss case. We find that the superposition input states are \emph{distorted} towards the basis state that is not affected by the loss with increasing loss probability, see Fig.~\ref{Fig:SingleQND_Bloch}a. This is a consequence to the asymmetry of the loss, occurring only from one basis state, as detailed in Eq.~\eqref{Eqn:SingleQND} in the Supplementary Material~\cite{supp_mat}.  Importantly, however, the states display no notable reduction in purity, regardless of the loss probability. More details are given in the Supplementary Material~\cite{supp_mat}.

The archetypal description of a qubit loss channel features symmetric loss, often referred to as a \emph{quantum erasure-channel}~\cite{Grassl1997}, where loss occurs with a given probability, irrespective of the qubit state, and the position of the lost qubit is known. Experimentally, we realize this quantum erasure-channel sequentially in two steps, by first inducing partial loss from $\ket{0}_q$ followed by its detection, and, conditional on detecting no loss in this first step, inducing the same amount of partial loss, but now from $\ket{1}_q$ in this second step. Experiments are conducted on a three ion string using a single system qubit (q) and two ancilla qubits $a_1$ and $a_2$ as depicted in Fig.~\ref{Fig:QInstrument}c. By observing the evolution of the Bloch vectors in Fig.~\ref{Fig:SingleQND_Bloch}b we find that the initial state is preserved up to experimental noise, as derived in Eq.~\eqref{Eqn:Erasure} in the Supplementary Material~\cite{supp_mat}. The purity is again found independent of the loss, see Supplementary Material~\cite{supp_mat}.

\begin{figure}[ht]
\centering
\includegraphics[width=0.5\textwidth]{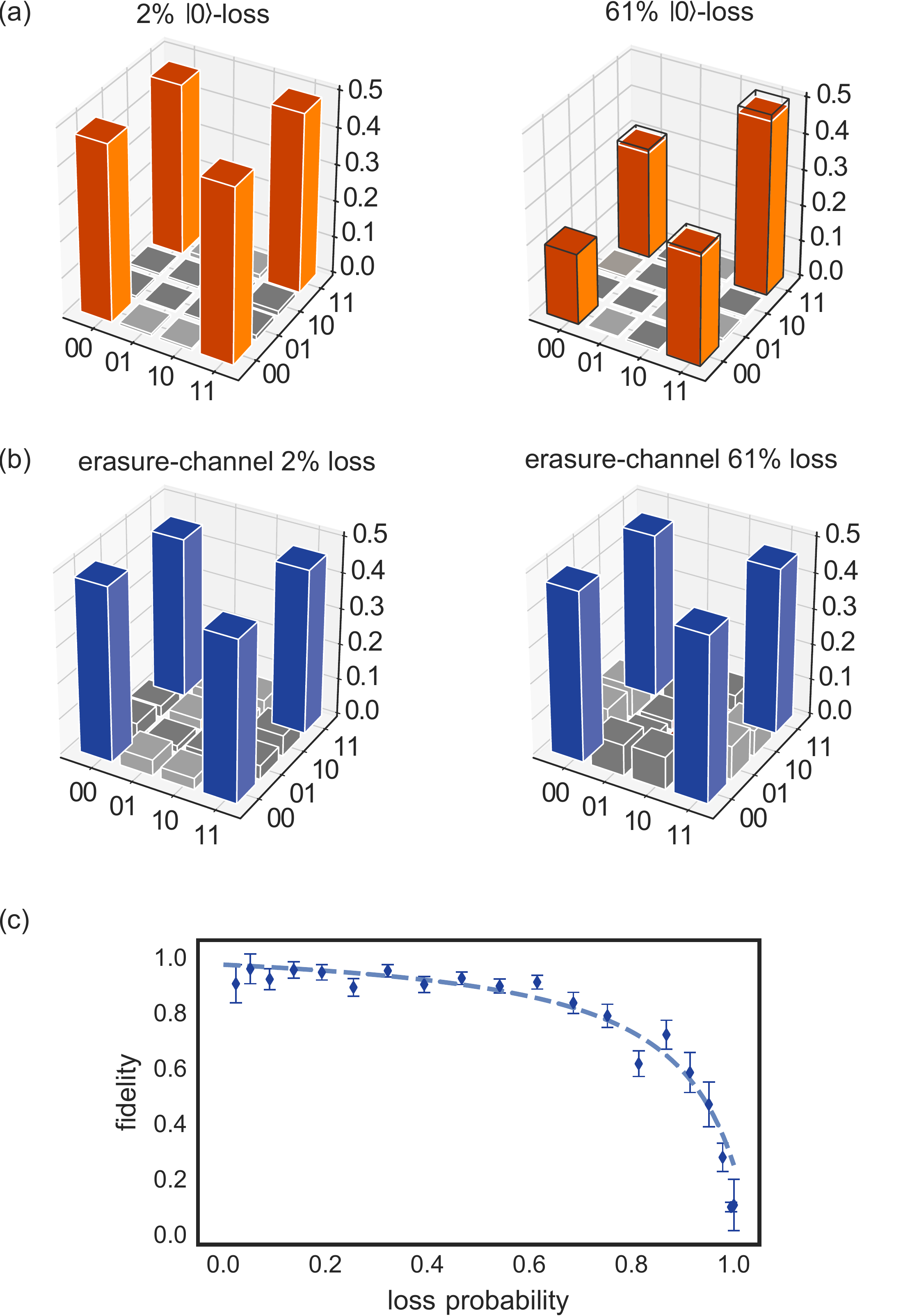}
\caption{\textbf{Tomographic reconstruction of the maps characterizing our quantum instrument}. \textbf{(a)} Single qubit Choi operators in the elementary basis $\lbrace\Ket{00}_q,...,\Ket{11}_q\rbrace$ describing the QND-detection under loss from $\ket{0}_q$. Process fidelities compared to the ideal map for $\SI{2}{\%}$ and $\SI{61}{\%}$ loss read $0.97(1)$ and $0.98(1)$ respectively. Black boxes denote the ideal operator in the higher loss case. \textbf{(b)} On the erasure-channel we receive the expected identity map for loss from both qubit states up to about \SI{50}{\%} before errors start to dominate. \textbf{(c)} Corresponding process fidelities compared with ideal maps together with the decay model (dashed-line) from Eq.~\eqref{Eqn:ErasureFidelity}.} 
\label{Fig:Choi_SingleQubit}
\end{figure}

These findings are further corroborated by quantum process tomography characterizing the map describing the system qubit dynamics by using the unconstrained reconstruction approach of Eq.~\eqref{eq:MLEopt}. In the case of the asymmetric loss previously discussed, the single qubit Choi operators for the map $\mathcal{E}_0$ are close to the identity only given little loss on the order of a few percent and clearly deviate for higher loss, revealing their non-unitary behavior, see Fig.~\ref{Fig:Choi_SingleQubit}a left for a low loss probability and right for a high loss probability. We note that a standard MLE approach would force unitary maps and thereby prevent the correct reconstruction not displaying this non-unitary behaviour. In contrast, for the quantum erasure-channel, for both $\SI{2}{\%}$ and $\SI{61}{\%}$ loss cases maps are found close to the identity following the theoretical predictions, depicted in Fig.~\ref{Fig:Choi_SingleQubit}b.

For higher loss rates, however, we observe a deviation of the reconstructed from the predicted channel, quantified by the fidelity between the reconstructed and ideal Choi operator shown in Fig.~\ref{Fig:Choi_SingleQubit}c. For high loss rates, only few experimental cycles remain in the no-loss case. As a result, error terms, such as state-preparation-and-measurement (SPAM) errors, as well as errors in the implementation of the loss process contribute with a higher relative weight. We can model these additional error terms as depolarizing noise at the level of the Choi operator as a function of the loss rate $p_\text{loss}$:
\begin{equation}
\begin{split}
\Lambda_\textsc{m}(p_\text{loss}) \propto &
(1-p_\text{loss})\cdot(1-p_\text{e})\cdot(1-p_\text{spam})\cdot \Lambda\\
&+p_\text{loss}\cdot p_\text{e}\cdot(1-p_\text{spam})\cdot \Ident/4 \\
&+p_\text{spam}\cdot \Ident/4 , 
\label{Eqn:ErasureFidelity}
\end{split}
\end{equation}
where $\Lambda$ denotes the ideal Choi operator of the no-loss channel, $\Ident$ is the identity matrix, representing a fully depolarizing channel, $p_\text{e}$ is a generic error rate of the erasure-channel and $p_\text{spam}$ is the error rate due to SPAM errors. The first term of Eq.~\eqref{Eqn:ErasureFidelity} describes the ideal channel where no loss happened and the QND detection worked, while the second term is a case where a loss happened, but the QND unit failed to detect it as such. The final term describes the contribution from SPAM errors. From a fit to the data, we find that $p_\text{e}=0.09$ and $p_\text{spam}=0.03$ captures the observed drop in fidelity well. From Fig.~\ref{Fig:Choi_SingleQubit}c we see that these effects become predominant for high loss rates, while for up to $\sim 60\%$ a faithful reconstruction of the experimental Choi matrix is possible. 

\begin{figure}[ht]
\includegraphics[width=0.42\textwidth]{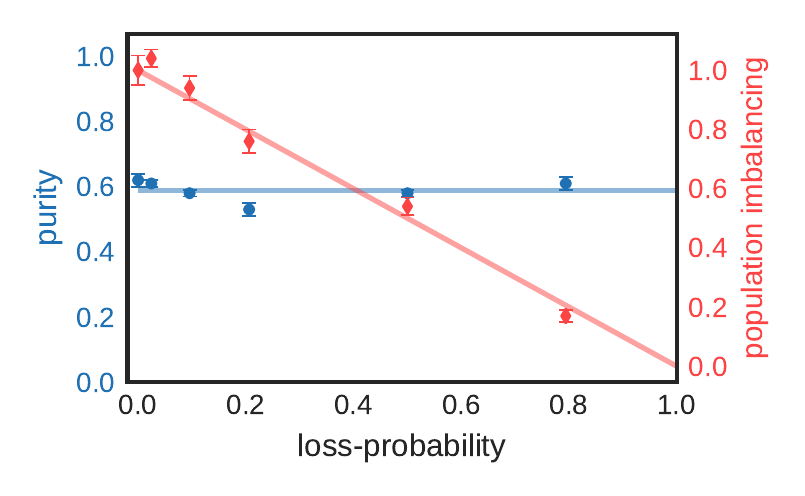}
\caption{\textbf{Multi qubit entangled state undergoing the QND-detection in the no-loss case}. As an input we chose the 4-qubit GHZ-state $(\ket{0000}_q+\ket{1111}_q)/\sqrt{2}$. Loss is induced from $\ket{0}_{q,1}$ on system qubit 1. Results for purity ($\bullet$) and population imbalancing between the GHZ basis states $\vert0000\rangle$ and $ \vert1111\rangle$ ($\scriptstyle\bLozenge$) in analogy to the Bloch-vector picture (Fig.~\ref{Fig:SingleQND_Bloch}) are shown. The purity is found constant, while the population imbalancing increases towards higher loss probabilities finally causing a \emph{distortion} to the state $\ket{1111}_q$ not affected by the loss. Errors correspond to one standard deviation of statistical uncertainty due to quantum projection noise.}
\label{Fig:GHZ_PurityFidelity}
\end{figure}

The results presented so far covered a single system qubit and revealed potential obstacles of our quantum instrument tomography, which are generally transferable to other experiments utilizing QND measurements. We now go one step further by analyzing these effects on a multi-qubit entangled state. Experiments are conducted using four system qubits, initialized in the GHZ-state $1/\sqrt{2}(\ket{0000}_q+\ket{1111}_q)$, accompanied by one ancilla. After state preparation, partial asymmetric loss from $\ket{0}_{q,1}$ on system qubit 1 is induced followed by its detection using the QND detection unit. The ``no-loss'' evolution $\mathcal{E}_0$ is analyzed by four-qubit quantum state tomography. In Fig.~\ref{Fig:GHZ_PurityFidelity} the states again show no significant reduction in purity ($\bullet$) over the range of measured loss probabilities and by that obscuring the non-unitary effect from our instrument. However, an asymmetric effect is displayed by computing the population ratio of the GHZ basis states $\vert0000\rangle_q$ and $\vert1111\rangle_q$ in Fig.~\ref{Fig:GHZ_PurityFidelity} referred to as population imbalancing ($\scriptstyle\bLozenge$) showing a \emph{distortion} towards the basis state not affected by loss in analogy to the Bloch-vectors in the single qubit case. The underlying theory curve follows $1-p_\text{loss}$ as can be seen from eqn.~\eqref{eq:no-loss-map}. The fidelity with the initial GHZ-state further remains above $\SI{50}{\%}$ within 1 standard deviation of statistical uncertainty, thus certifying multipartite entanglement independent of the loss probability.

\begin{figure*}[ht]
\centering
\includegraphics[width=0.9\textwidth]{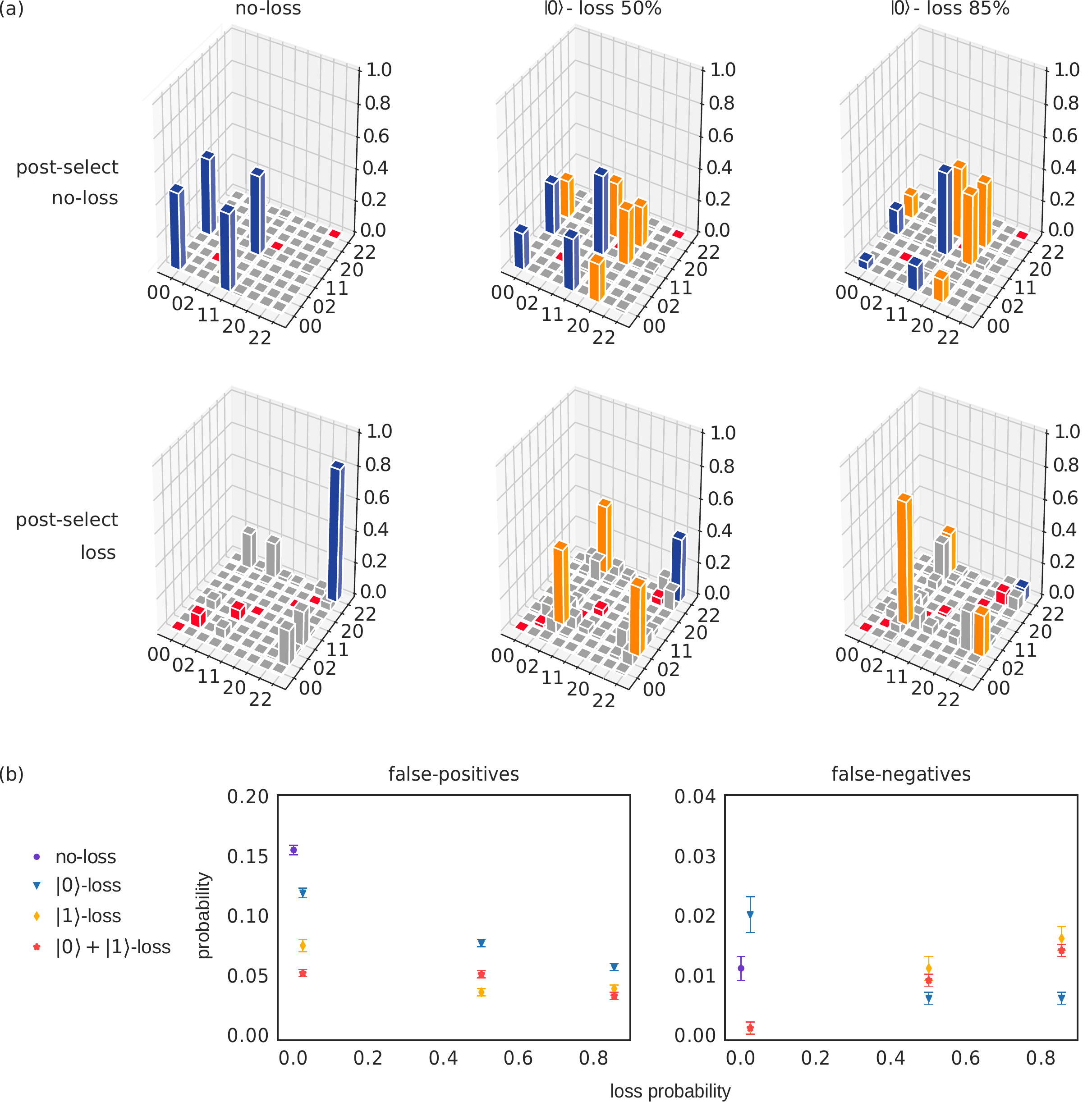}
\caption{\textbf{Full system dynamics from combined qutrit-ancilla quantum instrument tomography.} \textbf{(a)} Choi operator of the system qutrit evolution in the elementary basis $\lbrace\ket{00}_q,\ldots,\ket{22}_q\rbrace$ after post-selecting on the ancilla revealing either loss case (rows) examined for different loss probabilities from $\ket{0}_q$ (columns). The tricolor Choi operators show peaks in the absence of loss (blue), peaks occurring due to partial loss (orange) and erroneous peaks (red). The latter are only color coded on the diagonal for visualization purposes. Process fidelities with the ideal map from top left to bottom right read $\lbrace 0.97(1), 0.96(1), 0.95(1), 0.83(1), 0.86(1), 0.84(1)\rbrace$. \textbf{(b)} False-positive and false-negative rates extracted from raw data for loss states $\lbrace\ket{0}_q,\ket{1}_q,1/\sqrt{2}(\ket{0}_q+\ket{1}_q)\rbrace$ versus loss probability.}
\label{Fig:QutritChois}
\end{figure*}

The full dynamics of our coherent loss process can be reconstructed by explicitly taking the loss level $\ket{2}_q$ into account. The state of the system ion needs then to be represented by a qutrit with basis states $\lbrace\ket{0}_q,\ket{1}_q,\ket{2}_q\rbrace$. We perform quantum process tomography on the combined system of data qutrit and ancilla qubit. This allows us to study both loss cases by distinguishing the maps dependent upon the ancilla state, and provides more fine-grained information on the microscopic error processes. The reconstructed Choi operators for both ancilla outcomes and various loss probabilities are given in Fig.~\ref{Fig:QutritChois}a. For the sake of clarity the operators are color coded by peaks occurring in the absence of loss (blue), peaks denoting the partial loss rotation (orange) and erroneous peaks (red). The latter are restricted to the diagonal for simplicity. Note that these experimentally derived maps on the qutrit-level now follow unitary maps, see Eq.~\eqref{Eqn:SingleQutritQND} in the Supplementary Material~\cite{supp_mat}. 

One key piece of information gained from the full tomography are the dominant failure modes of the experimental realization of the QND detection unit. In the no-loss case, false-negatives are retrieved from diagonal elements $\lbrace\ket{02}_q,\ket{12}_q,\ket{22}_q\rbrace$ corresponding to undetected rotations to the level $\ket{2}_q$ outside the computational subspace. Likewise false-positives in the loss case are retrieved from the elements $\lbrace\ket{00}_q,\ket{01}_q,\ket{10}_q,\ket{11}_q,\ket{20}_q,\ket{21}_q\rbrace$ corresponding to qubit rotations mistakenly assigned as loss. Note that for standard tomography restricted to qubit levels such fine-grained analysis would be precluded for two main reasons. First, the true population in the loss state of the system qutrit cannot be estimated independently from the ancilla outcome in the qubit description. Thus, one cannot reliably assign false-positive and false-negative events by post-selecting on the ancilla since all erroneous population adds up to the main peaks $\lbrace\ket{00}_q,\ket{11}_q\rbrace$ blurring the information about the error origin. Second, when tracing over the ancilla, the loss state $\ket{2}_q$ would be incoherently added to the state $\ket{1}_q$ creating an unphysical bias under which tomography is likely to break; see Supplementary Material~\cite{supp_mat}. For a more quantitative analysis the corresponding false-positive and false-negative rates are depicted in Fig.~\ref{Fig:QutritChois}b. To avoid errors from the quantum instrument reconstruction, these rates were extracted from the raw-data for three different loss states: $\lbrace\ket{0}_q,\ket{1}_q,1/\sqrt{2}(\ket{0}_q+\ket{1}_q)\rbrace$. Notably, there is a significantly higher false-positive rate owing to their sensitivity on the entangling operation implementing a correlated two-qubit rotation. This operation shows a higher error rate compared to single-qubit operations~\cite{erhard_characterizing_2019} and only plays a role in the no-loss case: the reason is that as under loss the entangling operation, when it only acts on the ancilla qubit alone, its action is on purpose trivial and no longer induces a correlated qubit-qutrit flip process. Therefore, the loss map is left with the local bit-flip operations explaining why false-negatives are dominated by single-qubit errors, resulting in smaller rates. For loss detection in a QEC setting, we expect this asymmetry to be quite beneficial, as a false-positive event would merely trigger an unnecessary loss correction, while a false-negative event leads to an undetected loss, which can be catastrophic, i.e.~leading directly to uncorrectable logical errors, as will be discussed in the next section. 

We now build noise models to characterize the QND-detection unit, which can then be used to study implications on QEC. From the above phenomenological discussion, we assume that the dominant contributions will come from false-positive and false-negative events, where the latter in particular can have a severe impact. However, extracting the respective rates from tomography data as in Fig~\ref{Fig:QutritChois}b in the presence of SPAM errors can be unreliable if these contributions are of the same magnitude. A rough estimate of the SPAM errors from tomography of the identity yields a fidelity of 0.96(2), which indicates that this is indeed the parameter regime we are dealing with here. 

Hence, to describe imperfections in the QND loss detection unit we instead focus on a microscopic noise model $\mathcal{E}_\text{noise}$ defined as~\cite{supp_mat}
\begin{equation}
\rho \mapsto \mathcal{E}_\text{noise}(\rho) =  U_\text{noise} \rho U_\text{noise}^{\dag}
\end{equation}
where the unitary $U_\text{noise} = \text{MS}(\alpha)R^X(\beta)$ describes the dominating error source as correlated bit flips with a rate of $p_\text{corr} = \sin^2(\alpha/2)$ resulting from systematic miscalibrations in the two-ion MS-gate, and single qubit flips with a rate of $p_\text{single} = \sin^2(\beta/2)$ from errors in the collective local rotations. Fitting the channel $\mathcal{E}_\text{noise}$ to the experimental data returns values of $p_\text{corr.}=0.045$ and $p_\text{single}= 2.47\cdot 10^{-4}$, respectively, see Supplementary Material~\cite{supp_mat}. The fidelity of the experimental data with respect to this model in the no-loss case is 0.94, compared to 0.91 for the noiseless theory prediction.

In order to validate this model against generic hardware-agnostic noise models typically considered in the quantum information literature, we further add depolarizing and dephasing noise channels~\cite{NielsenChuang}. As discussed in detail in the Supplementary Material~\cite{supp_mat}, by fitting a model that includes all 4 error channels to the experimental data, we again find the correlated bit-flip error to be dominant. The contributions from depolarizing and dephasing noise are consistently on the order of $0.01$ and adding these terms does not significantly improve the fit to the data. From this analysis we conclude that the microscopic model is the most suitable description of our experimental noise and the resulting imperfections in the QND loss detection, and we will thus use this model in the following analysis of the impact of a faulty QND loss detection unit on QEC.

\section{Implication on quantum error correction}
\label{Sec:QECImplications}
In the context of QEC and the pursuit for robust and eventually fault-tolerant quantum computers, qubit leakage and loss errors are known to be particularly harmful to the performance of QEC codes, if they go unnoticed \cite{Fowler2013,Ghosh2013, Stace2018}. Dedicated protocols to fight qubit loss have been devised, including the 4-qubit quantum erasure code~\cite{Grassl1997}, which has been implemented in the form of post-selective state analysis protocols using photons~\cite{Ralph2005,Lu2008}. Moreover, protocols to cope with qubit loss in elementary quantum codes such as the 5-qubit code \cite{Laflamme1996} as well as topological QEC codes including the surface code~\cite{Kitaev2003} and color codes~\cite{Bombin2006, Bombin2007, Nigg2014} have been developed. 

Here, our aims are to (i) estimate the parameter regimes in which active qubit loss error correction and detection is expected to reach break-even, i.e.~to become beneficial for low-distance QEC codes as currently pursued in various efforts \cite{Knill2001, Chiaverini2004, Schindler1059, Reed2012, Nigg2014, Riste2015, Stricker2020, Varbanov2020}. (ii) Whereas most theory studies exclusively focus on the simple (and ideal) quantum erasure-channel to describe loss, we are interested in illustrating the effect of various qualitatively different imperfections in the loss detection process on QEC performance, highlighting the importance of microscopically informed noise models of the components used in QEC of qubit loss. (iii) Finally, to predict the performance of QEC protocols by numerical simulations, it is desirable to develop \textit{effective} few-parameter noise models, informed by experimental data, which can be simulated efficiently, e.g.~using stabilizer simulations, to predict the performance of large-scale QEC codes built from noisy components. Here, we are therefore interested in understanding to which extent the microscopic noise model of the QND loss detection can be reliably substituted by such efficiently simulatable noise models.

\begin{figure}[ht]\centering
\includegraphics[scale=0.75]{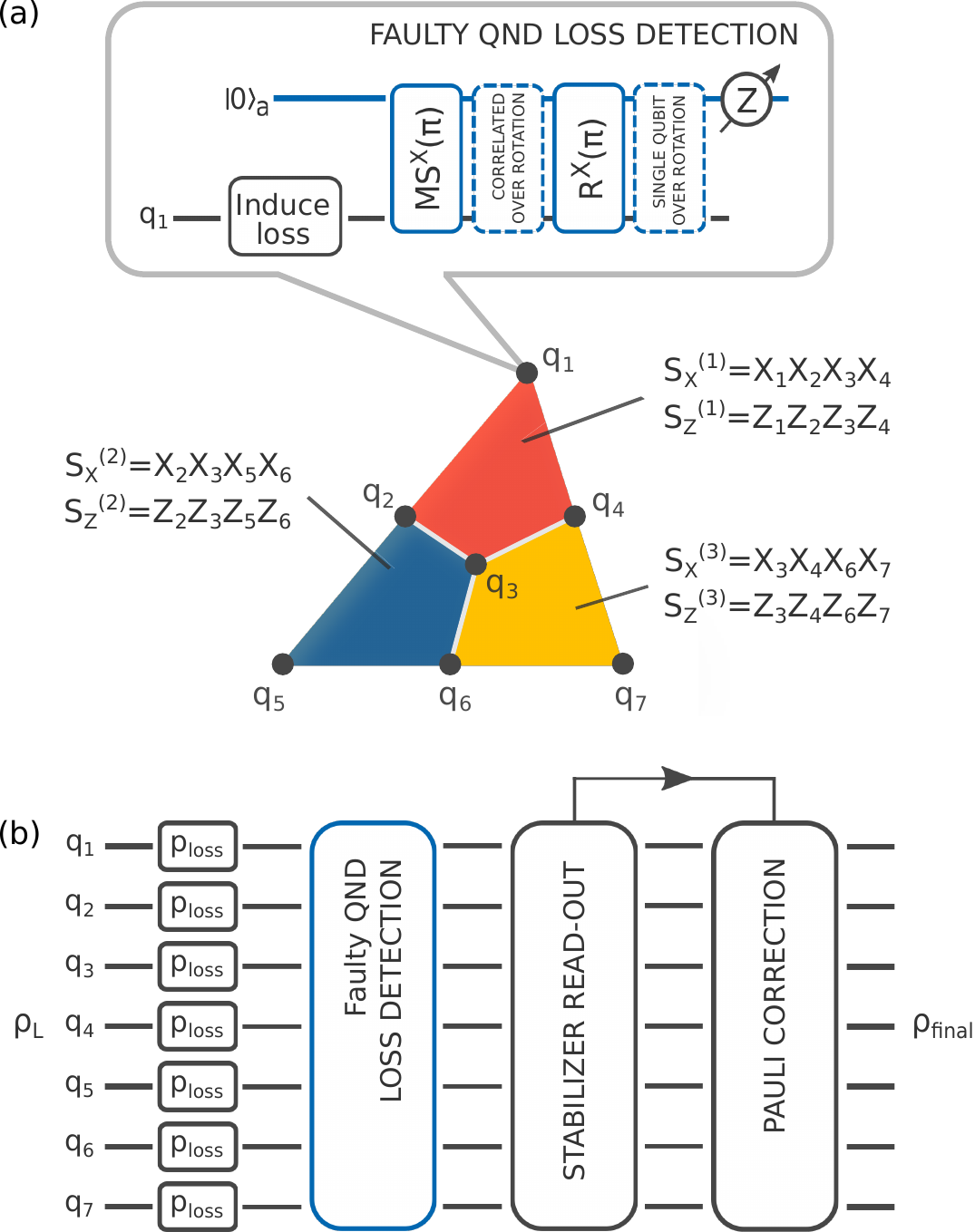}
\caption{\textbf{Simulations on the faulty QND loss detection embedded in the 7-qubit color code.} \textbf{(a)} A single logical qubit encoded on a triangular planar color code lattice formed of three interconnected plaquettes (lower part). The code space is formed by six stabilizer operators $S_x^{(i)}$ and $S_z^{(i)}$ each acting on a plaquette of four physical qubits~\cite{Nigg2014}. Loss is subsequently detected on all code qubits using a faulty QND circuit (top part). We model this taking into account both correlated and single qubit overrotations representing our leading error mechanisms by treating every qubit as a qutrit. \textbf{(b)} Single QEC cycle of qubit loss detection and correction including initial controlled induction of loss, followed by faulty QND loss detection operations on the qubit subspace of all physical qutrits and stabilizer measurements triggering respective conditional Pauli corrections.} \label{Fig:7QubitCode}
\end{figure}

To be concrete, we will focus on the smallest 2D color code \cite{Bombin2006}, a 7-qubit stabilizer code equivalent to the Steane code~\cite{Steane1996, Bombin2006}, which is at the focus of current experimental efforts to achieve the break-even point of beneficial and fault-tolerant QEC with low-distance QEC codes \cite{Bermudez2017, Bermudez2019, Gutierrez2019, Amaro2020}. The code is obtained by projecting the Hilbert space of seven qubits (Fig.~\ref{Fig:7QubitCode}) into the $+1$ eigenspace of six commuting stabilizer generators $S^{x}_i$ and $S^{z}_i$  ($i = 1, 2, 3$), see Fig.~\ref{Fig:7QubitCode}a, that define a two-dimensional code space hosting one logical qubit. Logical $X$ and $Z$ operators are defined as $X_L = \prod_{i=1}^{7} X_i$ and $Z_L = \prod_{i=1}^{7} Z_i$ and the logical basis states are $\ket{0_L} \propto \prod_{i=1}^{3}(\mathds{1} + S^{x}_i)\ket{0}^{\otimes7} $ and $\ket{1_L} = X_L\ket{0_L}$~\cite{supp_mat}. The code is a distance $d=3$ QEC code ($d=2n+1$, $n$ being the number of correctable computational errors), so that one arbitrary computational error (bit and/or phase flip error) on any of the physical qubits is correctable. Note that besides computational errors, this code also allows one to correct the loss of any two of the seven physical qubits, or even the loss of some, though not all subsets of three or even four qubits (see Supplementary Material for more details~\cite{supp_mat}). We note for each of the seven qubits forming the code, we incorporate the state $\ket{2}_q$, i.e.~adopt a qutrit-description, and use this additional level to induce loss of a controllable amount via the coherent rotation in the subspace $\{\ket{0}_q, \ket{2}_q\}$ of the quantum instrument depicted in Fig.~\ref{Fig:QInstrument}b. 

We then model one round of qubit loss error detection and correction, depicted in Fig.~\ref{Fig:7QubitCode}, as follows: starting from an ideal (noise free) logical state $\rho_L$ of the 7-qubit code, qubit loss is induced with an independent and equal probability $p_\mathrm{loss}$ on each of the physical qubits of the register. Subsequently, a noisy QND loss detection unit is sequentially applied to each of the 7 qubits, in order to detect the possible occurrence of loss. Each data qubit, for which the QND measurement indicates the occurrence of a loss, is replaced by a fresh qubit in the computational basis state $\ket{0}_q$. This is followed by one round of possibly faulty measurements of all six stabilizers of the code. For simplicity, since our focus lies on the QND loss detection, here we model imperfections in each stabilizer measurement by a phenomenological noise model, in which the stabilizer measurement outcome is assumed to be faulty with a probability $q$~\cite{Dennis2002}\footnote{Such a model could be refined by adopting a circuit-level description and specific compilations of the stabilizer readout into gates, based e.g.~on recently proposed flag-qubit based stabilizer readout protocols~\cite{Chao2018, Bermudez2019}.}. Since the four-qubit stabilizer operators are typically measured with a circuit involving (at least) four two-qubit gates, we work with four times the two-qubit error rate as the error rate of the stabilizer measurement, which results in $q= p_\text{corr}$, in what follows.
Based on the obtained syndrome ($\pm 1$ stabilizer eigenvalues) from the measurement of the stabilizers, Pauli corrections are applied if needed (such a Pauli frame update can be done on the software level and are thus modelled as error-free). Finally, to determine the logical error rate, it is checked whether the original logical state $\rho_L$ has been recovered or not, by evaluating the expectation value of the logical operator corresponding to the initially prepared encoded state.

\begin{figure}[ht]
    \centering
    \includegraphics[width=0.9\columnwidth]{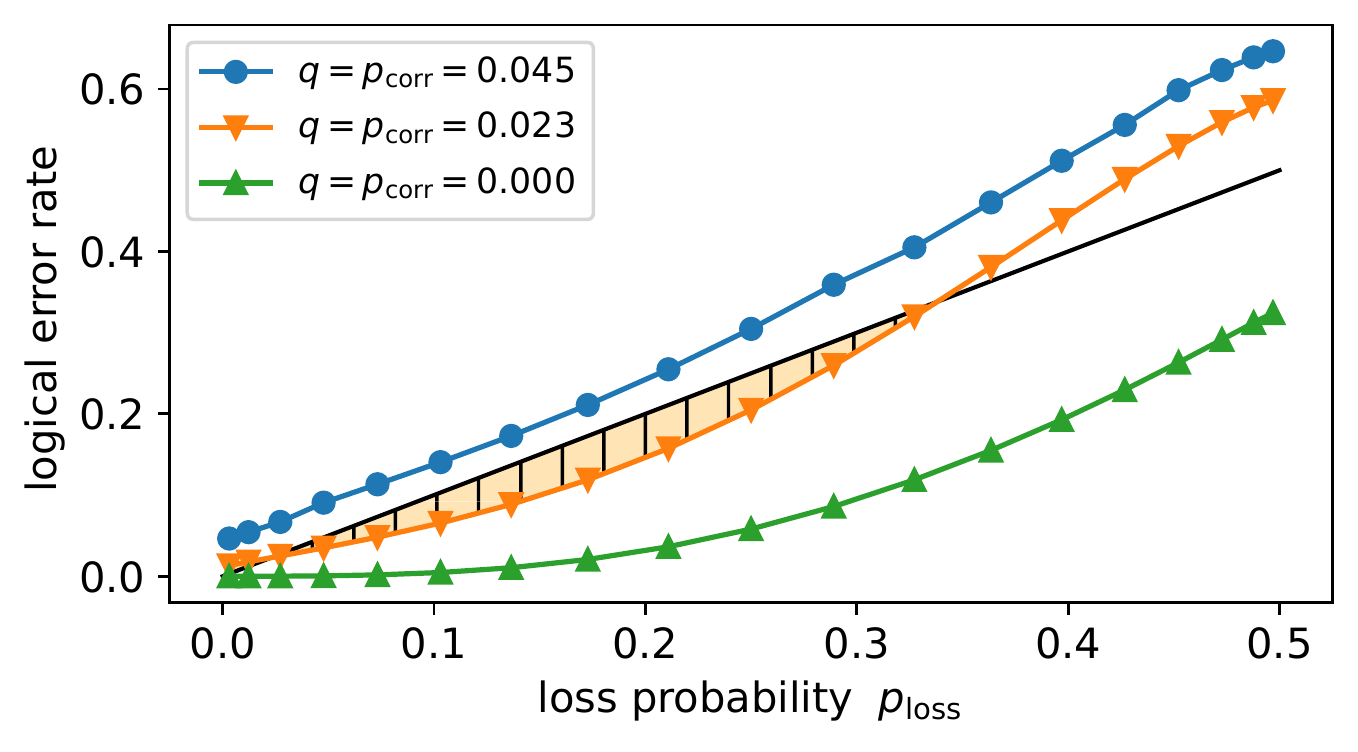}
    \caption{\textbf{Logical error rates simulated for a loss correction cycle of the 7-qubit color code with faulty stabilizer measurements}.
    The logical error rates are shown as a function of the loss probability $p_\text{loss}$ induced by the QND detection scheme of Fig.~\ref{Fig:7QubitCode} for different error rates $q$ in the stabilizer readout. The black line (with equation $1-p_\text{loss}$) represents the error rate when no encoding is performed.
    The logical error rates for the ideal case with no overrotation errors in the QND loss detection unit are shown in green $\blacktriangle$. Blue $\bullet$ show the logical error rates when the QND detection unit is simulated with overrotation parameters ($p_\text{corr} = 0.045$ resulting in a stabilizer measurement error rate $q=0.045$ and $p_\text{single} = 2.47\cdot 10^{-4}$)
    coming from the experimental data. Data simulated with $q=p_\text{corr} = 0.023$ corresponding to an improvement in the MS-gate fidelity is shown with orange $\blacktriangledown$. In the region with $0.03 \lesssim p_\text{loss} \lesssim 0.33$, error correction is beneficial in protecting the logical states with respect to storing information in an unencoded single physical qubit.}
    \label{fig_case_3_stab_errors}
\end{figure}

\begin{figure*}[ht]
    \centering
    \includegraphics[width=0.9\textwidth]{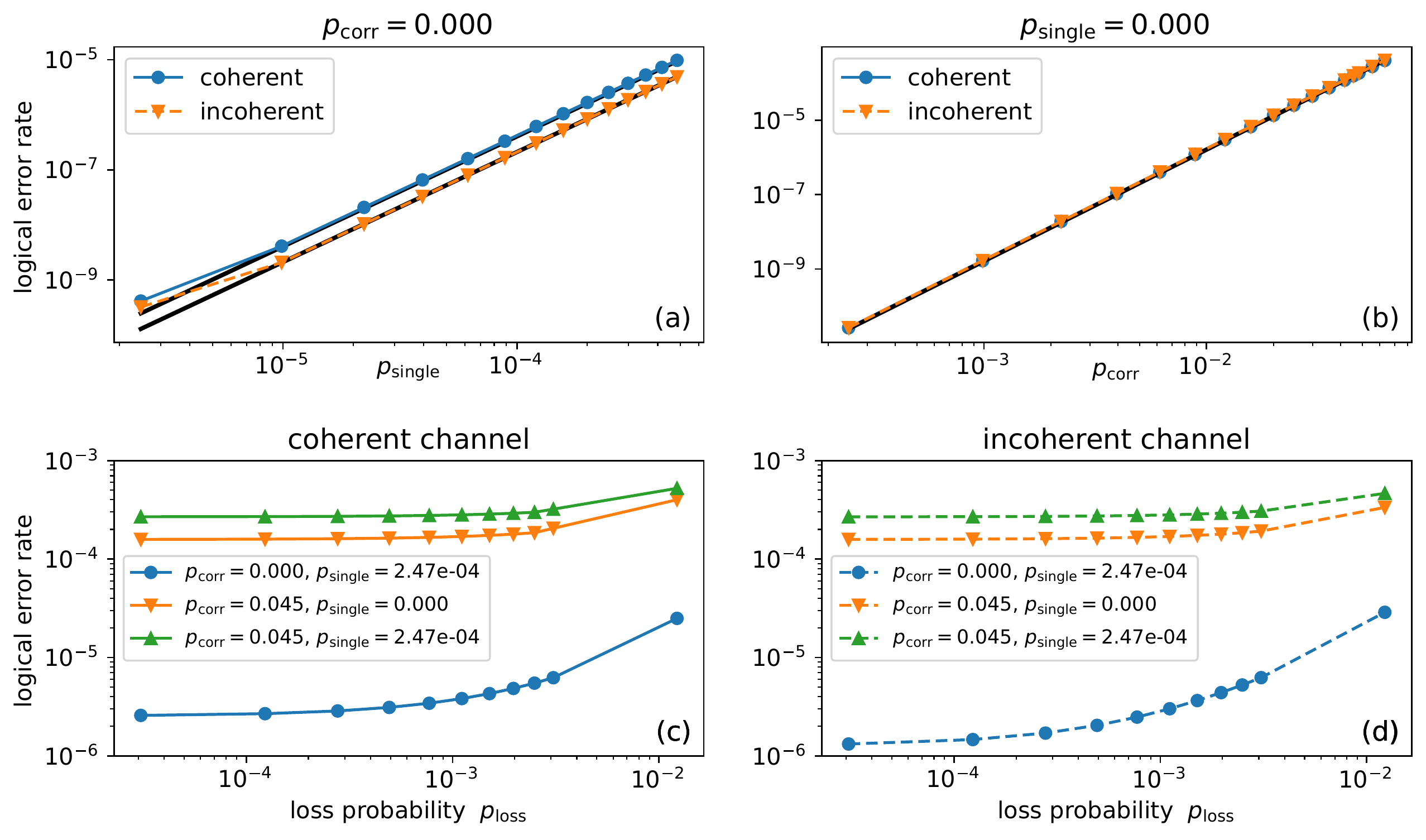}
    \caption{\textbf{Comparison between the coherent and incoherent implementations of the faulty QND loss-detection unit.}  \textbf{(a)}-\textbf{(b)}. Logical error rates  for $p_\text{loss} = 0$ as a function of  \textbf{(a)} the single qubit overrotation rate $p_\text{single}$ for  $p_\text{corr} = 0$ and \textbf{(b)}  the correlated  overrotation rate $p_\text{corr}$ for $p_\text{single} = 0$ after a round of error correction of the 7 qubit color code following the scheme in Fig.~\ref{Fig:7QubitCode} where the imperfections in the QND loss-detection unit are implemented either with a coherent or an incoherent   noise channel.  \textbf{(c)} Logical error rate as a function of the loss probability when the faulty QND loss-detection unit is modeled as a coherent channel. \textbf{(d)} same as (c), but when errors in the QND loss-detection are modeleded as an incoherent Clifford channel. }    
\label{fig_coherent_incoherent}
\end{figure*}

Figure \ref{fig_case_3_stab_errors} shows the predicted logical error rate of the  loss QEC cycle applied to all physical qubits as a function of the physical qubit loss rate~$p_\text{loss}$, for various error rates of faulty stabilizer measurements. At the current two-qubit gate infidelities and associated error rates, the regime of beneficial loss correction, when the logical error rate falls below the physical loss rate $p_\mathrm{loss}$, is not reachable. However, a moderate reduction of the two-qubit gate error rate by about 50\%, from $p_\text{corr} = 0.045$ to about $p_\text{corr} = 0.023$ suffices to enter the regime where applying a cycle of faulty loss QEC outperforms storing information in a single physical qubit that can suffer loss. 

Furthermore, Fig.~\ref{fig_coherent_incoherent}a-b shows the calculations of the logical error rate for the no-loss case $p_\text{loss}=0$, which highlights the effects resulting from imperfections in the QND loss detection unit itself in a full QEC cycle.  Here, the imperfections in the QND unit are implemented either with the coherent noise channel 
or an  effective incoherent few-parameter Clifford noise model (details on the error models are given in the Supplementary Material~\cite{supp_mat}). In Fig.~\ref{fig_coherent_incoherent}a the logical error rate is shown as a function of the single qubit overrotation rate $p_\text{single}$ for  $p_\text{corr} = 0$ and it goes to zero as $p_\text{single}^2$ (black lines) as expected, representing the rate of weight-two bit flip errors, which are uncorrectable by the distance-3 color code. In Fig.~\ref{fig_coherent_incoherent}b instead the logical error rate is shown as a function of the correlated  overrotation rate $p_\text{corr}$ for $p_\text{single} = 0$. In this case the error rate goes to zero as $p_\text{corr}^3$ (black line) representing the rate of three-bit flip errors. The bit-flip errors from the correlated overrotations result in false-positive events, where a non-lost qubit is substituted by a fresh qubit before the stabilizer measurement. Since two (detected) losses on any two qubits are correctable, some (detected) three-loss events are not, this results in the observed $p_\text{corr}^3$ scaling of the logical error rate. This highlights and explains the different sensitivity of the logical error rate to false positive and false negative events where the presence of false negative events, i.e.~overlooked losses, occurs for $p_\text{single} \neq 0$ and constitutes the more severe source of errors.

Finally, Figures~\ref{fig_coherent_incoherent}c-d show a comparison of the logical error rate for the two scenarios, where faults in the QND loss detection unit are modelled as coherent vs incoherent errors, respectively. When $p_\text{corr} \neq 0$ or $p_\text{single} \neq 0$, the logical error rate goes to a finite value when the loss probability $p_\text{loss} \to 0$ as error processes involving data qubit bit-flips arise and lead to a finite failure rate of the error correction cycle. Moreover, we observe that the incoherent approximation of the coherent error channel slightly underestimates the logical error rate, by a maximum relative factor of 0.51. This behavior is not unexpected, and has been observed also in other contexts, e.g.~for an incoherent approximation of coherent crosstalk errors \cite{Parrado2021}. Overall, the results therefore indicate the reliability of the incoherent approximation of the faulty QND loss detection unit in the QEC cycle. This is important as the latter incoherent model is efficiently simulatable and allows the study of faulty loss correction using stabilizer simulations of larger QEC codes. 

\section{Discussion \& Outlook}
\label{Sec:Outlook}
The techniques for characterizing quantum instruments that we introduced here can be applied to a wide range of scenarios beyond our example problem of qubit loss detection in QEC. Within QEC, other key operations, such as multi-qubit stabilizer measurements take a similar form where a classical outcome is used to feed-forward onto the post-measurement state, and are thus best described as quantum instruments. Characterizing such operations and producing reliable effective models will be crucial for understanding their effect onto QEC protocols. Beyond QEC, the same pattern is found in weak measurements, where a quantum system is measured in a way that minimizes the disturbance of the post-measurement state, while still extracting useful information on average. Furthermore quantum instruments are already used in quantum communication as the appropriate tool to describe the involved operations.

The presented techniques rely on tomographic reconstruction to guide the development of effective models for the studied quantum instruments. However, like any tomographic approach, such reconstructions are sensitive to SPAM errors, which can dominate in the low-error regime that is approached by state-of-the-art quantum devices. An interesting question for future research would thus be to generalize and validate SPAM-free characterization techniques such as randomized benchmarking and gate set tomography~\cite{Rudinger2021} with respect to quantum instruments with low error rates.

\newpage
\bibliographystyle{apsrev4-1new}
\bibliography{biblio}

\begin{thebibliography}{84}%
\makeatletter
\providecommand \@ifxundefined [1]{%
 \@ifx{#1\undefined}
}%
\providecommand \@ifnum [1]{%
 \ifnum #1\expandafter \@firstoftwo
 \else \expandafter \@secondoftwo
 \fi
}%
\providecommand \@ifx [1]{%
 \ifx #1\expandafter \@firstoftwo
 \else \expandafter \@secondoftwo
 \fi
}%
\providecommand \natexlab [1]{#1}%
\providecommand \enquote  [1]{``#1''}%
\providecommand \bibnamefont  [1]{#1}%
\providecommand \bibfnamefont [1]{#1}%
\providecommand \citenamefont [1]{#1}%
\providecommand \href@noop [0]{\@secondoftwo}%
\providecommand \href [0]{\begingroup \@sanitize@url \@href}%
\providecommand \@href[1]{\@@startlink{#1}\@@href}%
\providecommand \@@href[1]{\endgroup#1\@@endlink}%
\providecommand \@sanitize@url [0]{\catcode `\\12\catcode `\$12\catcode
  `\&12\catcode `\#12\catcode `\^12\catcode `\_12\catcode `\%12\relax}%
\providecommand \@@startlink[1]{}%
\providecommand \@@endlink[0]{}%
\providecommand \url  [0]{\begingroup\@sanitize@url \@url }%
\providecommand \@url [1]{\endgroup\@href {#1}{\urlprefix }}%
\providecommand \urlprefix  [0]{URL }%
\providecommand \Eprint [0]{\href }%
\providecommand \doibase [0]{http://dx.doi.org/}%
\providecommand \selectlanguage [0]{\@gobble}%
\providecommand \bibinfo  [0]{\@secondoftwo}%
\providecommand \bibfield  [0]{\@secondoftwo}%
\providecommand \translation [1]{[#1]}%
\providecommand \BibitemOpen [0]{}%
\providecommand \bibitemStop [0]{}%
\providecommand \bibitemNoStop [0]{.\EOS\space}%
\providecommand \EOS [0]{\spacefactor3000\relax}%
\providecommand \BibitemShut  [1]{\csname bibitem#1\endcsname}%
\let\auto@bib@innerbib\@empty
\bibitem [{\citenamefont {Stricker}\ \emph {et~al.}(2020)\citenamefont
  {Stricker}, \citenamefont {Vodola}, \citenamefont {Erhard}, \citenamefont
  {Postler}, \citenamefont {Meth}, \citenamefont {Ringbauer}, \citenamefont
  {Schindler}, \citenamefont {Monz}, \citenamefont {Müller},\ and\
  \citenamefont {Blatt}}]{Stricker2020}%
  \BibitemOpen
  \bibfield  {author} {\bibinfo {author} {\bibfnamefont {R.}~\bibnamefont
  {Stricker}}, \bibinfo {author} {\bibfnamefont {D.}~\bibnamefont {Vodola}},
  \bibinfo {author} {\bibfnamefont {A.}~\bibnamefont {Erhard}}, \bibinfo
  {author} {\bibfnamefont {L.}~\bibnamefont {Postler}}, \bibinfo {author}
  {\bibfnamefont {M.}~\bibnamefont {Meth}}, \bibinfo {author} {\bibfnamefont
  {M.}~\bibnamefont {Ringbauer}}, \bibinfo {author} {\bibfnamefont
  {P.}~\bibnamefont {Schindler}}, \bibinfo {author} {\bibfnamefont
  {T.}~\bibnamefont {Monz}}, \bibinfo {author} {\bibfnamefont {M.}~\bibnamefont
  {Müller}}, \ and\ \bibinfo {author} {\bibfnamefont {R.}~\bibnamefont
  {Blatt}},\ }\bibfield  {title} {\emph {\bibinfo {title} {Experimental
  deterministic correction of qubit loss},\ }}\href {\doibase
  10.1038/s41586-020-2667-0} {\bibfield  {journal} {\bibinfo  {journal}
  {Nature}\ }\textbf {\bibinfo {volume} {585}},\ \bibinfo {pages} {207}
  (\bibinfo {year} {2020})}\BibitemShut {NoStop}%
\bibitem [{\citenamefont {Davies}\ and\ \citenamefont
  {Lewis}(1970)}]{Davies1970}%
  \BibitemOpen
  \bibfield  {author} {\bibinfo {author} {\bibfnamefont {E.~B.}\ \bibnamefont
  {Davies}}\ and\ \bibinfo {author} {\bibfnamefont {J.~T.}\ \bibnamefont
  {Lewis}},\ }\bibfield  {title} {\emph {\bibinfo {title} {{An operational
  approach to quantum probability}},\ }}\href {\doibase 10.1007/BF01647093}
  {\bibfield  {journal} {\bibinfo  {journal} {Commun. Math. Phys.}\ }\textbf
  {\bibinfo {volume} {17}},\ \bibinfo {pages} {239} (\bibinfo {year}
  {1970})}\BibitemShut {NoStop}%
\bibitem [{\citenamefont {Ozawa}(1984)}]{Ozawa1984}%
  \BibitemOpen
  \bibfield  {author} {\bibinfo {author} {\bibfnamefont {M.}~\bibnamefont
  {Ozawa}},\ }\bibfield  {title} {\emph {\bibinfo {title} {{Quantum measuring
  processes of continuous observables}},\ }}\href {\doibase 10.1063/1.526000}
  {\bibfield  {journal} {\bibinfo  {journal} {J. Math. Phys.}\ }\textbf
  {\bibinfo {volume} {25}},\ \bibinfo {pages} {79} (\bibinfo {year}
  {1984})}\BibitemShut {NoStop}%
\bibitem [{\citenamefont {Dressel}\ and\ \citenamefont
  {Jordan}(2013)}]{Dressel2013}%
  \BibitemOpen
  \bibfield  {author} {\bibinfo {author} {\bibfnamefont {J.}~\bibnamefont
  {Dressel}}\ and\ \bibinfo {author} {\bibfnamefont {A.~N.}\ \bibnamefont
  {Jordan}},\ }\bibfield  {title} {\emph {\bibinfo {title} {{Quantum
  instruments as a foundation for both states and observables}},\ }}\href
  {\doibase 10.1103/PhysRevA.88.022107} {\bibfield  {journal} {\bibinfo
  {journal} {Phys. Rev. A}\ }\textbf {\bibinfo {volume} {88}},\ \bibinfo
  {pages} {022107} (\bibinfo {year} {2013})}\BibitemShut {NoStop}%
\bibitem [{\citenamefont {Chiribella}\ \emph {et~al.}(2009)\citenamefont
  {Chiribella}, \citenamefont {D'Ariano},\ and\ \citenamefont
  {Perinotti}}]{Chiribella2009}%
  \BibitemOpen
  \bibfield  {author} {\bibinfo {author} {\bibfnamefont {G.}~\bibnamefont
  {Chiribella}}, \bibinfo {author} {\bibfnamefont {G.~M.}\ \bibnamefont
  {D'Ariano}}, \ and\ \bibinfo {author} {\bibfnamefont {P.}~\bibnamefont
  {Perinotti}},\ }\bibfield  {title} {\emph {\bibinfo {title} {{Theoretical
  framework for quantum networks}},\ }}\href {\doibase
  10.1103/PhysRevA.80.022339} {\bibfield  {journal} {\bibinfo  {journal} {Phys.
  Rev. A}\ }\textbf {\bibinfo {volume} {80}},\ \bibinfo {pages} {022339}
  (\bibinfo {year} {2009})}\BibitemShut {NoStop}%
\bibitem [{\citenamefont {Oreshkov}\ \emph {et~al.}(2012)\citenamefont
  {Oreshkov}, \citenamefont {Costa},\ and\ \citenamefont
  {Brukner}}]{Oreshkov2012}%
  \BibitemOpen
  \bibfield  {author} {\bibinfo {author} {\bibfnamefont {O.}~\bibnamefont
  {Oreshkov}}, \bibinfo {author} {\bibfnamefont {F.}~\bibnamefont {Costa}}, \
  and\ \bibinfo {author} {\bibfnamefont {{\v{C}}.}~\bibnamefont {Brukner}},\
  }\bibfield  {title} {\emph {\bibinfo {title} {{Quantum correlations with no
  causal order}},\ }}\href {\doibase 10.1038/ncomms2076} {\bibfield  {journal}
  {\bibinfo  {journal} {Nat. Commun.}\ }\textbf {\bibinfo {volume} {3}},\
  \bibinfo {pages} {1092} (\bibinfo {year} {2012})}\BibitemShut {NoStop}%
\bibitem [{\citenamefont {Buscemi}\ and\ \citenamefont
  {Sacchi}(2006)}]{Buscemi2006}%
  \BibitemOpen
  \bibfield  {author} {\bibinfo {author} {\bibfnamefont {F.}~\bibnamefont
  {Buscemi}}\ and\ \bibinfo {author} {\bibfnamefont {M.~F.}\ \bibnamefont
  {Sacchi}},\ }\bibfield  {title} {\emph {\bibinfo {title}
  {{Information-disturbance trade-off in quantum-state discrimination}},\
  }}\href {\doibase 10.1103/PhysRevA.74.052320} {\bibfield  {journal} {\bibinfo
   {journal} {Phys. Rev. A}\ }\textbf {\bibinfo {volume} {74}},\ \bibinfo
  {pages} {052320} (\bibinfo {year} {2006})}\BibitemShut {NoStop}%
\bibitem [{\citenamefont {Knips}\ \emph {et~al.}(2018)\citenamefont {Knips},
  \citenamefont {Dziewior}, \citenamefont {Hashagen}, \citenamefont {Meinecke},
  \citenamefont {Weinfurter},\ and\ \citenamefont {Wolf}}]{Knips2018}%
  \BibitemOpen
  \bibfield  {author} {\bibinfo {author} {\bibfnamefont {L.}~\bibnamefont
  {Knips}}, \bibinfo {author} {\bibfnamefont {J.}~\bibnamefont {Dziewior}},
  \bibinfo {author} {\bibfnamefont {A.-L.~K.}\ \bibnamefont {Hashagen}},
  \bibinfo {author} {\bibfnamefont {J.~D.~A.}\ \bibnamefont {Meinecke}},
  \bibinfo {author} {\bibfnamefont {H.}~\bibnamefont {Weinfurter}}, \ and\
  \bibinfo {author} {\bibfnamefont {M.~M.}\ \bibnamefont {Wolf}},\ }\bibfield
  {title} {\emph {\bibinfo {title} {{Measurement-Disturbance Tradeoff
  Outperforming Optimal Cloning}},\ }}\href@noop {} {\bibfield  {journal}
  {\bibinfo  {journal} {arXiv preprint: 1808.07882}\ } (\bibinfo {year}
  {2018})}\BibitemShut {NoStop}%
\bibitem [{\citenamefont {Lloyd}\ and\ \citenamefont
  {Slotine}(2000)}]{Lloyd2000}%
  \BibitemOpen
  \bibfield  {author} {\bibinfo {author} {\bibfnamefont {S.}~\bibnamefont
  {Lloyd}}\ and\ \bibinfo {author} {\bibfnamefont {J.-J.~E.}\ \bibnamefont
  {Slotine}},\ }\bibfield  {title} {\emph {\bibinfo {title} {{Quantum feedback
  with weak measurements}},\ }}\href {\doibase 10.1103/PhysRevA.62.012307}
  {\bibfield  {journal} {\bibinfo  {journal} {Phys. Rev. A}\ }\textbf {\bibinfo
  {volume} {62}},\ \bibinfo {pages} {012307} (\bibinfo {year}
  {2000})}\BibitemShut {NoStop}%
\bibitem [{\citenamefont {Sun}\ \emph {et~al.}(2010)\citenamefont {Sun},
  \citenamefont {Al-Amri}, \citenamefont {Davidovich},\ and\ \citenamefont
  {Zubairy}}]{Sun2010}%
  \BibitemOpen
  \bibfield  {author} {\bibinfo {author} {\bibfnamefont {Q.}~\bibnamefont
  {Sun}}, \bibinfo {author} {\bibfnamefont {M.}~\bibnamefont {Al-Amri}},
  \bibinfo {author} {\bibfnamefont {L.}~\bibnamefont {Davidovich}}, \ and\
  \bibinfo {author} {\bibfnamefont {M.~S.}\ \bibnamefont {Zubairy}},\
  }\bibfield  {title} {\emph {\bibinfo {title} {{Reversing entanglement change
  by a weak measurement}},\ }}\href {\doibase 10.1103/PhysRevA.82.052323}
  {\bibfield  {journal} {\bibinfo  {journal} {Phys. Rev. A}\ }\textbf {\bibinfo
  {volume} {82}},\ \bibinfo {pages} {052323} (\bibinfo {year}
  {2010})}\BibitemShut {NoStop}%
\bibitem [{\citenamefont {Kim}\ \emph {et~al.}(2012)\citenamefont {Kim},
  \citenamefont {Lee}, \citenamefont {Kwon},\ and\ \citenamefont
  {Kim}}]{Kim2012}%
  \BibitemOpen
  \bibfield  {author} {\bibinfo {author} {\bibfnamefont {Y.-S.}\ \bibnamefont
  {Kim}}, \bibinfo {author} {\bibfnamefont {J.-C.}\ \bibnamefont {Lee}},
  \bibinfo {author} {\bibfnamefont {O.}~\bibnamefont {Kwon}}, \ and\ \bibinfo
  {author} {\bibfnamefont {Y.-H.}\ \bibnamefont {Kim}},\ }\bibfield  {title}
  {\emph {\bibinfo {title} {{Protecting entanglement from decoherence using
  weak measurement and quantum measurement reversal}},\ }}\href {\doibase
  10.1038/nphys2178} {\bibfield  {journal} {\bibinfo  {journal} {Nat. Phys.}\
  }\textbf {\bibinfo {volume} {8}},\ \bibinfo {pages} {117} (\bibinfo {year}
  {2012})}\BibitemShut {NoStop}%
\bibitem [{\citenamefont {Gottesman}(1998)}]{Gottesman1998}%
  \BibitemOpen
  \bibfield  {author} {\bibinfo {author} {\bibfnamefont {D.}~\bibnamefont
  {Gottesman}},\ }\bibfield  {title} {\emph {\bibinfo {title} {{Theory of
  fault-tolerant quantum computation}},\ }}\href {\doibase
  10.1103/PhysRevA.57.127} {\bibfield  {journal} {\bibinfo  {journal} {Phys.
  Rev. A}\ }\textbf {\bibinfo {volume} {57}},\ \bibinfo {pages} {127} (\bibinfo
  {year} {1998})}\BibitemShut {NoStop}%
\bibitem [{\citenamefont {Schindler}\ \emph {et~al.}(2011)\citenamefont
  {Schindler}, \citenamefont {Barreiro}, \citenamefont {Monz}, \citenamefont
  {Nebendahl}, \citenamefont {Nigg}, \citenamefont {Chwalla}, \citenamefont
  {Hennrich},\ and\ \citenamefont {Blatt}}]{Schindler1059}%
  \BibitemOpen
  \bibfield  {author} {\bibinfo {author} {\bibfnamefont {P.}~\bibnamefont
  {Schindler}}, \bibinfo {author} {\bibfnamefont {J.~T.}\ \bibnamefont
  {Barreiro}}, \bibinfo {author} {\bibfnamefont {T.}~\bibnamefont {Monz}},
  \bibinfo {author} {\bibfnamefont {V.}~\bibnamefont {Nebendahl}}, \bibinfo
  {author} {\bibfnamefont {D.}~\bibnamefont {Nigg}}, \bibinfo {author}
  {\bibfnamefont {M.}~\bibnamefont {Chwalla}}, \bibinfo {author} {\bibfnamefont
  {M.}~\bibnamefont {Hennrich}}, \ and\ \bibinfo {author} {\bibfnamefont
  {R.}~\bibnamefont {Blatt}},\ }\bibfield  {title} {\emph {\bibinfo {title}
  {Experimental repetitive quantum error correction},\ }}\href {\doibase
  10.1126/science.1203329} {\bibfield  {journal} {\bibinfo  {journal}
  {Science}\ }\textbf {\bibinfo {volume} {332}},\ \bibinfo {pages} {1059}
  (\bibinfo {year} {2011})}\BibitemShut {NoStop}%
\bibitem [{\citenamefont {Sun}\ \emph {et~al.}(2014)\citenamefont {Sun},
  \citenamefont {Petrenko}, \citenamefont {Leghtas}, \citenamefont {Vlastakis},
  \citenamefont {Kirchmair}, \citenamefont {Sliwa}, \citenamefont {Narla},
  \citenamefont {Hatridge}, \citenamefont {Shankar}, \citenamefont {Blumoff},
  \citenamefont {Frunzio}, \citenamefont {Mirrahimi}, \citenamefont {Devoret},\
  and\ \citenamefont {Schoelkopf}}]{sun_tracking_2014}%
  \BibitemOpen
  \bibfield  {author} {\bibinfo {author} {\bibfnamefont {L.}~\bibnamefont
  {Sun}}, \bibinfo {author} {\bibfnamefont {A.}~\bibnamefont {Petrenko}},
  \bibinfo {author} {\bibfnamefont {Z.}~\bibnamefont {Leghtas}}, \bibinfo
  {author} {\bibfnamefont {B.}~\bibnamefont {Vlastakis}}, \bibinfo {author}
  {\bibfnamefont {G.}~\bibnamefont {Kirchmair}}, \bibinfo {author}
  {\bibfnamefont {K.~M.}\ \bibnamefont {Sliwa}}, \bibinfo {author}
  {\bibfnamefont {A.}~\bibnamefont {Narla}}, \bibinfo {author} {\bibfnamefont
  {M.}~\bibnamefont {Hatridge}}, \bibinfo {author} {\bibfnamefont
  {S.}~\bibnamefont {Shankar}}, \bibinfo {author} {\bibfnamefont
  {J.}~\bibnamefont {Blumoff}}, \bibinfo {author} {\bibfnamefont
  {L.}~\bibnamefont {Frunzio}}, \bibinfo {author} {\bibfnamefont
  {M.}~\bibnamefont {Mirrahimi}}, \bibinfo {author} {\bibfnamefont {M.~H.}\
  \bibnamefont {Devoret}}, \ and\ \bibinfo {author} {\bibfnamefont {R.~J.}\
  \bibnamefont {Schoelkopf}},\ }\bibfield  {title} {\emph {\bibinfo {title}
  {Tracking photon jumps with repeated quantum non-demolition parity
  measurements},\ }}\href {\doibase 10.1038/nature13436} {\bibfield  {journal}
  {\bibinfo  {journal} {Nature}\ }\textbf {\bibinfo {volume} {511}},\ \bibinfo
  {pages} {444} (\bibinfo {year} {2014})}\BibitemShut {NoStop}%
\bibitem [{\citenamefont {Kelly}\ \emph {et~al.}(2015)\citenamefont {Kelly},
  \citenamefont {Barends}, \citenamefont {Fowler}, \citenamefont {Megrant},
  \citenamefont {Jeffrey}, \citenamefont {White}, \citenamefont {Sank},
  \citenamefont {Mutus}, \citenamefont {Campbell}, \citenamefont {Chen},
  \citenamefont {Chen}, \citenamefont {Chiaro}, \citenamefont {Dunsworth},
  \citenamefont {Hoi}, \citenamefont {Neill}, \citenamefont {O’Malley},
  \citenamefont {Quintana}, \citenamefont {Roushan}, \citenamefont
  {Vainsencher}, \citenamefont {Wenner}, \citenamefont {Cleland},\ and\
  \citenamefont {Martinis}}]{kelly_repetetive_2015}%
  \BibitemOpen
  \bibfield  {author} {\bibinfo {author} {\bibfnamefont {J.}~\bibnamefont
  {Kelly}}, \bibinfo {author} {\bibfnamefont {R.}~\bibnamefont {Barends}},
  \bibinfo {author} {\bibfnamefont {A.~G.}\ \bibnamefont {Fowler}}, \bibinfo
  {author} {\bibfnamefont {A.}~\bibnamefont {Megrant}}, \bibinfo {author}
  {\bibfnamefont {E.}~\bibnamefont {Jeffrey}}, \bibinfo {author} {\bibfnamefont
  {T.~C.}\ \bibnamefont {White}}, \bibinfo {author} {\bibfnamefont
  {D.}~\bibnamefont {Sank}}, \bibinfo {author} {\bibfnamefont {J.~Y.}\
  \bibnamefont {Mutus}}, \bibinfo {author} {\bibfnamefont {B.}~\bibnamefont
  {Campbell}}, \bibinfo {author} {\bibfnamefont {Y.}~\bibnamefont {Chen}},
  \bibinfo {author} {\bibfnamefont {Z.}~\bibnamefont {Chen}}, \bibinfo {author}
  {\bibfnamefont {B.}~\bibnamefont {Chiaro}}, \bibinfo {author} {\bibfnamefont
  {A.}~\bibnamefont {Dunsworth}}, \bibinfo {author} {\bibfnamefont {I.-C.}\
  \bibnamefont {Hoi}}, \bibinfo {author} {\bibfnamefont {C.}~\bibnamefont
  {Neill}}, \bibinfo {author} {\bibfnamefont {P.~J.~J.}\ \bibnamefont
  {O’Malley}}, \bibinfo {author} {\bibfnamefont {C.}~\bibnamefont
  {Quintana}}, \bibinfo {author} {\bibfnamefont {P.}~\bibnamefont {Roushan}},
  \bibinfo {author} {\bibfnamefont {A.}~\bibnamefont {Vainsencher}}, \bibinfo
  {author} {\bibfnamefont {J.}~\bibnamefont {Wenner}}, \bibinfo {author}
  {\bibfnamefont {A.~N.}\ \bibnamefont {Cleland}}, \ and\ \bibinfo {author}
  {\bibfnamefont {J.~M.}\ \bibnamefont {Martinis}},\ }\bibfield  {title} {\emph
  {\bibinfo {title} {State preservation by repetitive error detection in a
  superconducting quantum circuit},\ }}\href {\doibase 10.1038/nature14270}
  {\bibfield  {journal} {\bibinfo  {journal} {Nature}\ }\textbf {\bibinfo
  {volume} {519}},\ \bibinfo {pages} {66} (\bibinfo {year} {2015})}\BibitemShut
  {NoStop}%
\bibitem [{\citenamefont {Cramer}\ \emph {et~al.}(2016)\citenamefont {Cramer},
  \citenamefont {Kalb}, \citenamefont {Rol}, \citenamefont {Hensen},
  \citenamefont {Blok}, \citenamefont {Markham}, \citenamefont {Twitchen},
  \citenamefont {Hanson},\ and\ \citenamefont
  {Taminiau}}]{cramer_repeated_2016}%
  \BibitemOpen
  \bibfield  {author} {\bibinfo {author} {\bibfnamefont {J.}~\bibnamefont
  {Cramer}}, \bibinfo {author} {\bibfnamefont {N.}~\bibnamefont {Kalb}},
  \bibinfo {author} {\bibfnamefont {M.~A.}\ \bibnamefont {Rol}}, \bibinfo
  {author} {\bibfnamefont {B.}~\bibnamefont {Hensen}}, \bibinfo {author}
  {\bibfnamefont {M.~S.}\ \bibnamefont {Blok}}, \bibinfo {author}
  {\bibfnamefont {M.}~\bibnamefont {Markham}}, \bibinfo {author} {\bibfnamefont
  {D.~J.}\ \bibnamefont {Twitchen}}, \bibinfo {author} {\bibfnamefont
  {R.}~\bibnamefont {Hanson}}, \ and\ \bibinfo {author} {\bibfnamefont {T.~H.}\
  \bibnamefont {Taminiau}},\ }\bibfield  {title} {\emph {\bibinfo {title}
  {Repeated quantum error correction on a continuously encoded qubit by
  real-time feedback},\ }}\href {\doibase 10.1038/ncomms11526} {\bibfield
  {journal} {\bibinfo  {journal} {Nature Communications}\ }\textbf {\bibinfo
  {volume} {7}},\ \bibinfo {pages} {11526} (\bibinfo {year}
  {2016})}\BibitemShut {NoStop}%
\bibitem [{\citenamefont {Unden}\ \emph {et~al.}(2016)\citenamefont {Unden},
  \citenamefont {Balasubramanian}, \citenamefont {Louzon}, \citenamefont
  {Vinkler}, \citenamefont {Plenio}, \citenamefont {Markham}, \citenamefont
  {Twitchen}, \citenamefont {Stacey}, \citenamefont {Lovchinsky}, \citenamefont
  {Sushkov}, \citenamefont {Lukin}, \citenamefont {Retzker}, \citenamefont
  {Naydenov}, \citenamefont {McGuinness},\ and\ \citenamefont
  {Jelezko}}]{Jelezko2016}%
  \BibitemOpen
  \bibfield  {author} {\bibinfo {author} {\bibfnamefont {T.}~\bibnamefont
  {Unden}}, \bibinfo {author} {\bibfnamefont {P.}~\bibnamefont
  {Balasubramanian}}, \bibinfo {author} {\bibfnamefont {D.}~\bibnamefont
  {Louzon}}, \bibinfo {author} {\bibfnamefont {Y.}~\bibnamefont {Vinkler}},
  \bibinfo {author} {\bibfnamefont {M.~B.}\ \bibnamefont {Plenio}}, \bibinfo
  {author} {\bibfnamefont {M.}~\bibnamefont {Markham}}, \bibinfo {author}
  {\bibfnamefont {D.}~\bibnamefont {Twitchen}}, \bibinfo {author}
  {\bibfnamefont {A.}~\bibnamefont {Stacey}}, \bibinfo {author} {\bibfnamefont
  {I.}~\bibnamefont {Lovchinsky}}, \bibinfo {author} {\bibfnamefont {A.~O.}\
  \bibnamefont {Sushkov}}, \bibinfo {author} {\bibfnamefont {M.~D.}\
  \bibnamefont {Lukin}}, \bibinfo {author} {\bibfnamefont {A.}~\bibnamefont
  {Retzker}}, \bibinfo {author} {\bibfnamefont {B.}~\bibnamefont {Naydenov}},
  \bibinfo {author} {\bibfnamefont {L.~P.}\ \bibnamefont {McGuinness}}, \ and\
  \bibinfo {author} {\bibfnamefont {F.}~\bibnamefont {Jelezko}},\ }\bibfield
  {title} {\emph {\bibinfo {title} {Quantum metrology enhanced by repetitive
  quantum error correction},\ }}\href {\doibase 10.1103/PhysRevLett.116.230502}
  {\bibfield  {journal} {\bibinfo  {journal} {Phys. Rev. Lett.}\ }\textbf
  {\bibinfo {volume} {116}},\ \bibinfo {pages} {230502} (\bibinfo {year}
  {2016})}\BibitemShut {NoStop}%
\bibitem [{\citenamefont {Negnevitsky}\ \emph {et~al.}(2018)\citenamefont
  {Negnevitsky}, \citenamefont {Marinelli}, \citenamefont {Mehta},
  \citenamefont {Lo}, \citenamefont {Flühmann},\ and\ \citenamefont
  {Home}}]{negnevitsky_repeated_2018}%
  \BibitemOpen
  \bibfield  {author} {\bibinfo {author} {\bibfnamefont {V.}~\bibnamefont
  {Negnevitsky}}, \bibinfo {author} {\bibfnamefont {M.}~\bibnamefont
  {Marinelli}}, \bibinfo {author} {\bibfnamefont {K.~K.}\ \bibnamefont
  {Mehta}}, \bibinfo {author} {\bibfnamefont {H.-Y.}\ \bibnamefont {Lo}},
  \bibinfo {author} {\bibfnamefont {C.}~\bibnamefont {Flühmann}}, \ and\
  \bibinfo {author} {\bibfnamefont {J.~P.}\ \bibnamefont {Home}},\ }\bibfield
  {title} {\emph {\bibinfo {title} {Repeated multi-qubit readout and feedback
  with a mixed-species trapped-ion register},\ }}\href {\doibase
  10.1038/s41586-018-0668-z} {\bibfield  {journal} {\bibinfo  {journal}
  {Nature}\ }\textbf {\bibinfo {volume} {563}},\ \bibinfo {pages} {527}
  (\bibinfo {year} {2018})}\BibitemShut {NoStop}%
\bibitem [{\citenamefont {Unnikrishnan}(2015)}]{Unnikrishnan2015}%
  \BibitemOpen
  \bibfield  {author} {\bibinfo {author} {\bibfnamefont {C.}~\bibnamefont
  {Unnikrishnan}},\ }\bibfield  {title} {\emph {\bibinfo {title} {Quantum
  non-demolition measurements: Concepts, theory and practice},\ }}\href
  {\doibase 10.18520/v109/i11/2052-2060} {\bibfield  {journal} {\bibinfo
  {journal} {Current Science}\ }\textbf {\bibinfo {volume} {109}},\ \bibinfo
  {pages} {2052} (\bibinfo {year} {2015})}\BibitemShut {NoStop}%
\bibitem [{\citenamefont {Hume}\ \emph {et~al.}(2007)\citenamefont {Hume},
  \citenamefont {Rosenband},\ and\ \citenamefont {Wineland}}]{Hume2007}%
  \BibitemOpen
  \bibfield  {author} {\bibinfo {author} {\bibfnamefont {D.~B.}\ \bibnamefont
  {Hume}}, \bibinfo {author} {\bibfnamefont {T.}~\bibnamefont {Rosenband}}, \
  and\ \bibinfo {author} {\bibfnamefont {D.~J.}\ \bibnamefont {Wineland}},\
  }\bibfield  {title} {\emph {\bibinfo {title} {High-fidelity adaptive qubit
  detection through repetitive quantum nondemolition measurements},\ }}\href
  {\doibase 10.1103/PhysRevLett.99.120502} {\bibfield  {journal} {\bibinfo
  {journal} {Phys. Rev. Lett.}\ }\textbf {\bibinfo {volume} {99}},\ \bibinfo
  {pages} {120502} (\bibinfo {year} {2007})}\BibitemShut {NoStop}%
\bibitem [{\citenamefont {Sayrin}\ \emph {et~al.}(2011)\citenamefont {Sayrin},
  \citenamefont {Dotsenko}, \citenamefont {Zhou}, \citenamefont {Peaudecerf},
  \citenamefont {Rybarczyk}, \citenamefont {Gleyzes}, \citenamefont {Rouchon},
  \citenamefont {Mirrahimi}, \citenamefont {Amini}, \citenamefont {Brune},
  \citenamefont {Raimond},\ and\ \citenamefont
  {Haroche}}]{Sayrin_real-time_2011}%
  \BibitemOpen
  \bibfield  {author} {\bibinfo {author} {\bibfnamefont {C.}~\bibnamefont
  {Sayrin}}, \bibinfo {author} {\bibfnamefont {I.}~\bibnamefont {Dotsenko}},
  \bibinfo {author} {\bibfnamefont {X.}~\bibnamefont {Zhou}}, \bibinfo {author}
  {\bibfnamefont {B.}~\bibnamefont {Peaudecerf}}, \bibinfo {author}
  {\bibfnamefont {T.}~\bibnamefont {Rybarczyk}}, \bibinfo {author}
  {\bibfnamefont {S.}~\bibnamefont {Gleyzes}}, \bibinfo {author} {\bibfnamefont
  {P.}~\bibnamefont {Rouchon}}, \bibinfo {author} {\bibfnamefont
  {M.}~\bibnamefont {Mirrahimi}}, \bibinfo {author} {\bibfnamefont
  {H.}~\bibnamefont {Amini}}, \bibinfo {author} {\bibfnamefont
  {M.}~\bibnamefont {Brune}}, \bibinfo {author} {\bibfnamefont {J.-M.}\
  \bibnamefont {Raimond}}, \ and\ \bibinfo {author} {\bibfnamefont
  {S.}~\bibnamefont {Haroche}},\ }\bibfield  {title} {\emph {\bibinfo {title}
  {Real-time quantum feedback prepares and stabilizes photon number states},\
  }}\href {\doibase 10.1038/nature10376} {\bibfield  {journal} {\bibinfo
  {journal} {Nature}\ }\textbf {\bibinfo {volume} {477}},\ \bibinfo {pages}
  {73} (\bibinfo {year} {2011})}\BibitemShut {NoStop}%
\bibitem [{\citenamefont {Hatridge}\ \emph {et~al.}(2013)\citenamefont
  {Hatridge}, \citenamefont {Shankar}, \citenamefont {Mirrahimi}, \citenamefont
  {Schackert}, \citenamefont {Geerlings}, \citenamefont {Brecht}, \citenamefont
  {Sliwa}, \citenamefont {Abdo}, \citenamefont {Frunzio}, \citenamefont
  {Girvin}, \citenamefont {Schoelkopf},\ and\ \citenamefont
  {Devoret}}]{Hatridge2013}%
  \BibitemOpen
  \bibfield  {author} {\bibinfo {author} {\bibfnamefont {M.}~\bibnamefont
  {Hatridge}}, \bibinfo {author} {\bibfnamefont {S.}~\bibnamefont {Shankar}},
  \bibinfo {author} {\bibfnamefont {M.}~\bibnamefont {Mirrahimi}}, \bibinfo
  {author} {\bibfnamefont {F.}~\bibnamefont {Schackert}}, \bibinfo {author}
  {\bibfnamefont {K.}~\bibnamefont {Geerlings}}, \bibinfo {author}
  {\bibfnamefont {T.}~\bibnamefont {Brecht}}, \bibinfo {author} {\bibfnamefont
  {K.~M.}\ \bibnamefont {Sliwa}}, \bibinfo {author} {\bibfnamefont
  {B.}~\bibnamefont {Abdo}}, \bibinfo {author} {\bibfnamefont {L.}~\bibnamefont
  {Frunzio}}, \bibinfo {author} {\bibfnamefont {S.~M.}\ \bibnamefont {Girvin}},
  \bibinfo {author} {\bibfnamefont {R.~J.}\ \bibnamefont {Schoelkopf}}, \ and\
  \bibinfo {author} {\bibfnamefont {M.~H.}\ \bibnamefont {Devoret}},\
  }\bibfield  {title} {\emph {\bibinfo {title} {Quantum back-action of an
  individual variable-strength measurement},\ }}\href {\doibase
  10.1126/science.1226897} {\bibfield  {journal} {\bibinfo  {journal}
  {Science}\ }\textbf {\bibinfo {volume} {339}},\ \bibinfo {pages} {178}
  (\bibinfo {year} {2013})}\BibitemShut {NoStop}%
\bibitem [{\citenamefont {Blok}\ \emph {et~al.}(2014)\citenamefont {Blok},
  \citenamefont {Bonato}, \citenamefont {Markham}, \citenamefont {Twitchen},
  \citenamefont {Dobrovitski},\ and\ \citenamefont
  {Hanson}}]{Blok_backaction_2014}%
  \BibitemOpen
  \bibfield  {author} {\bibinfo {author} {\bibfnamefont {M.~S.}\ \bibnamefont
  {Blok}}, \bibinfo {author} {\bibfnamefont {C.}~\bibnamefont {Bonato}},
  \bibinfo {author} {\bibfnamefont {M.~L.}\ \bibnamefont {Markham}}, \bibinfo
  {author} {\bibfnamefont {D.~J.}\ \bibnamefont {Twitchen}}, \bibinfo {author}
  {\bibfnamefont {V.~V.}\ \bibnamefont {Dobrovitski}}, \ and\ \bibinfo {author}
  {\bibfnamefont {R.}~\bibnamefont {Hanson}},\ }\bibfield  {title} {\emph
  {\bibinfo {title} {Manipulating a qubit through the backaction of sequential
  partial measurements and real-time feedback},\ }}\href {\doibase
  10.1038/nphys2881} {\bibfield  {journal} {\bibinfo  {journal} {Nature
  Physics}\ }\textbf {\bibinfo {volume} {10}},\ \bibinfo {pages} {189}
  (\bibinfo {year} {2014})}\BibitemShut {NoStop}%
\bibitem [{\citenamefont {Rudinger}\ \emph {et~al.}(2021)\citenamefont
  {Rudinger}, \citenamefont {Ribeill}, \citenamefont {Govia}, \citenamefont
  {Ware}, \citenamefont {Nielsen}, \citenamefont {Young}, \citenamefont {Ohki},
  \citenamefont {Blume-Kohout},\ and\ \citenamefont {Proctor}}]{Rudinger2021}%
  \BibitemOpen
  \bibfield  {author} {\bibinfo {author} {\bibfnamefont {K.}~\bibnamefont
  {Rudinger}}, \bibinfo {author} {\bibfnamefont {G.}~\bibnamefont {Ribeill}},
  \bibinfo {author} {\bibfnamefont {L.}~\bibnamefont {Govia}}, \bibinfo
  {author} {\bibfnamefont {M.}~\bibnamefont {Ware}}, \bibinfo {author}
  {\bibfnamefont {E.}~\bibnamefont {Nielsen}}, \bibinfo {author} {\bibfnamefont
  {K.}~\bibnamefont {Young}}, \bibinfo {author} {\bibfnamefont
  {T.}~\bibnamefont {Ohki}}, \bibinfo {author} {\bibfnamefont {R.}~\bibnamefont
  {Blume-Kohout}}, \ and\ \bibinfo {author} {\bibfnamefont {T.}~\bibnamefont
  {Proctor}},\ }\bibfield  {title} {\emph {\bibinfo {title} {Characterizing
  mid-circuit measurements on a superconducting qubit using gate set
  tomography},\ }}\href@noop {} {\  (\bibinfo {year} {2021})}\BibitemShut
  {NoStop}%
\bibitem [{\citenamefont {Breuer}\ and\ \citenamefont
  {Petruccione}(2006)}]{Breuer2006}%
  \BibitemOpen
  \bibfield  {author} {\bibinfo {author} {\bibfnamefont {H.-P.}\ \bibnamefont
  {Breuer}}\ and\ \bibinfo {author} {\bibfnamefont {F.}~\bibnamefont
  {Petruccione}},\ }\href {\doibase 10.1093/acprof:oso/9780199213900.001.0001}
  {\emph {\bibinfo {title} {The Theory of Open Quantum Systems}}}\ (\bibinfo
  {publisher} {Oxford University Press},\ \bibinfo {year} {2006})\BibitemShut
  {NoStop}%
\bibitem [{\citenamefont {Ringbauer}\ \emph {et~al.}(2015)\citenamefont
  {Ringbauer}, \citenamefont {Wood}, \citenamefont {Modi}, \citenamefont
  {Gilchrist}, \citenamefont {White},\ and\ \citenamefont
  {Fedrizzi}}]{Ringbauer2015}%
  \BibitemOpen
  \bibfield  {author} {\bibinfo {author} {\bibfnamefont {M.}~\bibnamefont
  {Ringbauer}}, \bibinfo {author} {\bibfnamefont {C.~J.}\ \bibnamefont {Wood}},
  \bibinfo {author} {\bibfnamefont {K.}~\bibnamefont {Modi}}, \bibinfo {author}
  {\bibfnamefont {A.}~\bibnamefont {Gilchrist}}, \bibinfo {author}
  {\bibfnamefont {A.~G.}\ \bibnamefont {White}}, \ and\ \bibinfo {author}
  {\bibfnamefont {A.}~\bibnamefont {Fedrizzi}},\ }\bibfield  {title} {\emph
  {\bibinfo {title} {{Characterizing Quantum Dynamics with Initial
  System-Environment Correlations}},\ }}\href {\doibase
  10.1103/PhysRevLett.114.090402} {\bibfield  {journal} {\bibinfo  {journal}
  {Phys. Rev. Lett.}\ }\textbf {\bibinfo {volume} {114}},\ \bibinfo {pages}
  {090402} (\bibinfo {year} {2015})}\BibitemShut {NoStop}%
\bibitem [{\citenamefont {Rotter}\ and\ \citenamefont
  {Bird}(2015)}]{Rotter2015}%
  \BibitemOpen
  \bibfield  {author} {\bibinfo {author} {\bibfnamefont {I.}~\bibnamefont
  {Rotter}}\ and\ \bibinfo {author} {\bibfnamefont {J.~P.}\ \bibnamefont
  {Bird}},\ }\bibfield  {title} {\emph {\bibinfo {title} {A review of progress
  in the physics of open quantum systems: theory and experiment},\ }}\href
  {\doibase 10.1088/0034-4885/78/11/114001} {\bibfield  {journal} {\bibinfo
  {journal} {Reports on Progress in Physics}\ }\textbf {\bibinfo {volume}
  {78}},\ \bibinfo {pages} {114001} (\bibinfo {year} {2015})}\BibitemShut
  {NoStop}%
\bibitem [{\citenamefont {Barreiro}\ \emph {et~al.}(2011)\citenamefont
  {Barreiro}, \citenamefont {Müller}, \citenamefont {Schindler}, \citenamefont
  {Nigg}, \citenamefont {Monz}, \citenamefont {Chwalla}, \citenamefont
  {Hennrich}, \citenamefont {Roos}, \citenamefont {Zoller},\ and\ \citenamefont
  {Blatt}}]{barreiro_open-system_2011}%
  \BibitemOpen
  \bibfield  {author} {\bibinfo {author} {\bibfnamefont {J.~T.}\ \bibnamefont
  {Barreiro}}, \bibinfo {author} {\bibfnamefont {M.}~\bibnamefont {Müller}},
  \bibinfo {author} {\bibfnamefont {P.}~\bibnamefont {Schindler}}, \bibinfo
  {author} {\bibfnamefont {D.}~\bibnamefont {Nigg}}, \bibinfo {author}
  {\bibfnamefont {T.}~\bibnamefont {Monz}}, \bibinfo {author} {\bibfnamefont
  {M.}~\bibnamefont {Chwalla}}, \bibinfo {author} {\bibfnamefont
  {M.}~\bibnamefont {Hennrich}}, \bibinfo {author} {\bibfnamefont {C.~F.}\
  \bibnamefont {Roos}}, \bibinfo {author} {\bibfnamefont {P.}~\bibnamefont
  {Zoller}}, \ and\ \bibinfo {author} {\bibfnamefont {R.}~\bibnamefont
  {Blatt}},\ }\bibfield  {title} {\emph {\bibinfo {title} {An open-system
  quantum simulator with trapped ions},\ }}\href {\doibase 10.1038/nature09801}
  {\bibfield  {journal} {\bibinfo  {journal} {Nature}\ }\textbf {\bibinfo
  {volume} {470}},\ \bibinfo {pages} {486} (\bibinfo {year}
  {2011})}\BibitemShut {NoStop}%
\bibitem [{\citenamefont {Schindler}\ \emph
  {et~al.}(2013{\natexlab{a}})\citenamefont {Schindler}, \citenamefont
  {Müller}, \citenamefont {Nigg}, \citenamefont {Barreiro}, \citenamefont
  {Martinez}, \citenamefont {Hennrich}, \citenamefont {Monz}, \citenamefont
  {Diehl}, \citenamefont {Zoller},\ and\ \citenamefont
  {Blatt}}]{schindler_quantum_2013}%
  \BibitemOpen
  \bibfield  {author} {\bibinfo {author} {\bibfnamefont {P.}~\bibnamefont
  {Schindler}}, \bibinfo {author} {\bibfnamefont {M.}~\bibnamefont {Müller}},
  \bibinfo {author} {\bibfnamefont {D.}~\bibnamefont {Nigg}}, \bibinfo {author}
  {\bibfnamefont {J.~T.}\ \bibnamefont {Barreiro}}, \bibinfo {author}
  {\bibfnamefont {E.~A.}\ \bibnamefont {Martinez}}, \bibinfo {author}
  {\bibfnamefont {M.}~\bibnamefont {Hennrich}}, \bibinfo {author}
  {\bibfnamefont {T.}~\bibnamefont {Monz}}, \bibinfo {author} {\bibfnamefont
  {S.}~\bibnamefont {Diehl}}, \bibinfo {author} {\bibfnamefont
  {P.}~\bibnamefont {Zoller}}, \ and\ \bibinfo {author} {\bibfnamefont
  {R.}~\bibnamefont {Blatt}},\ }\bibfield  {title} {\emph {\bibinfo {title}
  {Quantum simulation of dynamical maps with trapped ions},\ }}\href {\doibase
  10.1038/nphys2630} {\bibfield  {journal} {\bibinfo  {journal} {Nature
  Physics}\ }\textbf {\bibinfo {volume} {9}},\ \bibinfo {pages} {361} (\bibinfo
  {year} {2013}{\natexlab{a}})}\BibitemShut {NoStop}%
\bibitem [{\citenamefont {Müller}\ \emph {et~al.}(2011)\citenamefont
  {Müller}, \citenamefont {Hammerer}, \citenamefont {Zhou}, \citenamefont
  {Roos},\ and\ \citenamefont {Zoller}}]{Mueller_2011}%
  \BibitemOpen
  \bibfield  {author} {\bibinfo {author} {\bibfnamefont {M.}~\bibnamefont
  {Müller}}, \bibinfo {author} {\bibfnamefont {K.}~\bibnamefont {Hammerer}},
  \bibinfo {author} {\bibfnamefont {Y.~L.}\ \bibnamefont {Zhou}}, \bibinfo
  {author} {\bibfnamefont {C.~F.}\ \bibnamefont {Roos}}, \ and\ \bibinfo
  {author} {\bibfnamefont {P.}~\bibnamefont {Zoller}},\ }\bibfield  {title}
  {\emph {\bibinfo {title} {Simulating open quantum systems: from many-body
  interactions to stabilizer pumping},\ }}\href {\doibase
  10.1088/1367-2630/13/8/085007} {\bibfield  {journal} {\bibinfo  {journal}
  {New Journal of Physics}\ }\textbf {\bibinfo {volume} {13}},\ \bibinfo
  {pages} {085007} (\bibinfo {year} {2011})}\BibitemShut {NoStop}%
\bibitem [{\citenamefont {Sieberer}\ \emph {et~al.}(2016)\citenamefont
  {Sieberer}, \citenamefont {Buchhold},\ and\ \citenamefont
  {Diehl}}]{Sieberer_2016}%
  \BibitemOpen
  \bibfield  {author} {\bibinfo {author} {\bibfnamefont {L.~M.}\ \bibnamefont
  {Sieberer}}, \bibinfo {author} {\bibfnamefont {M.}~\bibnamefont {Buchhold}},
  \ and\ \bibinfo {author} {\bibfnamefont {S.}~\bibnamefont {Diehl}},\
  }\bibfield  {title} {\emph {\bibinfo {title} {Keldysh field theory for driven
  open quantum systems},\ }}\href {\doibase 10.1088/0034-4885/79/9/096001}
  {\bibfield  {journal} {\bibinfo  {journal} {Reports on Progress in Physics}\
  }\textbf {\bibinfo {volume} {79}},\ \bibinfo {pages} {096001} (\bibinfo
  {year} {2016})}\BibitemShut {NoStop}%
\bibitem [{\citenamefont {Wagner}\ \emph {et~al.}(2020)\citenamefont {Wagner},
  \citenamefont {Bancal}, \citenamefont {Sangouard},\ and\ \citenamefont
  {Sekatski}}]{Wagner2020}%
  \BibitemOpen
  \bibfield  {author} {\bibinfo {author} {\bibfnamefont {S.}~\bibnamefont
  {Wagner}}, \bibinfo {author} {\bibfnamefont {J.-D.}\ \bibnamefont {Bancal}},
  \bibinfo {author} {\bibfnamefont {N.}~\bibnamefont {Sangouard}}, \ and\
  \bibinfo {author} {\bibfnamefont {P.}~\bibnamefont {Sekatski}},\ }\bibfield
  {title} {\emph {\bibinfo {title} {{Device-independent characterization of
  quantum instruments}},\ }}\href {\doibase 10.22331/q-2020-03-19-243}
  {\bibfield  {journal} {\bibinfo  {journal} {Quantum}\ }\textbf {\bibinfo
  {volume} {4}},\ \bibinfo {pages} {243} (\bibinfo {year} {2020})}\BibitemShut
  {NoStop}%
\bibitem [{\citenamefont {Miklin}\ \emph {et~al.}(2019)\citenamefont {Miklin},
  \citenamefont {Borka{\l}a},\ and\ \citenamefont {Paw{\l}owski}}]{Miklin2019}%
  \BibitemOpen
  \bibfield  {author} {\bibinfo {author} {\bibfnamefont {N.}~\bibnamefont
  {Miklin}}, \bibinfo {author} {\bibfnamefont {J.~J.}\ \bibnamefont
  {Borka{\l}a}}, \ and\ \bibinfo {author} {\bibfnamefont {M.}~\bibnamefont
  {Paw{\l}owski}},\ }\bibfield  {title} {\emph {\bibinfo {title} {{Self-testing
  of unsharp measurements}},\ }}\href@noop {} {\  (\bibinfo {year}
  {2019})}\BibitemShut {NoStop}%
\bibitem [{\citenamefont {Mohan}\ \emph {et~al.}(2019)\citenamefont {Mohan},
  \citenamefont {Tavakoli},\ and\ \citenamefont {Brunner}}]{Mohan2019}%
  \BibitemOpen
  \bibfield  {author} {\bibinfo {author} {\bibfnamefont {K.}~\bibnamefont
  {Mohan}}, \bibinfo {author} {\bibfnamefont {A.}~\bibnamefont {Tavakoli}}, \
  and\ \bibinfo {author} {\bibfnamefont {N.}~\bibnamefont {Brunner}},\
  }\bibfield  {title} {\emph {\bibinfo {title} {{Sequential random access codes
  and self-testing of quantum measurement instruments}},\ }}\href {\doibase
  10.1088/1367-2630/ab3773} {\bibfield  {journal} {\bibinfo  {journal} {New J.
  Phys.}\ }\textbf {\bibinfo {volume} {21}},\ \bibinfo {pages} {083034}
  (\bibinfo {year} {2019})}\BibitemShut {NoStop}%
\bibitem [{\citenamefont {G{\'{o}}mez}\ \emph {et~al.}(2016)\citenamefont
  {G{\'{o}}mez}, \citenamefont {G{\'{o}}mez}, \citenamefont {Gonz{\'{a}}lez},
  \citenamefont {Ca{\~{n}}as}, \citenamefont {Barra}, \citenamefont {Delgado},
  \citenamefont {Xavier}, \citenamefont {Cabello}, \citenamefont {Kleinmann},
  \citenamefont {V{\'{e}}rtesi},\ and\ \citenamefont {Lima}}]{Gomez2016}%
  \BibitemOpen
  \bibfield  {author} {\bibinfo {author} {\bibfnamefont {E.~S.}\ \bibnamefont
  {G{\'{o}}mez}}, \bibinfo {author} {\bibfnamefont {S.}~\bibnamefont
  {G{\'{o}}mez}}, \bibinfo {author} {\bibfnamefont {P.}~\bibnamefont
  {Gonz{\'{a}}lez}}, \bibinfo {author} {\bibfnamefont {G.}~\bibnamefont
  {Ca{\~{n}}as}}, \bibinfo {author} {\bibfnamefont {J.~F.}\ \bibnamefont
  {Barra}}, \bibinfo {author} {\bibfnamefont {A.}~\bibnamefont {Delgado}},
  \bibinfo {author} {\bibfnamefont {G.~B.}\ \bibnamefont {Xavier}}, \bibinfo
  {author} {\bibfnamefont {A.}~\bibnamefont {Cabello}}, \bibinfo {author}
  {\bibfnamefont {M.}~\bibnamefont {Kleinmann}}, \bibinfo {author}
  {\bibfnamefont {T.}~\bibnamefont {V{\'{e}}rtesi}}, \ and\ \bibinfo {author}
  {\bibfnamefont {G.}~\bibnamefont {Lima}},\ }\bibfield  {title} {\emph
  {\bibinfo {title} {{Device-Independent Certification of a Nonprojective Qubit
  Measurement}},\ }}\href {\doibase 10.1103/PhysRevLett.117.260401} {\bibfield
  {journal} {\bibinfo  {journal} {Phys. Rev. Lett.}\ }\textbf {\bibinfo
  {volume} {117}},\ \bibinfo {pages} {260401} (\bibinfo {year}
  {2016})}\BibitemShut {NoStop}%
\bibitem [{\citenamefont {Smania}\ \emph {et~al.}(2020)\citenamefont {Smania},
  \citenamefont {Mironowicz}, \citenamefont {Nawareg}, \citenamefont
  {Paw{\l}owski}, \citenamefont {Cabello},\ and\ \citenamefont
  {Bourennane}}]{Smania2020}%
  \BibitemOpen
  \bibfield  {author} {\bibinfo {author} {\bibfnamefont {M.}~\bibnamefont
  {Smania}}, \bibinfo {author} {\bibfnamefont {P.}~\bibnamefont {Mironowicz}},
  \bibinfo {author} {\bibfnamefont {M.}~\bibnamefont {Nawareg}}, \bibinfo
  {author} {\bibfnamefont {M.}~\bibnamefont {Paw{\l}owski}}, \bibinfo {author}
  {\bibfnamefont {A.}~\bibnamefont {Cabello}}, \ and\ \bibinfo {author}
  {\bibfnamefont {M.}~\bibnamefont {Bourennane}},\ }\bibfield  {title} {\emph
  {\bibinfo {title} {{Experimental certification of an informationally complete
  quantum measurement in a device-independent protocol}},\ }}\href {\doibase
  10.1364/OPTICA.377959} {\bibfield  {journal} {\bibinfo  {journal} {Optica}\
  }\textbf {\bibinfo {volume} {7}},\ \bibinfo {pages} {123} (\bibinfo {year}
  {2020})}\BibitemShut {NoStop}%
\bibitem [{\citenamefont {Helsen}\ \emph {et~al.}(2020)\citenamefont {Helsen},
  \citenamefont {Roth}, \citenamefont {Onorati}, \citenamefont {Werner},\ and\
  \citenamefont {Eisert}}]{helsen2020RB}%
  \BibitemOpen
  \bibfield  {author} {\bibinfo {author} {\bibfnamefont {J.}~\bibnamefont
  {Helsen}}, \bibinfo {author} {\bibfnamefont {I.}~\bibnamefont {Roth}},
  \bibinfo {author} {\bibfnamefont {E.}~\bibnamefont {Onorati}}, \bibinfo
  {author} {\bibfnamefont {A.~H.}\ \bibnamefont {Werner}}, \ and\ \bibinfo
  {author} {\bibfnamefont {J.}~\bibnamefont {Eisert}},\ }\href@noop {}
  {\bibinfo {title} {A general framework for randomized benchmarking},\ }
  (\bibinfo {year} {2020}),\ \Eprint {http://arxiv.org/abs/2010.07974}
  {arXiv:2010.07974} \BibitemShut {NoStop}%
\bibitem [{\citenamefont {Knill}\ \emph {et~al.}(2008)\citenamefont {Knill},
  \citenamefont {Leibfried}, \citenamefont {Reichle}, \citenamefont {Britton},
  \citenamefont {Blakestad}, \citenamefont {Jost}, \citenamefont {Langer},
  \citenamefont {Ozeri}, \citenamefont {Seidelin},\ and\ \citenamefont
  {Wineland}}]{Wineland2008}%
  \BibitemOpen
  \bibfield  {author} {\bibinfo {author} {\bibfnamefont {E.}~\bibnamefont
  {Knill}}, \bibinfo {author} {\bibfnamefont {D.}~\bibnamefont {Leibfried}},
  \bibinfo {author} {\bibfnamefont {R.}~\bibnamefont {Reichle}}, \bibinfo
  {author} {\bibfnamefont {J.}~\bibnamefont {Britton}}, \bibinfo {author}
  {\bibfnamefont {R.~B.}\ \bibnamefont {Blakestad}}, \bibinfo {author}
  {\bibfnamefont {J.~D.}\ \bibnamefont {Jost}}, \bibinfo {author}
  {\bibfnamefont {C.}~\bibnamefont {Langer}}, \bibinfo {author} {\bibfnamefont
  {R.}~\bibnamefont {Ozeri}}, \bibinfo {author} {\bibfnamefont
  {S.}~\bibnamefont {Seidelin}}, \ and\ \bibinfo {author} {\bibfnamefont
  {D.~J.}\ \bibnamefont {Wineland}},\ }\bibfield  {title} {\emph {\bibinfo
  {title} {Randomized benchmarking of quantum gates},\ }}\href {\doibase
  10.1103/PhysRevA.77.012307} {\bibfield  {journal} {\bibinfo  {journal} {Phys.
  Rev. A}\ }\textbf {\bibinfo {volume} {77}},\ \bibinfo {pages} {012307}
  (\bibinfo {year} {2008})}\BibitemShut {NoStop}%
\bibitem [{\citenamefont {Gaebler}\ \emph {et~al.}(2016)\citenamefont
  {Gaebler}, \citenamefont {Tan}, \citenamefont {Lin}, \citenamefont {Wan},
  \citenamefont {Bowler}, \citenamefont {Keith}, \citenamefont {Glancy},
  \citenamefont {Coakley}, \citenamefont {Knill}, \citenamefont {Leibfried},\
  and\ \citenamefont {Wineland}}]{Gaebler2016}%
  \BibitemOpen
  \bibfield  {author} {\bibinfo {author} {\bibfnamefont {J.~P.}\ \bibnamefont
  {Gaebler}}, \bibinfo {author} {\bibfnamefont {T.~R.}\ \bibnamefont {Tan}},
  \bibinfo {author} {\bibfnamefont {Y.}~\bibnamefont {Lin}}, \bibinfo {author}
  {\bibfnamefont {Y.}~\bibnamefont {Wan}}, \bibinfo {author} {\bibfnamefont
  {R.}~\bibnamefont {Bowler}}, \bibinfo {author} {\bibfnamefont {A.~C.}\
  \bibnamefont {Keith}}, \bibinfo {author} {\bibfnamefont {S.}~\bibnamefont
  {Glancy}}, \bibinfo {author} {\bibfnamefont {K.}~\bibnamefont {Coakley}},
  \bibinfo {author} {\bibfnamefont {E.}~\bibnamefont {Knill}}, \bibinfo
  {author} {\bibfnamefont {D.}~\bibnamefont {Leibfried}}, \ and\ \bibinfo
  {author} {\bibfnamefont {D.~J.}\ \bibnamefont {Wineland}},\ }\bibfield
  {title} {\emph {\bibinfo {title} {High-fidelity universal gate set for
  ${^{9}\mathrm{Be}}^{+}$ ion qubits},\ }}\href {\doibase
  10.1103/PhysRevLett.117.060505} {\bibfield  {journal} {\bibinfo  {journal}
  {Phys. Rev. Lett.}\ }\textbf {\bibinfo {volume} {117}},\ \bibinfo {pages}
  {060505} (\bibinfo {year} {2016})}\BibitemShut {NoStop}%
\bibitem [{\citenamefont {Blume-Kohout}\ \emph {et~al.}(2017)\citenamefont
  {Blume-Kohout}, \citenamefont {Gamble}, \citenamefont {Nielsen},
  \citenamefont {Rudinger}, \citenamefont {Mizrahi}, \citenamefont {Fortier},\
  and\ \citenamefont {Maunz}}]{blume-kohout_gst_2017}%
  \BibitemOpen
  \bibfield  {author} {\bibinfo {author} {\bibfnamefont {R.}~\bibnamefont
  {Blume-Kohout}}, \bibinfo {author} {\bibfnamefont {J.~K.}\ \bibnamefont
  {Gamble}}, \bibinfo {author} {\bibfnamefont {E.}~\bibnamefont {Nielsen}},
  \bibinfo {author} {\bibfnamefont {K.}~\bibnamefont {Rudinger}}, \bibinfo
  {author} {\bibfnamefont {J.}~\bibnamefont {Mizrahi}}, \bibinfo {author}
  {\bibfnamefont {K.}~\bibnamefont {Fortier}}, \ and\ \bibinfo {author}
  {\bibfnamefont {P.}~\bibnamefont {Maunz}},\ }\bibfield  {title} {\emph
  {\bibinfo {title} {Demonstration of qubit operations below a rigorous fault
  tolerance threshold with gate set tomography},\ }}\href {\doibase
  10.1038/ncomms14485} {\bibfield  {journal} {\bibinfo  {journal} {Nature
  Communications}\ }\textbf {\bibinfo {volume} {8}},\ \bibinfo {pages} {14485}
  (\bibinfo {year} {2017})}\BibitemShut {NoStop}%
\bibitem [{\citenamefont {Erhard}\ \emph {et~al.}(2019)\citenamefont {Erhard},
  \citenamefont {Wallman}, \citenamefont {Postler}, \citenamefont {Meth},
  \citenamefont {Stricker}, \citenamefont {Martinez}, \citenamefont
  {Schindler}, \citenamefont {Monz}, \citenamefont {Emerson},\ and\
  \citenamefont {Blatt}}]{erhard_characterizing_2019}%
  \BibitemOpen
  \bibfield  {author} {\bibinfo {author} {\bibfnamefont {A.}~\bibnamefont
  {Erhard}}, \bibinfo {author} {\bibfnamefont {J.~J.}\ \bibnamefont {Wallman}},
  \bibinfo {author} {\bibfnamefont {L.}~\bibnamefont {Postler}}, \bibinfo
  {author} {\bibfnamefont {M.}~\bibnamefont {Meth}}, \bibinfo {author}
  {\bibfnamefont {R.}~\bibnamefont {Stricker}}, \bibinfo {author}
  {\bibfnamefont {E.~A.}\ \bibnamefont {Martinez}}, \bibinfo {author}
  {\bibfnamefont {P.}~\bibnamefont {Schindler}}, \bibinfo {author}
  {\bibfnamefont {T.}~\bibnamefont {Monz}}, \bibinfo {author} {\bibfnamefont
  {J.}~\bibnamefont {Emerson}}, \ and\ \bibinfo {author} {\bibfnamefont
  {R.}~\bibnamefont {Blatt}},\ }\bibfield  {title} {\emph {\bibinfo {title}
  {Characterizing large-scale quantum computers via cycle benchmarking},\
  }}\href {\doibase 10.1038/s41467-019-13068-7} {\bibfield  {journal} {\bibinfo
   {journal} {Nature Communications}\ }\textbf {\bibinfo {volume} {10}},\
  \bibinfo {pages} {5347} (\bibinfo {year} {2019})}\BibitemShut {NoStop}%
\bibitem [{\citenamefont {McKay}\ \emph {et~al.}(2019)\citenamefont {McKay},
  \citenamefont {Sheldon}, \citenamefont {Smolin}, \citenamefont {Chow},\ and\
  \citenamefont {Gambetta}}]{MCKay2019}%
  \BibitemOpen
  \bibfield  {author} {\bibinfo {author} {\bibfnamefont {D.~C.}\ \bibnamefont
  {McKay}}, \bibinfo {author} {\bibfnamefont {S.}~\bibnamefont {Sheldon}},
  \bibinfo {author} {\bibfnamefont {J.~A.}\ \bibnamefont {Smolin}}, \bibinfo
  {author} {\bibfnamefont {J.~M.}\ \bibnamefont {Chow}}, \ and\ \bibinfo
  {author} {\bibfnamefont {J.~M.}\ \bibnamefont {Gambetta}},\ }\bibfield
  {title} {\emph {\bibinfo {title} {Three-qubit randomized benchmarking},\
  }}\href {\doibase 10.1103/PhysRevLett.122.200502} {\bibfield  {journal}
  {\bibinfo  {journal} {Phys. Rev. Lett.}\ }\textbf {\bibinfo {volume} {122}},\
  \bibinfo {pages} {200502} (\bibinfo {year} {2019})}\BibitemShut {NoStop}%
\bibitem [{\citenamefont {Hughes}\ \emph {et~al.}(2020)\citenamefont {Hughes},
  \citenamefont {Sch\"afer}, \citenamefont {Thirumalai}, \citenamefont
  {Nadlinger}, \citenamefont {Woodrow}, \citenamefont {Lucas},\ and\
  \citenamefont {Ballance}}]{Ballance2020}%
  \BibitemOpen
  \bibfield  {author} {\bibinfo {author} {\bibfnamefont {A.~C.}\ \bibnamefont
  {Hughes}}, \bibinfo {author} {\bibfnamefont {V.~M.}\ \bibnamefont
  {Sch\"afer}}, \bibinfo {author} {\bibfnamefont {K.}~\bibnamefont
  {Thirumalai}}, \bibinfo {author} {\bibfnamefont {D.~P.}\ \bibnamefont
  {Nadlinger}}, \bibinfo {author} {\bibfnamefont {S.~R.}\ \bibnamefont
  {Woodrow}}, \bibinfo {author} {\bibfnamefont {D.~M.}\ \bibnamefont {Lucas}},
  \ and\ \bibinfo {author} {\bibfnamefont {C.~J.}\ \bibnamefont {Ballance}},\
  }\bibfield  {title} {\emph {\bibinfo {title} {Benchmarking a high-fidelity
  mixed-species entangling gate},\ }}\href {\doibase
  10.1103/PhysRevLett.125.080504} {\bibfield  {journal} {\bibinfo  {journal}
  {Phys. Rev. Lett.}\ }\textbf {\bibinfo {volume} {125}},\ \bibinfo {pages}
  {080504} (\bibinfo {year} {2020})}\BibitemShut {NoStop}%
\bibitem [{\citenamefont {Fowler}(2013)}]{Fowler2013}%
  \BibitemOpen
  \bibfield  {author} {\bibinfo {author} {\bibfnamefont {A.~G.}\ \bibnamefont
  {Fowler}},\ }\bibfield  {title} {\emph {\bibinfo {title} {Coping with qubit
  leakage in topological codes},\ }}\href {\doibase 10.1103/PhysRevA.88.042308}
  {\bibfield  {journal} {\bibinfo  {journal} {Phys. Rev. A}\ }\textbf {\bibinfo
  {volume} {88}},\ \bibinfo {pages} {042308} (\bibinfo {year}
  {2013})}\BibitemShut {NoStop}%
\bibitem [{\citenamefont {Ghosh}\ \emph {et~al.}(2013)\citenamefont {Ghosh},
  \citenamefont {Fowler}, \citenamefont {Martinis},\ and\ \citenamefont
  {Geller}}]{Ghosh2013}%
  \BibitemOpen
  \bibfield  {author} {\bibinfo {author} {\bibfnamefont {J.}~\bibnamefont
  {Ghosh}}, \bibinfo {author} {\bibfnamefont {A.~G.}\ \bibnamefont {Fowler}},
  \bibinfo {author} {\bibfnamefont {J.~M.}\ \bibnamefont {Martinis}}, \ and\
  \bibinfo {author} {\bibfnamefont {M.~R.}\ \bibnamefont {Geller}},\ }\bibfield
   {title} {\emph {\bibinfo {title} {Understanding the effects of leakage in
  superconducting quantum-error-detection circuits},\ }}\href {\doibase
  10.1103/PhysRevA.88.062329} {\bibfield  {journal} {\bibinfo  {journal} {Phys.
  Rev. A}\ }\textbf {\bibinfo {volume} {88}},\ \bibinfo {pages} {062329}
  (\bibinfo {year} {2013})}\BibitemShut {NoStop}%
\bibitem [{\citenamefont {Stace}\ \emph {et~al.}(2009)\citenamefont {Stace},
  \citenamefont {Barrett},\ and\ \citenamefont {Doherty}}]{Stace2018}%
  \BibitemOpen
  \bibfield  {author} {\bibinfo {author} {\bibfnamefont {T.~M.}\ \bibnamefont
  {Stace}}, \bibinfo {author} {\bibfnamefont {S.~D.}\ \bibnamefont {Barrett}},
  \ and\ \bibinfo {author} {\bibfnamefont {A.~C.}\ \bibnamefont {Doherty}},\
  }\bibfield  {title} {\emph {\bibinfo {title} {Thresholds for topological
  codes in the presence of loss},\ }}\href {\doibase
  10.1103/PhysRevLett.102.200501} {\bibfield  {journal} {\bibinfo  {journal}
  {Phys. Rev. Lett.}\ }\textbf {\bibinfo {volume} {102}},\ \bibinfo {pages}
  {200501} (\bibinfo {year} {2009})}\BibitemShut {NoStop}%
\bibitem [{\citenamefont {Roos}\ \emph {et~al.}(2006)\citenamefont {Roos},
  \citenamefont {Chwalla}, \citenamefont {Kim}, \citenamefont {Riebe},\ and\
  \citenamefont {Blatt}}]{Roos2006}%
  \BibitemOpen
  \bibfield  {author} {\bibinfo {author} {\bibfnamefont {C.}~\bibnamefont
  {Roos}}, \bibinfo {author} {\bibfnamefont {M.}~\bibnamefont {Chwalla}},
  \bibinfo {author} {\bibfnamefont {K.}~\bibnamefont {Kim}}, \bibinfo {author}
  {\bibfnamefont {M.}~\bibnamefont {Riebe}}, \ and\ \bibinfo {author}
  {\bibfnamefont {R.}~\bibnamefont {Blatt}},\ }\bibfield  {title} {\emph
  {\bibinfo {title} {‘designer atoms’ for quantum metrology},\ }}\href
  {\doibase 10.1038/nature05101} {\bibfield  {journal} {\bibinfo  {journal}
  {Nature}\ }\textbf {\bibinfo {volume} {443}},\ \bibinfo {pages} {316}
  (\bibinfo {year} {2006})}\BibitemShut {NoStop}%
\bibitem [{\citenamefont {Chuang}\ and\ \citenamefont
  {Nielsen}(1997)}]{Chuang1997}%
  \BibitemOpen
  \bibfield  {author} {\bibinfo {author} {\bibfnamefont {I.~L.}\ \bibnamefont
  {Chuang}}\ and\ \bibinfo {author} {\bibfnamefont {M.~A.}\ \bibnamefont
  {Nielsen}},\ }\bibfield  {title} {\emph {\bibinfo {title} {Prescription for
  experimental determination of the dynamics of a quantum black box},\ }}\href
  {\doibase 10.1080/09500349708231894} {\bibfield  {journal} {\bibinfo
  {journal} {Journal of Modern Optics}\ }\textbf {\bibinfo {volume} {44}},\
  \bibinfo {pages} {2455} (\bibinfo {year} {1997})}\BibitemShut {NoStop}%
\bibitem [{\citenamefont {Hradil}(1997)}]{Hradil1997}%
  \BibitemOpen
  \bibfield  {author} {\bibinfo {author} {\bibfnamefont {Z.}~\bibnamefont
  {Hradil}},\ }\bibfield  {title} {\emph {\bibinfo {title} {{Quantum-state
  estimation}},\ }}\href {\doibase 10.1103/PhysRevA.55.R1561} {\bibfield
  {journal} {\bibinfo  {journal} {Phys. Rev. A}\ }\textbf {\bibinfo {volume}
  {55}},\ \bibinfo {pages} {R1561} (\bibinfo {year} {1997})}\BibitemShut
  {NoStop}%
\bibitem [{\citenamefont {James}\ \emph {et~al.}(2001)\citenamefont {James},
  \citenamefont {Kwiat}, \citenamefont {Munro},\ and\ \citenamefont
  {White}}]{James2001}%
  \BibitemOpen
  \bibfield  {author} {\bibinfo {author} {\bibfnamefont {D.~F.~V.}\
  \bibnamefont {James}}, \bibinfo {author} {\bibfnamefont {P.~G.}\ \bibnamefont
  {Kwiat}}, \bibinfo {author} {\bibfnamefont {W.~J.}\ \bibnamefont {Munro}}, \
  and\ \bibinfo {author} {\bibfnamefont {A.~G.}\ \bibnamefont {White}},\
  }\bibfield  {title} {\emph {\bibinfo {title} {{Measurement of qubits}},\
  }}\href {\doibase 10.1103/PhysRevA.64.052312} {\bibfield  {journal} {\bibinfo
   {journal} {Phys. Rev. A}\ }\textbf {\bibinfo {volume} {64}},\ \bibinfo
  {pages} {052312} (\bibinfo {year} {2001})}\BibitemShut {NoStop}%
\bibitem [{\citenamefont {Maciel}\ \emph {et~al.}(2015)\citenamefont {Maciel},
  \citenamefont {Vianna}, \citenamefont {Sarthour},\ and\ \citenamefont
  {Oliveira}}]{Maciel2015}%
  \BibitemOpen
  \bibfield  {author} {\bibinfo {author} {\bibfnamefont {T.~O.}\ \bibnamefont
  {Maciel}}, \bibinfo {author} {\bibfnamefont {R.~O.}\ \bibnamefont {Vianna}},
  \bibinfo {author} {\bibfnamefont {R.~S.}\ \bibnamefont {Sarthour}}, \ and\
  \bibinfo {author} {\bibfnamefont {I.~S.}\ \bibnamefont {Oliveira}},\
  }\bibfield  {title} {\emph {\bibinfo {title} {Quantum process tomography with
  informational incomplete data of {twoJ}-coupled heterogeneous spins
  relaxation in a time window much greater {thanT}1},\ }}\href {\doibase
  10.1088/1367-2630/17/11/113012} {\bibfield  {journal} {\bibinfo  {journal}
  {New Journal of Physics}\ }\textbf {\bibinfo {volume} {17}},\ \bibinfo
  {pages} {113012} (\bibinfo {year} {2015})}\BibitemShut {NoStop}%
\bibitem [{\citenamefont {Bongioanni}\ \emph {et~al.}(2010)\citenamefont
  {Bongioanni}, \citenamefont {Sansoni}, \citenamefont {Sciarrino},
  \citenamefont {Vallone},\ and\ \citenamefont {Mataloni}}]{Bongioanni2010}%
  \BibitemOpen
  \bibfield  {author} {\bibinfo {author} {\bibfnamefont {I.}~\bibnamefont
  {Bongioanni}}, \bibinfo {author} {\bibfnamefont {L.}~\bibnamefont {Sansoni}},
  \bibinfo {author} {\bibfnamefont {F.}~\bibnamefont {Sciarrino}}, \bibinfo
  {author} {\bibfnamefont {G.}~\bibnamefont {Vallone}}, \ and\ \bibinfo
  {author} {\bibfnamefont {P.}~\bibnamefont {Mataloni}},\ }\bibfield  {title}
  {\emph {\bibinfo {title} {Experimental quantum process tomography of
  non-trace-preserving maps},\ }}\href {\doibase 10.1103/PhysRevA.82.042307}
  {\bibfield  {journal} {\bibinfo  {journal} {Phys. Rev. A}\ }\textbf {\bibinfo
  {volume} {82}},\ \bibinfo {pages} {042307} (\bibinfo {year}
  {2010})}\BibitemShut {NoStop}%
\bibitem [{\citenamefont {Choi}(1975)}]{choi1975}%
  \BibitemOpen
  \bibfield  {author} {\bibinfo {author} {\bibfnamefont {M.-D.}\ \bibnamefont
  {Choi}},\ }\bibfield  {title} {\emph {\bibinfo {title} {Completely positive
  linear maps on complex matrices},\ }}\href {\doibase
  https://doi.org/10.1016/0024-3795(75)90075-0} {\bibfield  {journal} {\bibinfo
   {journal} {Linear Algebra and its Applications}\ }\textbf {\bibinfo {volume}
  {10}},\ \bibinfo {pages} {285 } (\bibinfo {year} {1975})}\BibitemShut
  {NoStop}%
\bibitem [{\citenamefont {Schindler}\ \emph
  {et~al.}(2013{\natexlab{b}})\citenamefont {Schindler}, \citenamefont {Nigg},
  \citenamefont {Monz}, \citenamefont {Barreiro}, \citenamefont {Martinez},
  \citenamefont {Wang}, \citenamefont {Quint}, \citenamefont {Brandl},
  \citenamefont {Nebendahl}, \citenamefont {Roos}, \citenamefont {Chwalla},
  \citenamefont {Hennrich},\ and\ \citenamefont {Blatt}}]{toolbox}%
  \BibitemOpen
  \bibfield  {author} {\bibinfo {author} {\bibfnamefont {P.}~\bibnamefont
  {Schindler}}, \bibinfo {author} {\bibfnamefont {D.}~\bibnamefont {Nigg}},
  \bibinfo {author} {\bibfnamefont {T.}~\bibnamefont {Monz}}, \bibinfo {author}
  {\bibfnamefont {J.~T.}\ \bibnamefont {Barreiro}}, \bibinfo {author}
  {\bibfnamefont {E.}~\bibnamefont {Martinez}}, \bibinfo {author}
  {\bibfnamefont {S.~X.}\ \bibnamefont {Wang}}, \bibinfo {author}
  {\bibfnamefont {S.}~\bibnamefont {Quint}}, \bibinfo {author} {\bibfnamefont
  {M.~F.}\ \bibnamefont {Brandl}}, \bibinfo {author} {\bibfnamefont
  {V.}~\bibnamefont {Nebendahl}}, \bibinfo {author} {\bibfnamefont {C.~F.}\
  \bibnamefont {Roos}}, \bibinfo {author} {\bibfnamefont {M.}~\bibnamefont
  {Chwalla}}, \bibinfo {author} {\bibfnamefont {M.}~\bibnamefont {Hennrich}}, \
  and\ \bibinfo {author} {\bibfnamefont {R.}~\bibnamefont {Blatt}},\ }\bibfield
   {title} {\emph {\bibinfo {title} {A quantum information processor with
  trapped ions},\ }}\href {\doibase 10.1088/1367-2630/15/12/123012} {\bibfield
  {journal} {\bibinfo  {journal} {New J. Phys.}\ }\textbf {\bibinfo {volume}
  {15}},\ \bibinfo {pages} {123012} (\bibinfo {year}
  {2013}{\natexlab{b}})}\BibitemShut {NoStop}%
\bibitem [{\citenamefont {M\o{}lmer}\ and\ \citenamefont
  {S\o{}rensen}(1999)}]{MSgate}%
  \BibitemOpen
  \bibfield  {author} {\bibinfo {author} {\bibfnamefont {K.}~\bibnamefont
  {M\o{}lmer}}\ and\ \bibinfo {author} {\bibfnamefont {A.}~\bibnamefont
  {S\o{}rensen}},\ }\bibfield  {title} {\emph {\bibinfo {title} {Multiparticle
  entanglement of hot trapped ions},\ }}\href {\doibase
  10.1103/PhysRevLett.82.1835} {\bibfield  {journal} {\bibinfo  {journal}
  {Phys. Rev. Lett.}\ }\textbf {\bibinfo {volume} {82}},\ \bibinfo {pages}
  {1835} (\bibinfo {year} {1999})}\BibitemShut {NoStop}%
\bibitem [{sup()}]{supp_mat}%
  \BibitemOpen
  \href@noop {} {\bibinfo {title} {See supplementary materials},\ }\BibitemShut
  {NoStop}%
\bibitem [{\citenamefont {Kreuter}\ \emph {et~al.}(2004)\citenamefont
  {Kreuter}, \citenamefont {Becher}, \citenamefont {Lancaster}, \citenamefont
  {Mundt}, \citenamefont {Russo}, \citenamefont {H\"affner}, \citenamefont
  {Roos}, \citenamefont {Eschner}, \citenamefont {Schmidt-Kaler},\ and\
  \citenamefont {Blatt}}]{Kreuter2004}%
  \BibitemOpen
  \bibfield  {author} {\bibinfo {author} {\bibfnamefont {A.}~\bibnamefont
  {Kreuter}}, \bibinfo {author} {\bibfnamefont {C.}~\bibnamefont {Becher}},
  \bibinfo {author} {\bibfnamefont {G.~P.~T.}\ \bibnamefont {Lancaster}},
  \bibinfo {author} {\bibfnamefont {A.~B.}\ \bibnamefont {Mundt}}, \bibinfo
  {author} {\bibfnamefont {C.}~\bibnamefont {Russo}}, \bibinfo {author}
  {\bibfnamefont {H.}~\bibnamefont {H\"affner}}, \bibinfo {author}
  {\bibfnamefont {C.}~\bibnamefont {Roos}}, \bibinfo {author} {\bibfnamefont
  {J.}~\bibnamefont {Eschner}}, \bibinfo {author} {\bibfnamefont
  {F.}~\bibnamefont {Schmidt-Kaler}}, \ and\ \bibinfo {author} {\bibfnamefont
  {R.}~\bibnamefont {Blatt}},\ }\bibfield  {title} {\emph {\bibinfo {title}
  {Spontaneous emission lifetime of a single trapped ${\mathrm{ca}}^{+}$ ion in
  a high finesse cavity},\ }}\href {\doibase 10.1103/PhysRevLett.92.203002}
  {\bibfield  {journal} {\bibinfo  {journal} {Phys. Rev. Lett.}\ }\textbf
  {\bibinfo {volume} {92}},\ \bibinfo {pages} {203002} (\bibinfo {year}
  {2004})}\BibitemShut {NoStop}%
\bibitem [{\citenamefont {Hayes}\ \emph {et~al.}(2020)\citenamefont {Hayes},
  \citenamefont {Stack}, \citenamefont {Bjork}, \citenamefont {Potter},
  \citenamefont {Baldwin},\ and\ \citenamefont {Stutz}}]{Hayes2020}%
  \BibitemOpen
  \bibfield  {author} {\bibinfo {author} {\bibfnamefont {D.}~\bibnamefont
  {Hayes}}, \bibinfo {author} {\bibfnamefont {D.}~\bibnamefont {Stack}},
  \bibinfo {author} {\bibfnamefont {B.}~\bibnamefont {Bjork}}, \bibinfo
  {author} {\bibfnamefont {A.~C.}\ \bibnamefont {Potter}}, \bibinfo {author}
  {\bibfnamefont {C.~H.}\ \bibnamefont {Baldwin}}, \ and\ \bibinfo {author}
  {\bibfnamefont {R.~P.}\ \bibnamefont {Stutz}},\ }\bibfield  {title} {\emph
  {\bibinfo {title} {Eliminating leakage errors in hyperfine qubits},\ }}\href
  {\doibase 10.1103/PhysRevLett.124.170501} {\bibfield  {journal} {\bibinfo
  {journal} {Physical Review Letters}\ }\textbf {\bibinfo {volume} {124}}
  (\bibinfo {year} {2020}),\ 10.1103/PhysRevLett.124.170501}\BibitemShut
  {NoStop}%
\bibitem [{\citenamefont {Nigg}\ \emph {et~al.}(2014)\citenamefont {Nigg},
  \citenamefont {M{\"u}ller}, \citenamefont {Martinez}, \citenamefont
  {Schindler}, \citenamefont {Hennrich}, \citenamefont {Monz}, \citenamefont
  {Martin-Delgado},\ and\ \citenamefont {Blatt}}]{Nigg2014}%
  \BibitemOpen
  \bibfield  {author} {\bibinfo {author} {\bibfnamefont {D.}~\bibnamefont
  {Nigg}}, \bibinfo {author} {\bibfnamefont {M.}~\bibnamefont {M{\"u}ller}},
  \bibinfo {author} {\bibfnamefont {E.~A.}\ \bibnamefont {Martinez}}, \bibinfo
  {author} {\bibfnamefont {P.}~\bibnamefont {Schindler}}, \bibinfo {author}
  {\bibfnamefont {M.}~\bibnamefont {Hennrich}}, \bibinfo {author}
  {\bibfnamefont {T.}~\bibnamefont {Monz}}, \bibinfo {author} {\bibfnamefont
  {M.~A.}\ \bibnamefont {Martin-Delgado}}, \ and\ \bibinfo {author}
  {\bibfnamefont {R.}~\bibnamefont {Blatt}},\ }\bibfield  {title} {\emph
  {\bibinfo {title} {Quantum computations on a topologically encoded qubit},\
  }}\href {\doibase 10.1126/science.1253742} {\bibfield  {journal} {\bibinfo
  {journal} {Science}\ }\textbf {\bibinfo {volume} {345}},\ \bibinfo {pages}
  {302} (\bibinfo {year} {2014})}\BibitemShut {NoStop}%
\bibitem [{\citenamefont {Niemietz}\ \emph {et~al.}(2021)\citenamefont
  {Niemietz}, \citenamefont {Farrera}, \citenamefont {Langenfeld},\ and\
  \citenamefont {Rempe}}]{Niemietz2021}%
  \BibitemOpen
  \bibfield  {author} {\bibinfo {author} {\bibfnamefont {D.}~\bibnamefont
  {Niemietz}}, \bibinfo {author} {\bibfnamefont {P.}~\bibnamefont {Farrera}},
  \bibinfo {author} {\bibfnamefont {S.}~\bibnamefont {Langenfeld}}, \ and\
  \bibinfo {author} {\bibfnamefont {G.}~\bibnamefont {Rempe}},\ }\bibfield
  {title} {\emph {\bibinfo {title} {Nondestructive detection of photonic
  qubits},\ }}\href {\doibase 10.1038/s41586-021-03290-z} {\bibfield  {journal}
  {\bibinfo  {journal} {Nature}\ }\textbf {\bibinfo {volume} {591}},\ \bibinfo
  {pages} {570} (\bibinfo {year} {2021})}\BibitemShut {NoStop}%
\bibitem [{\citenamefont {Varbanov}\ \emph {et~al.}(2020)\citenamefont
  {Varbanov}, \citenamefont {Battistel}, \citenamefont {Tarasinski},
  \citenamefont {Ostroukh}, \citenamefont {O'Brien}, \citenamefont {DiCarlo},\
  and\ \citenamefont {Terhal}}]{Varbanov2020}%
  \BibitemOpen
  \bibfield  {author} {\bibinfo {author} {\bibfnamefont {B.~M.}\ \bibnamefont
  {Varbanov}}, \bibinfo {author} {\bibfnamefont {F.}~\bibnamefont {Battistel}},
  \bibinfo {author} {\bibfnamefont {B.~M.}\ \bibnamefont {Tarasinski}},
  \bibinfo {author} {\bibfnamefont {V.~P.}\ \bibnamefont {Ostroukh}}, \bibinfo
  {author} {\bibfnamefont {T.~E.}\ \bibnamefont {O'Brien}}, \bibinfo {author}
  {\bibfnamefont {L.}~\bibnamefont {DiCarlo}}, \ and\ \bibinfo {author}
  {\bibfnamefont {B.~M.}\ \bibnamefont {Terhal}},\ }\bibfield  {title} {\emph
  {\bibinfo {title} {Leakage detection for a transmon-based surface code},\
  }}\href {\doibase 10.1038/s41534-020-00330-w} {\bibfield  {journal} {\bibinfo
   {journal} {npj Quantum Information}\ }\textbf {\bibinfo {volume} {6}},\
  \bibinfo {pages} {102} (\bibinfo {year} {2020})}\BibitemShut {NoStop}%
\bibitem [{\citenamefont {Vala}\ \emph {et~al.}(2005)\citenamefont {Vala},
  \citenamefont {Whaley},\ and\ \citenamefont {Weiss}}]{Vala2005}%
  \BibitemOpen
  \bibfield  {author} {\bibinfo {author} {\bibfnamefont {J.}~\bibnamefont
  {Vala}}, \bibinfo {author} {\bibfnamefont {K.~B.}\ \bibnamefont {Whaley}}, \
  and\ \bibinfo {author} {\bibfnamefont {D.~S.}\ \bibnamefont {Weiss}},\
  }\bibfield  {title} {\emph {\bibinfo {title} {Quantum error correction of a
  qubit loss in an addressable atomic system},\ }}\href {\doibase
  10.1103/PhysRevA.72.052318} {\bibfield  {journal} {\bibinfo  {journal} {Phys.
  Rev. A}\ }\textbf {\bibinfo {volume} {72}},\ \bibinfo {pages} {052318}
  (\bibinfo {year} {2005})}\BibitemShut {NoStop}%
\bibitem [{\citenamefont {Grassl}\ \emph {et~al.}(1997)\citenamefont {Grassl},
  \citenamefont {Beth},\ and\ \citenamefont {Pellizzari}}]{Grassl1997}%
  \BibitemOpen
  \bibfield  {author} {\bibinfo {author} {\bibfnamefont {M.}~\bibnamefont
  {Grassl}}, \bibinfo {author} {\bibfnamefont {T.}~\bibnamefont {Beth}}, \ and\
  \bibinfo {author} {\bibfnamefont {T.}~\bibnamefont {Pellizzari}},\ }\bibfield
   {title} {\emph {\bibinfo {title} {Codes for the quantum erasure channel},\
  }}\href {\doibase 10.1103/PhysRevA.56.33} {\bibfield  {journal} {\bibinfo
  {journal} {Phys. Rev. A}\ }\textbf {\bibinfo {volume} {56}},\ \bibinfo
  {pages} {33} (\bibinfo {year} {1997})}\BibitemShut {NoStop}%
\bibitem [{\citenamefont {Nielsen}\ and\ \citenamefont
  {Chuang}(2011)}]{NielsenChuang}%
  \BibitemOpen
  \bibfield  {author} {\bibinfo {author} {\bibfnamefont {M.~A.}\ \bibnamefont
  {Nielsen}}\ and\ \bibinfo {author} {\bibfnamefont {I.~L.}\ \bibnamefont
  {Chuang}},\ }\href@noop {} {\emph {\bibinfo {title} {Quantum Computation and
  Quantum Information: 10th Anniversary Edition}}},\ \bibinfo {edition} {10th}\
  ed.\ (\bibinfo  {publisher} {Cambridge University Press},\ \bibinfo {address}
  {New York, USA},\ \bibinfo {year} {2011})\BibitemShut {NoStop}%
\bibitem [{\citenamefont {Ralph}\ \emph {et~al.}(2005)\citenamefont {Ralph},
  \citenamefont {Hayes},\ and\ \citenamefont {Gilchrist}}]{Ralph2005}%
  \BibitemOpen
  \bibfield  {author} {\bibinfo {author} {\bibfnamefont {T.~C.}\ \bibnamefont
  {Ralph}}, \bibinfo {author} {\bibfnamefont {A.~J.~F.}\ \bibnamefont {Hayes}},
  \ and\ \bibinfo {author} {\bibfnamefont {A.}~\bibnamefont {Gilchrist}},\
  }\bibfield  {title} {\emph {\bibinfo {title} {Loss-tolerant optical qubits},\
  }}\href {\doibase 10.1103/PhysRevLett.95.100501} {\bibfield  {journal}
  {\bibinfo  {journal} {Phys. Rev. Lett.}\ }\textbf {\bibinfo {volume} {95}},\
  \bibinfo {pages} {100501} (\bibinfo {year} {2005})}\BibitemShut {NoStop}%
\bibitem [{\citenamefont {Lu}\ \emph {et~al.}(2008)\citenamefont {Lu},
  \citenamefont {Gao}, \citenamefont {Zhang}, \citenamefont {Zhou},
  \citenamefont {Yang},\ and\ \citenamefont {Pan}}]{Lu2008}%
  \BibitemOpen
  \bibfield  {author} {\bibinfo {author} {\bibfnamefont {C.-Y.}\ \bibnamefont
  {Lu}}, \bibinfo {author} {\bibfnamefont {W.-B.}\ \bibnamefont {Gao}},
  \bibinfo {author} {\bibfnamefont {J.}~\bibnamefont {Zhang}}, \bibinfo
  {author} {\bibfnamefont {X.-Q.}\ \bibnamefont {Zhou}}, \bibinfo {author}
  {\bibfnamefont {T.}~\bibnamefont {Yang}}, \ and\ \bibinfo {author}
  {\bibfnamefont {J.-W.}\ \bibnamefont {Pan}},\ }\bibfield  {title} {\emph
  {\bibinfo {title} {Experimental quantum coding against qubit loss error},\
  }}\href {\doibase 10.1073/pnas.0800740105} {\bibfield  {journal} {\bibinfo
  {journal} {Proceedings of the National Academy of Sciences}\ }\textbf
  {\bibinfo {volume} {105}},\ \bibinfo {pages} {11050} (\bibinfo {year}
  {2008})}\BibitemShut {NoStop}%
\bibitem [{\citenamefont {Laflamme}\ \emph {et~al.}(1996)\citenamefont
  {Laflamme}, \citenamefont {Miquel}, \citenamefont {Paz},\ and\ \citenamefont
  {Zurek}}]{Laflamme1996}%
  \BibitemOpen
  \bibfield  {author} {\bibinfo {author} {\bibfnamefont {R.}~\bibnamefont
  {Laflamme}}, \bibinfo {author} {\bibfnamefont {C.}~\bibnamefont {Miquel}},
  \bibinfo {author} {\bibfnamefont {J.~P.}\ \bibnamefont {Paz}}, \ and\
  \bibinfo {author} {\bibfnamefont {W.~H.}\ \bibnamefont {Zurek}},\ }\bibfield
  {title} {\emph {\bibinfo {title} {Perfect quantum error correcting code},\
  }}\href {\doibase 10.1103/PhysRevLett.77.198} {\bibfield  {journal} {\bibinfo
   {journal} {Phys. Rev. Lett.}\ }\textbf {\bibinfo {volume} {77}},\ \bibinfo
  {pages} {198} (\bibinfo {year} {1996})}\BibitemShut {NoStop}%
\bibitem [{\citenamefont {Kitaev}(2003)}]{Kitaev2003}%
  \BibitemOpen
  \bibfield  {author} {\bibinfo {author} {\bibfnamefont {A.}~\bibnamefont
  {Kitaev}},\ }\bibfield  {title} {\emph {\bibinfo {title} {Fault-tolerant
  quantum computation by anyons},\ }}\href {\doibase
  10.1016/S0003-4916(02)00018-0} {\bibfield  {journal} {\bibinfo  {journal}
  {Ann. Phys.}\ }\textbf {\bibinfo {volume} {303}},\ \bibinfo {pages} {2 }
  (\bibinfo {year} {2003})}\BibitemShut {NoStop}%
\bibitem [{\citenamefont {Bombin}\ and\ \citenamefont
  {Martin-Delgado}(2006)}]{Bombin2006}%
  \BibitemOpen
  \bibfield  {author} {\bibinfo {author} {\bibfnamefont {H.}~\bibnamefont
  {Bombin}}\ and\ \bibinfo {author} {\bibfnamefont {M.~A.}\ \bibnamefont
  {Martin-Delgado}},\ }\bibfield  {title} {\emph {\bibinfo {title} {Topological
  quantum distillation},\ }}\href {\doibase 10.1103/PhysRevLett.97.180501}
  {\bibfield  {journal} {\bibinfo  {journal} {Phys. Rev. Lett.}\ }\textbf
  {\bibinfo {volume} {97}},\ \bibinfo {pages} {180501} (\bibinfo {year}
  {2006})}\BibitemShut {NoStop}%
\bibitem [{\citenamefont {Bombin}\ and\ \citenamefont
  {Martin-Delgado}(2007)}]{Bombin2007}%
  \BibitemOpen
  \bibfield  {author} {\bibinfo {author} {\bibfnamefont {H.}~\bibnamefont
  {Bombin}}\ and\ \bibinfo {author} {\bibfnamefont {M.~A.}\ \bibnamefont
  {Martin-Delgado}},\ }\bibfield  {title} {\emph {\bibinfo {title} {Topological
  computation without braiding},\ }}\href {\doibase
  10.1103/PhysRevLett.98.160502} {\bibfield  {journal} {\bibinfo  {journal}
  {Phys. Rev. Lett.}\ }\textbf {\bibinfo {volume} {98}},\ \bibinfo {pages}
  {160502} (\bibinfo {year} {2007})}\BibitemShut {NoStop}%
\bibitem [{\citenamefont {Knill}\ \emph {et~al.}(2001)\citenamefont {Knill},
  \citenamefont {Laflamme}, \citenamefont {Martinez},\ and\ \citenamefont
  {Negrevergne}}]{Knill2001}%
  \BibitemOpen
  \bibfield  {author} {\bibinfo {author} {\bibfnamefont {E.}~\bibnamefont
  {Knill}}, \bibinfo {author} {\bibfnamefont {R.}~\bibnamefont {Laflamme}},
  \bibinfo {author} {\bibfnamefont {R.}~\bibnamefont {Martinez}}, \ and\
  \bibinfo {author} {\bibfnamefont {C.}~\bibnamefont {Negrevergne}},\
  }\bibfield  {title} {\emph {\bibinfo {title} {Benchmarking quantum computers:
  The five-qubit error correcting code},\ }}\href {\doibase
  10.1103/PhysRevLett.86.5811} {\bibfield  {journal} {\bibinfo  {journal}
  {Phys. Rev. Lett.}\ }\textbf {\bibinfo {volume} {86}},\ \bibinfo {pages}
  {5811} (\bibinfo {year} {2001})}\BibitemShut {NoStop}%
\bibitem [{\citenamefont {Chiaverini}\ \emph {et~al.}(2004)\citenamefont
  {Chiaverini}, \citenamefont {Leibfried}, \citenamefont {Schaetz},
  \citenamefont {Barrett}, \citenamefont {Blakestad}, \citenamefont {Britton},
  \citenamefont {Itano}, \citenamefont {Jost}, \citenamefont {Knill},
  \citenamefont {Langer}, \citenamefont {Ozeri},\ and\ \citenamefont
  {Wineland}}]{Chiaverini2004}%
  \BibitemOpen
  \bibfield  {author} {\bibinfo {author} {\bibfnamefont {J.}~\bibnamefont
  {Chiaverini}}, \bibinfo {author} {\bibfnamefont {D.}~\bibnamefont
  {Leibfried}}, \bibinfo {author} {\bibfnamefont {T.}~\bibnamefont {Schaetz}},
  \bibinfo {author} {\bibfnamefont {M.}~\bibnamefont {Barrett}}, \bibinfo
  {author} {\bibfnamefont {R.}~\bibnamefont {Blakestad}}, \bibinfo {author}
  {\bibfnamefont {J.}~\bibnamefont {Britton}}, \bibinfo {author} {\bibfnamefont
  {W.}~\bibnamefont {Itano}}, \bibinfo {author} {\bibfnamefont
  {J.}~\bibnamefont {Jost}}, \bibinfo {author} {\bibfnamefont {E.}~\bibnamefont
  {Knill}}, \bibinfo {author} {\bibfnamefont {C.}~\bibnamefont {Langer}},
  \bibinfo {author} {\bibfnamefont {R.}~\bibnamefont {Ozeri}}, \ and\ \bibinfo
  {author} {\bibfnamefont {D.}~\bibnamefont {Wineland}},\ }\bibfield  {title}
  {\emph {\bibinfo {title} {Realization of quantum error correction},\ }}\href
  {\doibase 10.1038/nature03074} {\bibfield  {journal} {\bibinfo  {journal}
  {Nature}\ }\textbf {\bibinfo {volume} {432}},\ \bibinfo {pages} {602—605}
  (\bibinfo {year} {2004})}\BibitemShut {NoStop}%
\bibitem [{\citenamefont {Reed}\ \emph {et~al.}(2012)\citenamefont {Reed},
  \citenamefont {DiCarlo}, \citenamefont {Nigg}, \citenamefont {Sun},
  \citenamefont {Frunzio}, \citenamefont {Girvin},\ and\ \citenamefont
  {Schoelkopf}}]{Reed2012}%
  \BibitemOpen
  \bibfield  {author} {\bibinfo {author} {\bibfnamefont {M.~D.}\ \bibnamefont
  {Reed}}, \bibinfo {author} {\bibfnamefont {L.}~\bibnamefont {DiCarlo}},
  \bibinfo {author} {\bibfnamefont {S.~E.}\ \bibnamefont {Nigg}}, \bibinfo
  {author} {\bibfnamefont {L.}~\bibnamefont {Sun}}, \bibinfo {author}
  {\bibfnamefont {L.}~\bibnamefont {Frunzio}}, \bibinfo {author} {\bibfnamefont
  {S.~M.}\ \bibnamefont {Girvin}}, \ and\ \bibinfo {author} {\bibfnamefont
  {R.~J.}\ \bibnamefont {Schoelkopf}},\ }\bibfield  {title} {\emph {\bibinfo
  {title} {Realization of three-qubit quantum error correction with
  superconducting circuits},\ }}\href {\doibase 10.1038/nature10786} {\bibfield
   {journal} {\bibinfo  {journal} {Nature}\ }\textbf {\bibinfo {volume}
  {482}},\ \bibinfo {pages} {382} (\bibinfo {year} {2012})}\BibitemShut
  {NoStop}%
\bibitem [{\citenamefont {Rist{\`e}}\ \emph {et~al.}(2015)\citenamefont
  {Rist{\`e}}, \citenamefont {Poletto}, \citenamefont {Huang}, \citenamefont
  {Bruno}, \citenamefont {Vesterinen}, \citenamefont {Saira},\ and\
  \citenamefont {DiCarlo}}]{Riste2015}%
  \BibitemOpen
  \bibfield  {author} {\bibinfo {author} {\bibfnamefont {D.}~\bibnamefont
  {Rist{\`e}}}, \bibinfo {author} {\bibfnamefont {S.}~\bibnamefont {Poletto}},
  \bibinfo {author} {\bibfnamefont {M.~Z.}\ \bibnamefont {Huang}}, \bibinfo
  {author} {\bibfnamefont {A.}~\bibnamefont {Bruno}}, \bibinfo {author}
  {\bibfnamefont {V.}~\bibnamefont {Vesterinen}}, \bibinfo {author}
  {\bibfnamefont {O.~P.}\ \bibnamefont {Saira}}, \ and\ \bibinfo {author}
  {\bibfnamefont {L.}~\bibnamefont {DiCarlo}},\ }\bibfield  {title} {\emph
  {\bibinfo {title} {Detecting bit-flip errors in a logical qubit using
  stabilizer measurements},\ }}\href {\doibase 10.1038/ncomms7983} {\bibfield
  {journal} {\bibinfo  {journal} {Nature Communications}\ }\textbf {\bibinfo
  {volume} {6}},\ \bibinfo {pages} {6983} (\bibinfo {year} {2015})}\BibitemShut
  {NoStop}%
\bibitem [{\citenamefont {Steane}(1996)}]{Steane1996}%
  \BibitemOpen
  \bibfield  {author} {\bibinfo {author} {\bibfnamefont {A.~M.}\ \bibnamefont
  {Steane}},\ }\bibfield  {title} {\emph {\bibinfo {title} {Error correcting
  codes in quantum theory},\ }}\href {\doibase 10.1103/PhysRevLett.77.793}
  {\bibfield  {journal} {\bibinfo  {journal} {Phys. Rev. Lett.}\ }\textbf
  {\bibinfo {volume} {77}},\ \bibinfo {pages} {793} (\bibinfo {year}
  {1996})}\BibitemShut {NoStop}%
\bibitem [{\citenamefont {Bermudez}\ \emph {et~al.}(2017)\citenamefont
  {Bermudez}, \citenamefont {Xu}, \citenamefont {Nigmatullin}, \citenamefont
  {O'Gorman}, \citenamefont {Negnevitsky}, \citenamefont {Schindler},
  \citenamefont {Monz}, \citenamefont {Poschinger}, \citenamefont {Hempel},
  \citenamefont {Home}, \citenamefont {Schmidt-Kaler}, \citenamefont {Biercuk},
  \citenamefont {Blatt}, \citenamefont {Benjamin},\ and\ \citenamefont
  {M\"uller}}]{Bermudez2017}%
  \BibitemOpen
  \bibfield  {author} {\bibinfo {author} {\bibfnamefont {A.}~\bibnamefont
  {Bermudez}}, \bibinfo {author} {\bibfnamefont {X.}~\bibnamefont {Xu}},
  \bibinfo {author} {\bibfnamefont {R.}~\bibnamefont {Nigmatullin}}, \bibinfo
  {author} {\bibfnamefont {J.}~\bibnamefont {O'Gorman}}, \bibinfo {author}
  {\bibfnamefont {V.}~\bibnamefont {Negnevitsky}}, \bibinfo {author}
  {\bibfnamefont {P.}~\bibnamefont {Schindler}}, \bibinfo {author}
  {\bibfnamefont {T.}~\bibnamefont {Monz}}, \bibinfo {author} {\bibfnamefont
  {U.~G.}\ \bibnamefont {Poschinger}}, \bibinfo {author} {\bibfnamefont
  {C.}~\bibnamefont {Hempel}}, \bibinfo {author} {\bibfnamefont
  {J.}~\bibnamefont {Home}}, \bibinfo {author} {\bibfnamefont {F.}~\bibnamefont
  {Schmidt-Kaler}}, \bibinfo {author} {\bibfnamefont {M.}~\bibnamefont
  {Biercuk}}, \bibinfo {author} {\bibfnamefont {R.}~\bibnamefont {Blatt}},
  \bibinfo {author} {\bibfnamefont {S.}~\bibnamefont {Benjamin}}, \ and\
  \bibinfo {author} {\bibfnamefont {M.}~\bibnamefont {M\"uller}},\ }\bibfield
  {title} {\emph {\bibinfo {title} {Assessing the progress of trapped-ion
  processors towards fault-tolerant quantum computation},\ }}\href {\doibase
  10.1103/PhysRevX.7.041061} {\bibfield  {journal} {\bibinfo  {journal} {Phys.
  Rev. X}\ }\textbf {\bibinfo {volume} {7}},\ \bibinfo {pages} {041061}
  (\bibinfo {year} {2017})}\BibitemShut {NoStop}%
\bibitem [{\citenamefont {Bermudez}\ \emph {et~al.}(2019)\citenamefont
  {Bermudez}, \citenamefont {Xu}, \citenamefont {Guti\'errez}, \citenamefont
  {Benjamin},\ and\ \citenamefont {M\"uller}}]{Bermudez2019}%
  \BibitemOpen
  \bibfield  {author} {\bibinfo {author} {\bibfnamefont {A.}~\bibnamefont
  {Bermudez}}, \bibinfo {author} {\bibfnamefont {X.}~\bibnamefont {Xu}},
  \bibinfo {author} {\bibfnamefont {M.}~\bibnamefont {Guti\'errez}}, \bibinfo
  {author} {\bibfnamefont {S.~C.}\ \bibnamefont {Benjamin}}, \ and\ \bibinfo
  {author} {\bibfnamefont {M.}~\bibnamefont {M\"uller}},\ }\bibfield  {title}
  {\emph {\bibinfo {title} {Fault-tolerant protection of near-term trapped-ion
  topological qubits under realistic noise sources},\ }}\href {\doibase
  10.1103/PhysRevA.100.062307} {\bibfield  {journal} {\bibinfo  {journal}
  {Phys. Rev. A}\ }\textbf {\bibinfo {volume} {100}},\ \bibinfo {pages}
  {062307} (\bibinfo {year} {2019})}\BibitemShut {NoStop}%
\bibitem [{\citenamefont {Guti\'errez}\ \emph {et~al.}(2019)\citenamefont
  {Guti\'errez}, \citenamefont {M\"uller},\ and\ \citenamefont
  {Berm\'udez}}]{Gutierrez2019}%
  \BibitemOpen
  \bibfield  {author} {\bibinfo {author} {\bibfnamefont {M.}~\bibnamefont
  {Guti\'errez}}, \bibinfo {author} {\bibfnamefont {M.}~\bibnamefont
  {M\"uller}}, \ and\ \bibinfo {author} {\bibfnamefont {A.}~\bibnamefont
  {Berm\'udez}},\ }\bibfield  {title} {\emph {\bibinfo {title} {Transversality
  and lattice surgery: Exploring realistic routes toward coupled logical qubits
  with trapped-ion quantum processors},\ }}\href {\doibase
  10.1103/PhysRevA.99.022330} {\bibfield  {journal} {\bibinfo  {journal} {Phys.
  Rev. A}\ }\textbf {\bibinfo {volume} {99}},\ \bibinfo {pages} {022330}
  (\bibinfo {year} {2019})}\BibitemShut {NoStop}%
\bibitem [{\citenamefont {Amaro}\ \emph {et~al.}(2020)\citenamefont {Amaro},
  \citenamefont {Bennett}, \citenamefont {Vodola},\ and\ \citenamefont
  {M\"uller}}]{Amaro2020}%
  \BibitemOpen
  \bibfield  {author} {\bibinfo {author} {\bibfnamefont {D.}~\bibnamefont
  {Amaro}}, \bibinfo {author} {\bibfnamefont {J.}~\bibnamefont {Bennett}},
  \bibinfo {author} {\bibfnamefont {D.}~\bibnamefont {Vodola}}, \ and\ \bibinfo
  {author} {\bibfnamefont {M.}~\bibnamefont {M\"uller}},\ }\bibfield  {title}
  {\emph {\bibinfo {title} {Analytical percolation theory for topological color
  codes under qubit loss},\ }}\href {\doibase 10.1103/PhysRevA.101.032317}
  {\bibfield  {journal} {\bibinfo  {journal} {Phys. Rev. A}\ }\textbf {\bibinfo
  {volume} {101}},\ \bibinfo {pages} {032317} (\bibinfo {year}
  {2020})}\BibitemShut {NoStop}%
\bibitem [{\citenamefont {Dennis}\ \emph {et~al.}(2002)\citenamefont {Dennis},
  \citenamefont {Kitaev}, \citenamefont {Landahl},\ and\ \citenamefont
  {Preskill}}]{Dennis2002}%
  \BibitemOpen
  \bibfield  {author} {\bibinfo {author} {\bibfnamefont {E.}~\bibnamefont
  {Dennis}}, \bibinfo {author} {\bibfnamefont {A.}~\bibnamefont {Kitaev}},
  \bibinfo {author} {\bibfnamefont {A.}~\bibnamefont {Landahl}}, \ and\
  \bibinfo {author} {\bibfnamefont {J.}~\bibnamefont {Preskill}},\ }\bibfield
  {title} {\emph {\bibinfo {title} {Topological quantum memory},\ }}\href
  {\doibase 10.1063/1.1499754} {\bibfield  {journal} {\bibinfo  {journal}
  {Journal of Mathematical Physics}\ }\textbf {\bibinfo {volume} {43}},\
  \bibinfo {pages} {4452} (\bibinfo {year} {2002})}\BibitemShut {NoStop}%
\bibitem [{\citenamefont {Chao}\ and\ \citenamefont
  {Reichardt}(2018)}]{Chao2018}%
  \BibitemOpen
  \bibfield  {author} {\bibinfo {author} {\bibfnamefont {R.}~\bibnamefont
  {Chao}}\ and\ \bibinfo {author} {\bibfnamefont {B.~W.}\ \bibnamefont
  {Reichardt}},\ }\bibfield  {title} {\emph {\bibinfo {title} {Quantum error
  correction with only two extra qubits},\ }}\href {\doibase
  10.1103/PhysRevLett.121.050502} {\bibfield  {journal} {\bibinfo  {journal}
  {Phys. Rev. Lett.}\ }\textbf {\bibinfo {volume} {121}},\ \bibinfo {pages}
  {050502} (\bibinfo {year} {2018})}\BibitemShut {NoStop}%
\bibitem [{\citenamefont {Parrado-Rodriguez}\ \emph {et~al.}(2021)\citenamefont
  {Parrado-Rodriguez}, \citenamefont {Ryan-Anderson}, \citenamefont
  {Bermudez},\ and\ \citenamefont {M\"uller}}]{Parrado2021}%
  \BibitemOpen
  \bibfield  {author} {\bibinfo {author} {\bibfnamefont {P.}~\bibnamefont
  {Parrado-Rodriguez}}, \bibinfo {author} {\bibfnamefont {C.}~\bibnamefont
  {Ryan-Anderson}}, \bibinfo {author} {\bibfnamefont {A.}~\bibnamefont
  {Bermudez}}, \ and\ \bibinfo {author} {\bibfnamefont {M.}~\bibnamefont
  {M\"uller}},\ }\bibfield  {title} {\emph {\bibinfo {title} {Crosstalk
  suppression for fault-tolerant quantum error correction with trapped ions},\
  }}\href {\doibase 10.22331/q-2021-06-29-487} {\bibfield  {journal} {\bibinfo
  {journal} {Quantum}\ }\textbf {\bibinfo {volume} {5}},\ \bibinfo {pages}
  {487} (\bibinfo {year} {2021})}\BibitemShut {NoStop}%
\bibitem [{\citenamefont {Ringbauer}\ \emph {et~al.}(2021)\citenamefont
  {Ringbauer}, \citenamefont {Meth}, \citenamefont {Postler}, \citenamefont
  {Stricker}, \citenamefont {Blatt}, \citenamefont {Schindler},\ and\
  \citenamefont {Monz}}]{Ringbauer2021}%
  \BibitemOpen
  \bibfield  {author} {\bibinfo {author} {\bibfnamefont {M.}~\bibnamefont
  {Ringbauer}}, \bibinfo {author} {\bibfnamefont {M.}~\bibnamefont {Meth}},
  \bibinfo {author} {\bibfnamefont {L.}~\bibnamefont {Postler}}, \bibinfo
  {author} {\bibfnamefont {R.}~\bibnamefont {Stricker}}, \bibinfo {author}
  {\bibfnamefont {R.}~\bibnamefont {Blatt}}, \bibinfo {author} {\bibfnamefont
  {P.}~\bibnamefont {Schindler}}, \ and\ \bibinfo {author} {\bibfnamefont
  {T.}~\bibnamefont {Monz}},\ }\href@noop {} {\bibinfo {title} {A universal
  qudit quantum processor with trapped ions},\ } (\bibinfo {year} {2021}),\
  \Eprint {http://arxiv.org/abs/2109.06903} {arXiv:2109.06903 [quant-ph]}
  \BibitemShut {NoStop}%
\bibitem [{\citenamefont {Wei}\ \emph {et~al.}(2013)\citenamefont {Wei},
  \citenamefont {Ren},\ and\ \citenamefont {Deng}}]{Wei2013}%
  \BibitemOpen
  \bibfield  {author} {\bibinfo {author} {\bibfnamefont {H.-R.}\ \bibnamefont
  {Wei}}, \bibinfo {author} {\bibfnamefont {B.-C.}\ \bibnamefont {Ren}}, \ and\
  \bibinfo {author} {\bibfnamefont {F.-G.}\ \bibnamefont {Deng}},\ }\bibfield
  {title} {\emph {\bibinfo {title} {Geometric measure of quantum discord for a
  two-parameter class of states in a qubit--qutrit system under various
  dissipative channels},\ }}\href {\doibase 10.1007/s11128-012-0458-8}
  {\bibfield  {journal} {\bibinfo  {journal} {Quantum Information Processing}\
  }\textbf {\bibinfo {volume} {12}},\ \bibinfo {pages} {1109} (\bibinfo {year}
  {2013})}\BibitemShut {NoStop}%
\end{thebibliography}%


%

\vspace{0.25cm}
\noindent
\textbf{Acknowledgements} We gratefully acknowledge funding by the U.S. ARO Grant No. W911NF-21-1-0007. We also acknowledge funding by the Austrian Science Fund (FWF), through the SFB BeyondC (FWF Project No. F7109), by the Austrian Research Promotion Agency (FFG) contract 872766, by the EU H2020-FETFLAG-2018-03 under Grant Agreement no. 820495, and by the Office of the Director of National Intelligence (ODNI), Intelligence Advanced Research Projects Activity (IARPA), via the U.S. ARO Grant No. W911NF-16-1-0070. This project has received funding from the European Union’s Horizon 2020 research and innovation programme under the Marie Sk{\l}odowska-Curie grant agreement No 840450, as well as by the European Research Council (ERC) via ERC Starting Grant QNets Grant Number 804247. It reflects only the author's view, the EU Agency is not responsible for any use that may be made of the information it contains. We acknowledge the use of computational resources from the parallel computing cluster of the Open Physics Hub at the Physics and Astronomy Department in Bologna.\\
\textbf{Supplementary Material} is available for this paper. \\
\textbf{Author contributions} DV and MM derived the theory results. RS, AE, LP, MiM, MR, PS, and TM performed the experiments. RS analyzed the data. PS, TM, MM, and RB supervised the project. All authors contributed to writing the manuscript. \\
\textbf{Competing interests} The authors declare no competing interests. \\
\textbf{Materials \& Correspondence} Requests for materials and correspondence should be addressed to RS (email: roman.stricker@uibk.ac.at).
\normalsize

\onecolumngrid
\setcounter{figure}{0}
\setcounter{equation}{0}
\setcounter{table}{0}
\setcounter{section}{0}
\makeatletter 
\renewcommand{\theequation}{S\@arabic\c@equation}
\renewcommand{\thefigure}{S\@arabic\c@figure}
\renewcommand{\thetable}{S\@arabic\c@table}
\renewcommand{\RS}[1]{{\color{BLUE}{[RS]: #1}}}
\makeatother

\begin{center}
{\bf \large Supplementary Material: \\
Characterizing quantum instruments: \\
from non-demolition measurements to quantum error correction}
\label{SI}
\end{center}
\medskip
The additional information presented here aims at providing further experimental and theoretical results supporting our findings into more detail. We start by giving a broader insight on the experimental setup in Sec.~\ref{SISec:Toolbox}. Upon those building blocks we will be able to derive the maps underlying our quantum non-demolition (QND) loss detection unit for both asymmetric loss and the quantum erasure-channel. This will be complemented by further experimental data, both presented in section Sec.~\ref{SISec:QND}. We continue in Sec.~\ref{SISec:NoiseModel} by developing a noise model giving a well-founded description to our experimental limitations. Thereafter, those noise models form the basic building blocks to numerical simulations studying the implications of the loss detection in respect of quantum error correcting (QEC) codes. We conclude with Sec.~\ref{SISec:ColorCode} by giving more detailed derivations covering the loss treatment in the 7-qubit color code.

\section{Experimental details and toolbox}
\label{SISec:Toolbox}
Our ion-trap based quantum computer works on a chain of $^{40}$Ca$^+$ ions stably confined within ultra high vacuum using a linear Paul trap~\cite{toolbox}. Each ion represents a physical qubit encoded in the electronic levels $S_{1/2}(m=-1/2) = \ket{0}$ and $D_{5/2}(m=-1/2) = \ket{1}$ denoting the computational subspace. Qubit manipulation is realized by coherent laser-ion interaction. Thereby the setup is capable of implementing a universal set of quantum gate operations consisting of addressed single-qubit rotations with an angle $\theta$ around the x- and the y-axis of the form $\text{R}^\sigma(\theta) = \exp(-i{\theta}\sum_j {\sigma_j}/{2})$ with $\sigma = X$ or $Y$ together with two-qubit M\o{}lmer-S\o{}renson entangling gate operations $\text{MS}_{i,j}(\theta) = \exp(-i{\theta}\sum_{i,j} X_i X_j /2)$~\cite{MSgate}. Multiple addressed laser beams, coherent among themselves, allow for arbitrary two-qubit connectivity across the entire ion string. Read-out is performed by driving a dipole transition only resonant to the lower qubit level $\ket{0}$ and simultaneously collecting its scattered photons revealing the qubit's state after multiple shots. This dipole laser collectively covers the entire ion string. However, we are able to read out only a subset of the qubits within the register, by shelving electronic populations of specific qubits in the upper D-state manifold, referred to as addressed read out. Apart from quantum non-demolition measurements this becomes essential for reading out quantum trits (qutrits) demanding for two subsequent detections separating both levels $D_{5/2}(m=-1/2) = \ket{1}$ and $D_{5/2}(m=+1/2) = \ket{2}$, see~\cite{Ringbauer2021}. 

\section{Quantum Instrument: QND loss detection}
\label{SISec:QND}
This section gives a more thorough introduction to the QND-detection, serving as our quantum instrument working example, by deriving all maps relevant to our studies. Then, additional experiments are presented addressing the demonstration of the QND-detection's principal working ability complemented by results on the higher dimensional process tomography fully characterizing its underlying maps.

As loss on our setup naturally occurs on rates similar to those of single-qubit errors we introduce it in controlled fashion. For instance from the system qutrit's (q) state $\ket{0}_q$ loss can be induced by coherently transferring part of the population outside the computational subspace into $D_{5/2}(m=+1/2)=\ket{2}_q$ via the rotation
\begin{align}
\begin{split}
\mathrm{R}_\text{loss}(\phi) &= \ket{1}\bra{1}_q + \cos(\phi/2) (\ket{0}\bra{0}_q+\ket{2}\bra{2}_q) +\sin(\phi/2)(\ket{0}\bra{2}_q - \ket{2}\bra{0}_q).
\end{split}
\end{align}
The loss rate $\phi$ relates to the loss probability via $p_\text{loss}=\sin^2(\phi/2)$. Note that loss in general can be induced through an arbitrary state $\alpha\ket{0}_q+\beta\ket{1}_q$ with $\abs{\alpha}^2+\abs{\beta}^2=1$ using a single coherent rotation on the system qutrit before and its inverse after the loss rotation $R_\text{loss}(\phi)$. To detect loss two full entangling  $\text{MS}^X(\phi/2)\cdot\text{MS}^X(\phi/2)=\text{MS}^X(\phi)$ couple to ancilla and system qutrit and realize a collective bit-flip only if both qubits are present to their computational subspace:
 \begin{equation}
 \begin{split}
\text{MS}^X(\phi) &= \exp\big( - i\frac{\phi}{2} X_a X_q\big) =\Big(\cos(\phi/2)(\mathds{1}_a \otimes \mathds{1}_q - \ket{2}\bra{2}_q) 
-i \sin(\phi/2) X_a X_q\Big)+\ket{2}\bra{2}_q
\end{split}
\end{equation}
with 
\begin{align}
\begin{split}
\mathds{1}_a &=
\begin{pmatrix}
1&0\\
0&1\\
\end{pmatrix},
\hspace{0.3cm} \mathds{1}_q =
\begin{pmatrix}
1&0&0\\
0&1&0\\
0&0&1\\
\end{pmatrix},
\hspace{0.3cm}X_{a} =
\begin{pmatrix}
0&1\\
1&0\\
\end{pmatrix},\hspace{0.3cm} X_{q} =
\begin{pmatrix}
0&1&0\\
1&0&0\\
0&0&0\\
\end{pmatrix}.
\end{split}    
\end{align}
On the other hand if the system qutrit occupies a state outside the computational subspace, for instance in $\ket{2}_q$, the MS-gate is subject to an identity operation, which can be seen from the argument of its exponential $X_i X_i = \mathds{I}$ acting merely on the ancilla qubit. This follows a collective bit-flip
\begin{align}
\begin{split}
\mathrm{R}^X_a(\pi) &= -i (\ket{0}\bra{1}_a + \ket{1}\bra{0}_a ) \\ 
\mathrm{R}^X_q(\pi) &=\ket{2}\bra{2}_q -i (\ket{0}\bra{1}_q + \ket{1}\bra{0}_q )
\end{split}
\end{align}
Consequently, in the absence of loss the effect of the MS-gate is undone whereas under loss the ancilla qubit gets excited by the final bit-flip signaling the event of loss. The overall unitary combining loss operation and QND detection is given by
\begin{align}
\begin{split}
    U &= \mathrm{R}^X_a(\pi) \mathrm{R}^X_q(\pi)\mathrm{MS}^X(\pi)\mathrm{R_\text{loss}}(\phi)\\
    &=\mathds{1}_a \otimes U^{(0)}   + X_a \otimes U^{(1)}
\label{Eqn:SingleQutritQND}
\end{split}
\end{align}
with 
\begin{align}
\begin{split}
    U_q^{(0)} & =\ket{1}\bra{1}_q + \cos(\phi/2) \ket{0}\bra{0}_q + \sin(\phi/2) \ket{0}\bra{2}_q,   \\
    U_q^{(1)} & = \sin(\phi/2)\ket{2}\bra{0}_q - \cos(\phi/2) \ket{2}\bra{2}_q.
\end{split}
\end{align}
Taking the additional loss state $\ket{2}_q$ on the system qubit into account and by that extending the view from qubit to qutrit one ends up with two unitary processes fully describing this quantum instrument. We emphasize that on the qutrit level the entire dynamics of our detection unit can be captured, which is well exploited by the experiments from Fig.~\ref{Fig:QutritChois} in the main text of the paper.

However, to pick up the discussion on the non-unitary effects potentially leading to unwanted and erroneous mechanisms, we restrict our view again to the qubit level and further assume that no population was initially present in $\ket{2}_q$. Hence, the unitary operators $U_q^{(0)}$ and  $U_q^{(1)}$ reduce to the non-unitary ones
\begin{align}
\begin{split}
    A_q^{(0)} & =\ket{1}\bra{1}_q + \cos(\phi/2) \ket{0}\bra{0}_q, \\
    A_q^{(1)} & = \sin(\phi/2)\ket{2}\bra{0}_q.
\end{split}
\end{align}
leading to single qubit processes describing the QND detection restricted to the system qubit. We can describe both maps $\lbrace A_q^{(0)},A_q^{(1)}\rbrace$ by two trace non-increasing completely-positive (CP) maps $\mathcal{E}_0$ and $\mathcal{E}_1$.
\begin{align}
\begin{split}
    \mathcal{E}_0\colon \rho \mapsto A_q^{(0)}\rho A_q^{(0)\dag} \\
    \mathcal{E}_1\colon \rho \mapsto A_q^{(1)}\rho_q A_q^{(1)\dag} 
\end{split}
\end{align}
acting on the system qubits as 
\begin{equation}
    \rho \mapsto \ket{0}\bra{0}_a \otimes \mathcal{E}_0(\rho)   +   \ket{1}\bra{1}_a \otimes \mathcal{E}_1(\rho)
\label{Eqn:SingleQND}
\end{equation}
where the two maps are together unitary again. It is noteworthy that the no-loss map $\mathcal{E}_0$ initially starting from the superposition state $1/\sqrt{2}(\ket{0}_q+\ket{1}_q)$ would be transitioning to $\ket{1}_q$ as the loss probability from $\ket{0}_q$ increases, which is subject to Fig.~\ref{Fig:SingleQND_Bloch}a in the main text. Only for very little loss $\phi\sim 0$ the no-loss map converges to an identity operation. 

Next to having loss asymmetrically with respect to either computational basis state $\lbrace\ket{0}_q,\Ket{1}_q\rbrace$, we follow a different, often utilized, scenario called the quantum erasure-channel~\cite{Grassl1997}. Its circuit is depicted in Fig.~\ref{Fig:QInstrument}c from the main text. First, partial loss is induced from $\ket{0}_q$ followed by its detection. The protocol only continues in the absence of loss by inducing the same partial loss from the other qubit state $\ket{1}_q$ together with its detection. The second part of the map can be expressed via $\lbrace\tilde{A}_q^{(0)},\tilde{A}_q^{(1)}\rbrace$ where we swap the role of $\ket{0}_q$ and $\ket{1}_q$. Thus, the quantum erasure-channel can be described using the following map
\begin{align}
\begin{split}
    \rho \mapsto  & (1-p_L)  (1-\widetilde{p}_L)\widetilde{A}_q^{(0)} A_q^{(0)}  \rho A_q^{(0)\dag} \widetilde{A}_q^{(0)\dag}\\
    &+ (1-p_L) \widetilde{p}_L \widetilde{A}_q^{(1)}  \rho \widetilde{A}_q^{(1)\dag}  + p_L A_q^{(1)}  \rho A_q^{(1)\dag}
\end{split}
\end{align}
with probabilities $p_L$ and $\widetilde{p}_L$ for any arbitrary input state $\alpha\ket{0}_q + \beta\ket{1}_q $
\begin{equation}
p_L = \abs{\alpha}^2 \sin^2(\phi/2) \qquad
\widetilde{p}_L = \frac{\abs{\beta}^2 \sin^2(\phi/2)}{(\abs{\alpha}^2\cos^2(\phi/2) + \abs{\beta}^2)}.
\end{equation}
In this case the process reduces to
\begin{equation}
    \rho \mapsto  \cos^2(\phi/2) \rho + \sin^2(\phi/2) \ket{2}\bra{2}_q,
\label{Eqn:Erasure}
\end{equation}  
where the effect of the loss is proportional to the arbitrary input state $\rho$ indicating that after normalization the initial state can be retrieved independent of the loss probability.

\begin{figure}[ht]
\centering
\includegraphics[width=\textwidth]{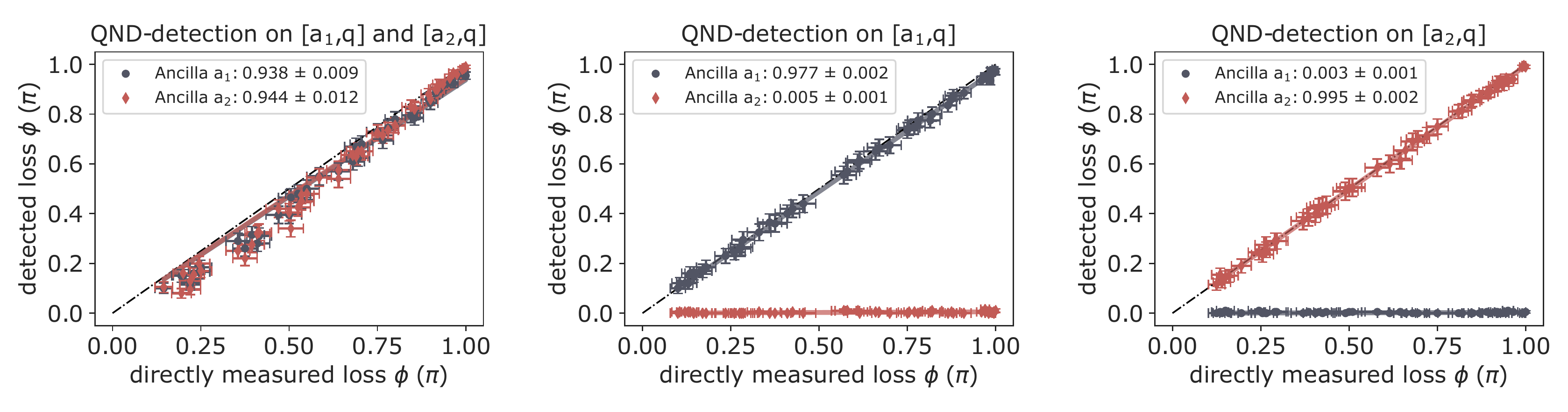}
\caption{\textbf{Investigating the performance of the QND-detection unit according to Fig.~\ref{Fig:QInstrument}b in the main text.} Population in the D$_{5/2}$ state for the system qubit (directly measured loss) vs. transferred excitation on the ancilla qubit (detected loss) in case of detecting loss repeatedly using both ancilla qubits $a_1$ and $a_2$ (left), solely with ancilla $a_1$ (middle) and ancilla $a_2$ (right). The imprinted detection efficiencies demonstrate reliable loss mapping onto the ancilla qubit and its read out by means of QND. Errors correspond to one standard deviation of statistical uncertainty due to quantum projection noise.}
\label{SIFig:QNDTest}
\end{figure}

The basic idea of this quantum instrument is the detection of qubit loss, i.e unwanted leakage to levels outside the computational subspace, that in a realistic scenario would be followed by its correction representing the scope of our foregoing work~\cite{Stricker2020}. To give the rather formal discussion a physical meaning, we demonstrate the unit's working principle. Partial loss induced from $\ket{0}_q$ via the loss transition $R_\text{loss}(\phi)$ is continuously increased and subsequently detected. Note that both qubits are read out denoting their population in the upper $\ket{D}$-state manifold referring to directly measured loss in case of the system qubit and detected loss for the ancilla qubit. In Fig.~\ref{SIFig:QNDTest} results are presented for individual and repeated loss detection employing up to two ancilla qubits. Slopes extracted from the linear-fit in the repeated detection read $0.938(9)$ and $0.944(12)$ for ancilla $a_1$ and $a_2$ respectively. On the individual read-outs we get $0.977(2)$ and $0.995(2)$ with a resonant cross-talk to the ancilla not participating of $0.005(1)$ and $0.003(1)$ respectively. When utilizing $a_2$ we end up with a higher detection efficiency because of a better performing MS-gate on the particular ion-pair. 200 cycles were taken on this measurement. Errors correspond to one standard deviation of statistical uncertainty due to quantum projection noise.

\begin{figure}[ht]
\includegraphics[width=0.8\textwidth]{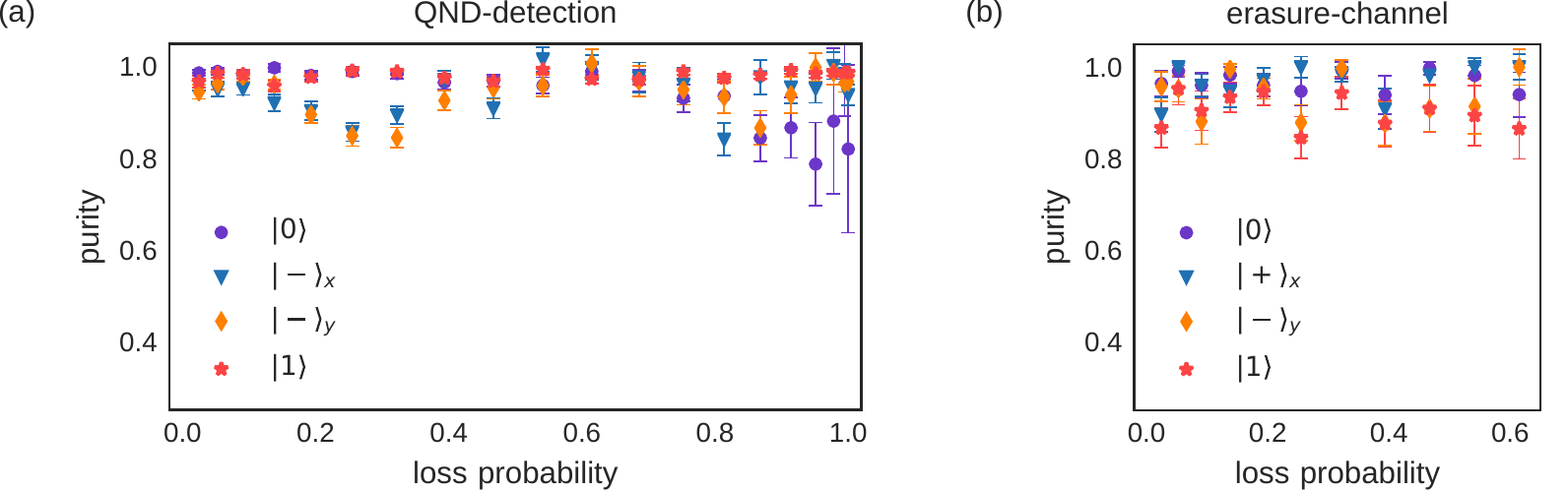}
\caption{\textbf{Purity of a single system qubit after undergoing the QND-detection in the no-loss case for different loss channels}. \textbf{(a)} After a single QND-detection with loss from $\ket{0}_q$ we find purity values unaffected by the amount of loss for all of the given input states. At high loss probabilities, tomography becomes unreliable due to the low count rates. \textbf{(b)} The erasure-channel is realized by consecutively inducing the same partial from $\ket{0}_q$ followed by $\ket{1}_q$ and post-selecting to both ancilla $\ket{0}_a$ outcomes. The purity of the output state is again unaffected by loss for any of the probed input states. Errors correspond to one standard deviation of statistical uncertainty due to quantum projection noise.}
\label{SIFig:QNDPurity}
\end{figure}

Next, we complement the results from  Fig.~\ref{Fig:SingleQND_Bloch}a in the main part revealing a pull towards the state not affected from asymmetric loss by further demonstrating that the purity $\Tr(\rho^2)$ of the associated reconstructed states remains constant across the entire loss probability range; see Fig.~\ref{SIFig:QNDPurity}a. The purity value is found independent of the loss and therefore underlining at first glance a correct experimental outcome whereas only in the Bloch sphere picture (Fig.~\ref{Fig:SingleQND_Bloch}a) deviations due to the non-unitary map become visible. Likewise considerations have been done on the erasure-channel, previously discussed in Fig.~~\ref{Fig:SingleQND_Bloch}b and similarly producing purity values independent of loss, as can be seen in Fig~\ref{SIFig:QNDPurity}b. 

\begin{figure}[ht]
\centering
\includegraphics[width=1.0\textwidth]{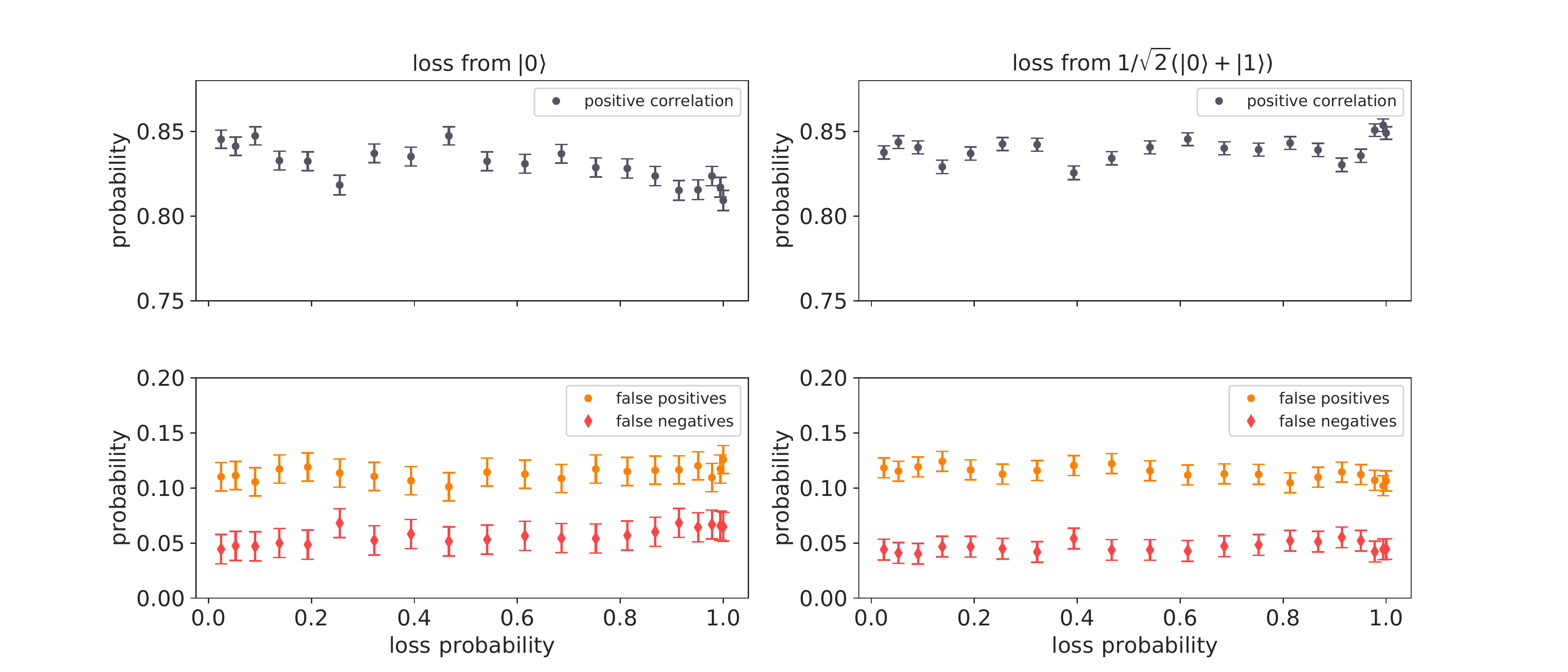}
\caption{\textbf{Correlations between two repeated QND detections according to Fig.~\ref{Fig:QInstrument}b in the main text.} Loss on the system qubit was induced from the imprinted states followed by two repeated detections using ancilla $a_1$ and $a_2$. A Positive correlation refers to successfully detecting the same loss event twice, whereas faulty assignments can be separated in false-positive and false-negative events; shown in the lower figure part. Errors correspond to one standard deviation of statistical uncertainty due to quantum projection noise.}
\label{SIFig:QNDRepeatability}
\end{figure}

Next, we estimate the detection correlation of a single loss event by two repeated detections. Such system capabilities emphasize the work on the erasure-channel and more generally become relevant in a realistic scenario demanding several consecutive read-outs especially when embedded in QEC codes. In Fig.~\ref{SIFig:QNDRepeatability} positive correlation occurs for a certain shot when both ancilla qubits agree upon a certain loss event. Further, the data on the repeated read-out allows us to quantify false-positive and false-negative rates manifesting important failure modes of our detection unit. Again, false-positive rates dominate owing to their strong sensitivity on the entangling MS-gate as was the case in Fig.~\ref{Fig:QutritChois}b in the main text. 100 cycles for $\ket{0}_q$-loss and 200 cycles for $1/\sqrt{2}(\ket{0}_q+\Ket{1}_q)$-loss were taken on this measurement. Errors correspond to one standard deviation of statistical uncertainty due to quantum projection noise.

\begin{figure}[ht]
\includegraphics[width=\textwidth]{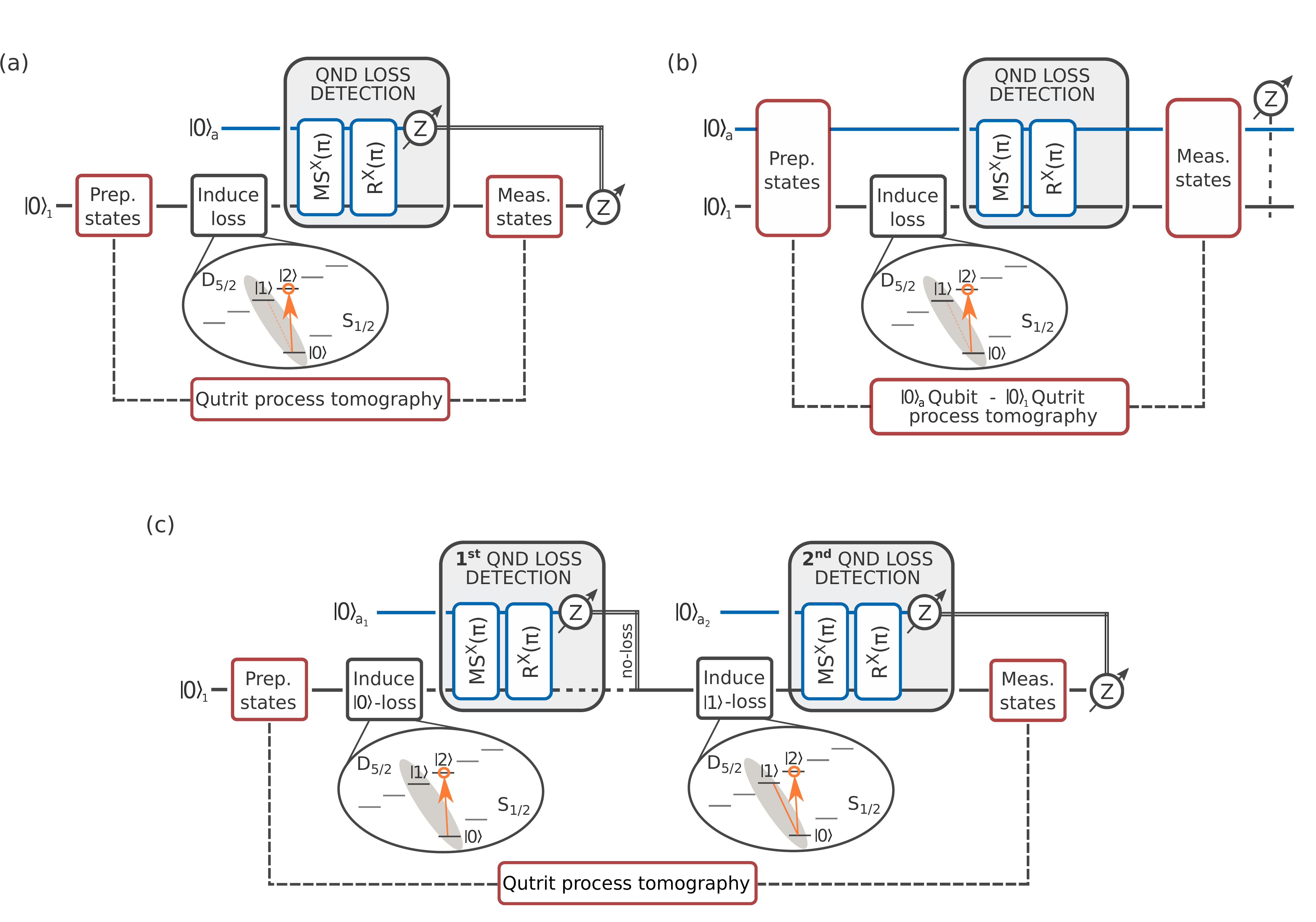}
\caption{\textbf{Schematics on higher dimensional process tomography}. \textbf{(a)} Qutrit process tomography solely covering the system qubit $(q)$ together with the loss level $\lbrace\ket{0}_q,\ket{1}_q,\ket{2}_q\rbrace$ undergoing the QND-detection unit by using 9 preparation settings together with 6 measurement settings resulting in 54 experiments each run. \textbf{(b)} Combined process tomography on ancilla $(a)$ and qutrit $(1)$ capturing the entire dynamics of this quantum instrument using 12 settings on the ancilla qubit (4 preparation settings and 3 measurement settings) alongside 54 settings on the system qutrit resulting in 648 experiments. \textbf{(c)} Qutrit process tomography on the erasure-channel focusing on the no-loss case, i.e. twice post-selecting the ancilla qubit's $\ket{0}_a$ outcome.}\label{SIFig:Tomography_Schematics}
\end{figure}

\begin{figure}[ht]
\centering
\includegraphics[width=0.8\textwidth]{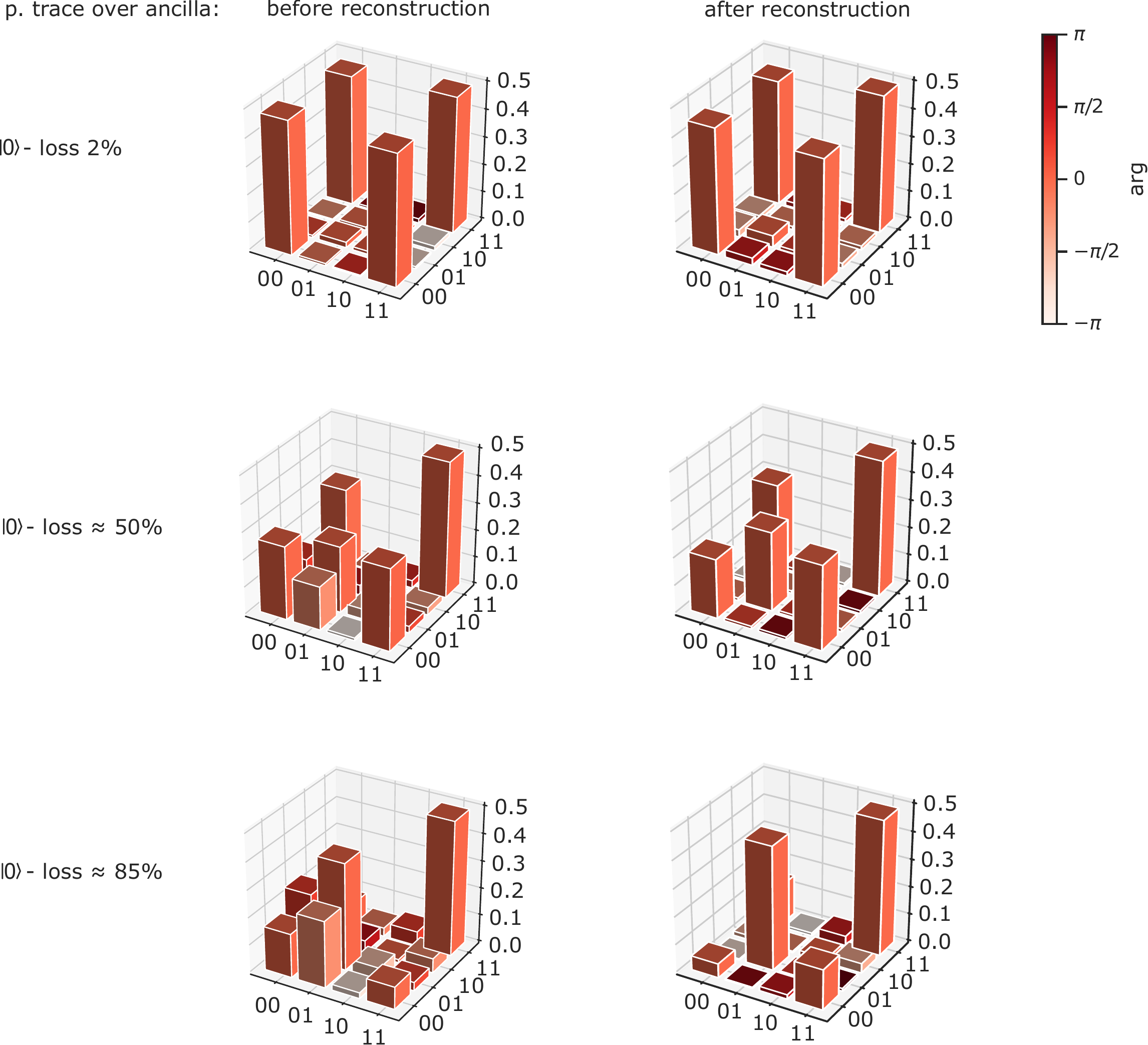}
\caption{\textbf{Potential risks on system qubit process reconstruction when partial tracing the ancilla qubit}. \textbf{(left column)} Partial tracing before reconstructing the map directly on the raw data, previously used in Fig.~\ref{Fig:QutritChois} of the main part. In this case, the loss level $\ket{2}_q$ is incoherently added onto state $\ket{1}_q$. Coherences present in $\ket{01}_q$ originate from the reconstruction technique forcing physical properties. \textbf{(right column)} Post-selecting from the already reconstructed qubit-qutrit maps presented in Fig.~\ref{SIFig:QubitQutrit}. In contrast to before coherences on $\ket{01}_q$ vanish, whereas the unphysical bias remains.}
\label{SIFig:QubitChoiTraced}
\end{figure}

We switch our consideration from qubit to qutrit level and resume the discussion on the process tomography covering both ancilla qubit and system qurtrit from Fig.~\ref{Fig:QutritChois}a in the main text. Thereby, all presented Choi-operators were post-selected upon the ancilla outcome denoting the qutrit maps separated by both loss-cases. This has the advantage of unitary operators describing the full dynamics of the system qutrit in either loss case that moreover gave an estimation on the QND detection's dominant failure mode, namely false-positive and false-negative rates. 
As discussed in detail in the main part of the paper standard tomography restricted to qubit level prevents us from getting such fine-grained analysis for mainly two reasons: First, reliably assigning false-positive and false-negative events is not possible when post-selecting by the ancilla qubit's measurement outcome. Second, when tracing over the ancilla the loss state $\ket{2}_q$ is incoherently added to the qubit state $\ket{1}_q$ creating an unphysical bias under which tomography is likely to break as demonstrated in Fig.~\ref{SIFig:QubitChoiTraced}. Here, we distinguish between tracing before and after tomography reconstruction. On the one hand, when directly tracing on the raw data and subsequently reconstructing the map it includes coherences owing to the reconstruction technique forcing physical properties. On the other hand, when tracing after process reconstruction coherences on $\ket{01}_q$ vanish. Both cases draw attention to potential risks on how commonly known process tomography fails to describe quantum instruments.

\begin{figure}[ht]
\centering
\includegraphics[width=\textwidth]{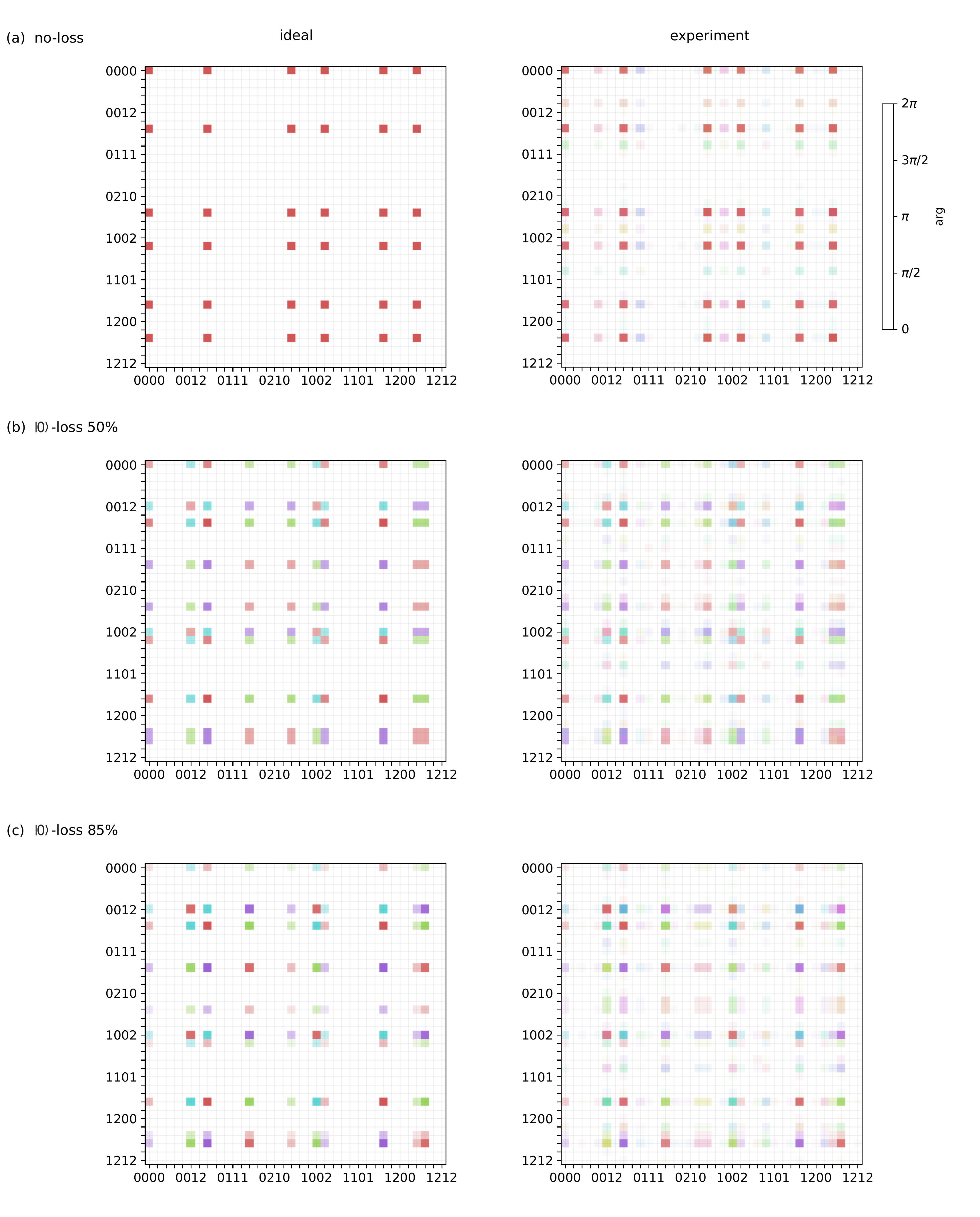}
\caption{\textbf{Combined process reconstruction on ancilla-qutrit according to Fig.~\ref{SIFig:Tomography_Schematics}b}. The resulting Choi-operators (right column) denoted in elementary basis ($\lbrace\ket{0000}_{a,q},...,\ket{1212}_{a,q}\rbrace$) describe the whole dynamics of the QND-detection unit under loss from $\ket{0}_q$. Hue relates to phase according to the top right color bar and saturation to the absolute value of the complex entries. Process fidelities with the ideal Choi-operators (left column) from top to bottom read $\lbrace 0.91(1), 0.89(1), 0.85(1)\rbrace$. Errors correspond to 1 standard deviation of statistical uncertainty due to quantum projection noise.}
\label{SIFig:QubitQutrit}
\end{figure}

In the context of the numerical simulations covering implications on QEC however, we made use of the full map capturing the combined ancilla-qutrit dynamics together with the noise-models; further discussed below. We present experimentally estimated ancilla-qutrit Choi-operators for various loss probabilities in Fig.~\ref{SIFig:QubitQutrit} using the elementary basis according to $\lbrace\ket{0000}_{a,q},...,\ket{1212}_{a,q}\rbrace$. The process tomography of every loss probability required $54\times12=648$ experimental settings. For the sake of clarity we plot ideal Choi-operators (left column) and the experimental ones (right column) for various loss probabilities separated by rows side by side. Color and saturation refer to argument and absolute value of the complex matrix entries. Process fidelities with the ideal Choi-operator from top to bottom read $\lbrace 0.91(1), 0.89(1), 0.85(1)\rbrace$ referring to the loss probabilities $\lbrace \SI{0}{\%}, \SI{50}{\%}, \SI{85}{\%}\rbrace$. 100 cycles were taken for each experiment. In the no-loss case the expected controlled $\hat{X}_a$ operation signaling a loss event whenever the system qubit occupies level $\ket{2}_q$ is clearly reproduced as expressed by Eq.~\eqref{Eqn:SingleQutritQND} derived in the beginning of this section.

\begin{figure}[ht]
\centering
\includegraphics[width=0.8\textwidth]{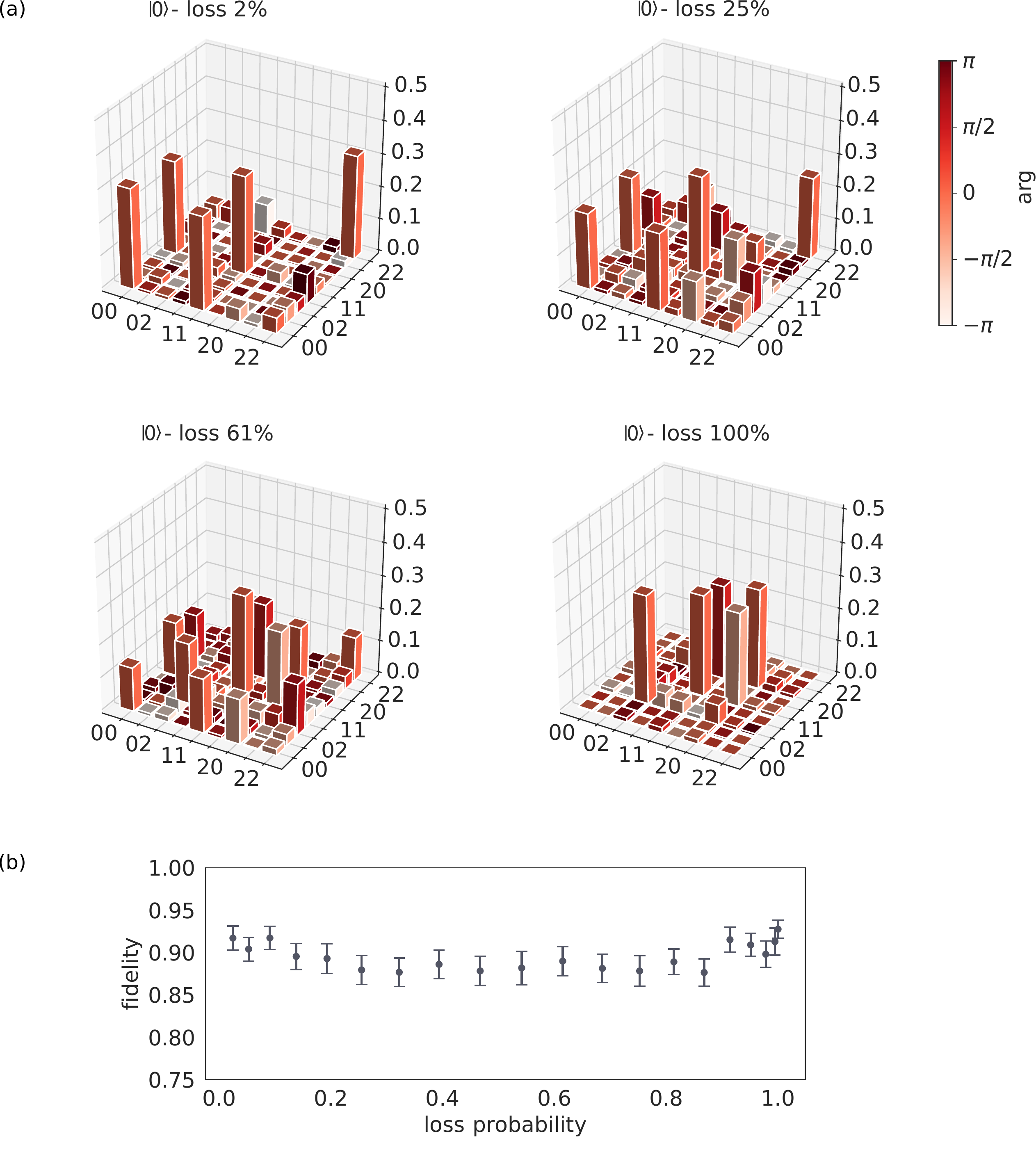}
\caption{\textbf{Qutrit process tomography characterizing the QND-detection unit for loss from $\ket{0}_q$ according to Fig.~\ref{SIFig:Tomography_Schematics}a}. \textbf{(a)} System qutrit's Choi-operator in elementary basis $\lbrace\ket{00}_q,...,\ket{22}_q\rbrace$ after tracing over the ancilla qubit and various loss probabilities denoting the effect of the loss-transition transferring population from $\ket{0}_q$ to $\ket{2}_q$. \textbf{(b)} The respective fidelities with the ideal Choi-operators covering the complete loss range.}
\label{SIFig:SingleQNDQutrit}
\end{figure}

\begin{figure}
\centering
\includegraphics[width=0.8\textwidth]{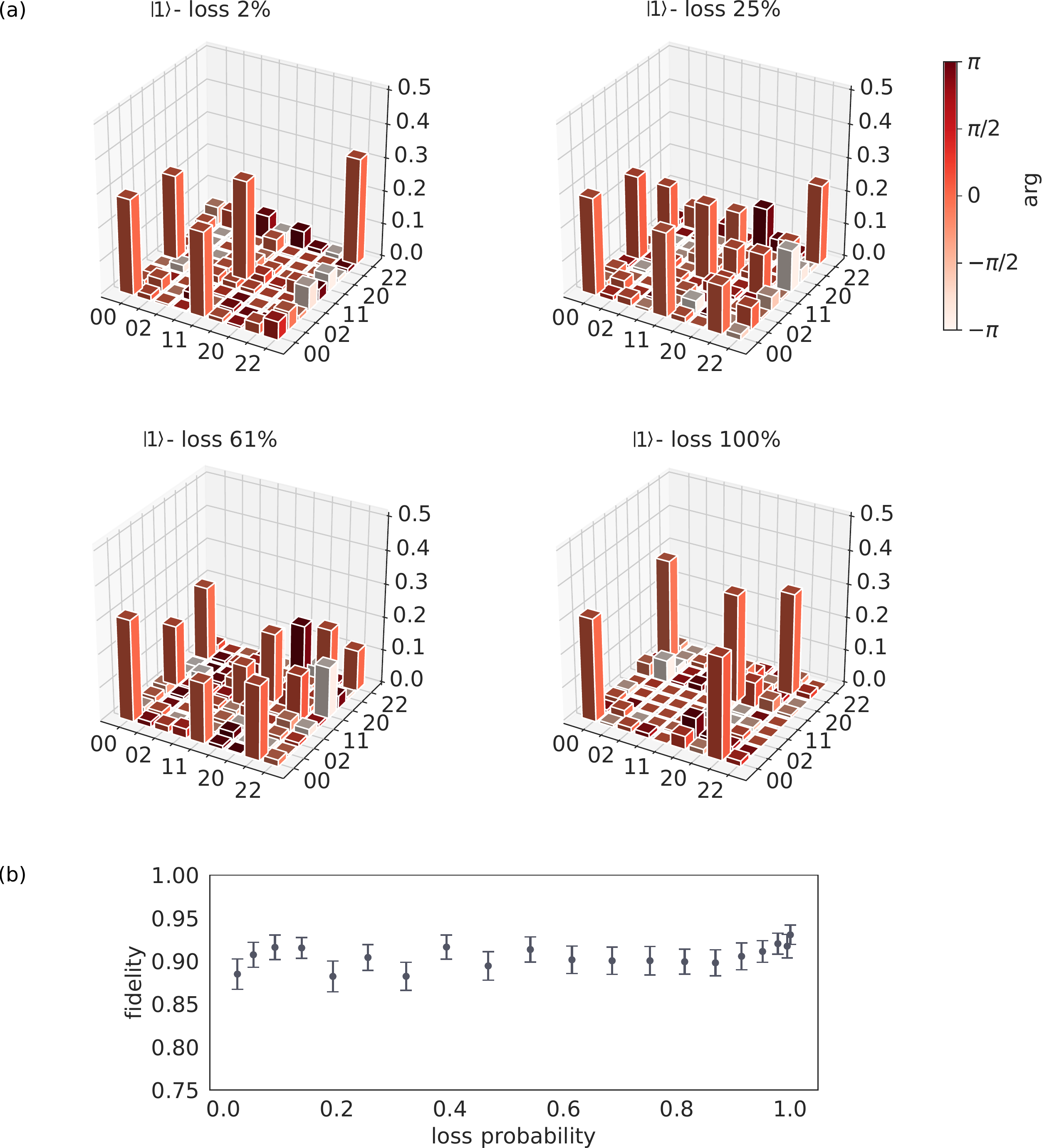}
\caption{\textbf{Qutrit process tomography characterizing the QND-detection unit for loss from $\ket{1}_q$ according to Fig.~\ref{SIFig:Tomography_Schematics}a}. \textbf{(a)} System qutrit's Choi-operator in elementary basis $\lbrace\ket{00}_q,...,\ket{22}_q\rbrace$ after tracing over the ancilla qubit and several different loss probabilities denoting the effect of the loss-transition transferring population from $\ket{1}_q$ to $\ket{2}_q$. \textbf{(b)} The respective fidelities compared to the ideal Choi-operators covering the complete loss range.}
\label{SIFig:SingleQNDQutritD}
\end{figure}

Finally, we present additional data on the system qutrit process tomography according to Fig.~\ref{SIFig:Tomography_Schematics}a and loss induced from $\ket{0}_q$ and $\ket{1}_q$ presented in Fig.~\ref{SIFig:SingleQNDQutrit} and Fig.~\ref{SIFig:SingleQNDQutritD} respectively. We emphasize that here similar to the qubit level certain coherences vanish when tracing over the ancilla, which is no longer covered by the process tomography. Still, the dynamics on the system qutrit clearly captures the population transfer from either basis state $\lbrace\ket{0}_q,\ket{1}_q\rbrace$ to the loss level $\ket{2}_q$. Further, a change in the asymmetric behaviour between loss from $\ket{0}_q$ and $\Ket{1}_q$ becomes distinctly visible in the qubit subspace.

\begin{figure}[ht]
\centering
\includegraphics[width=0.8\textwidth]{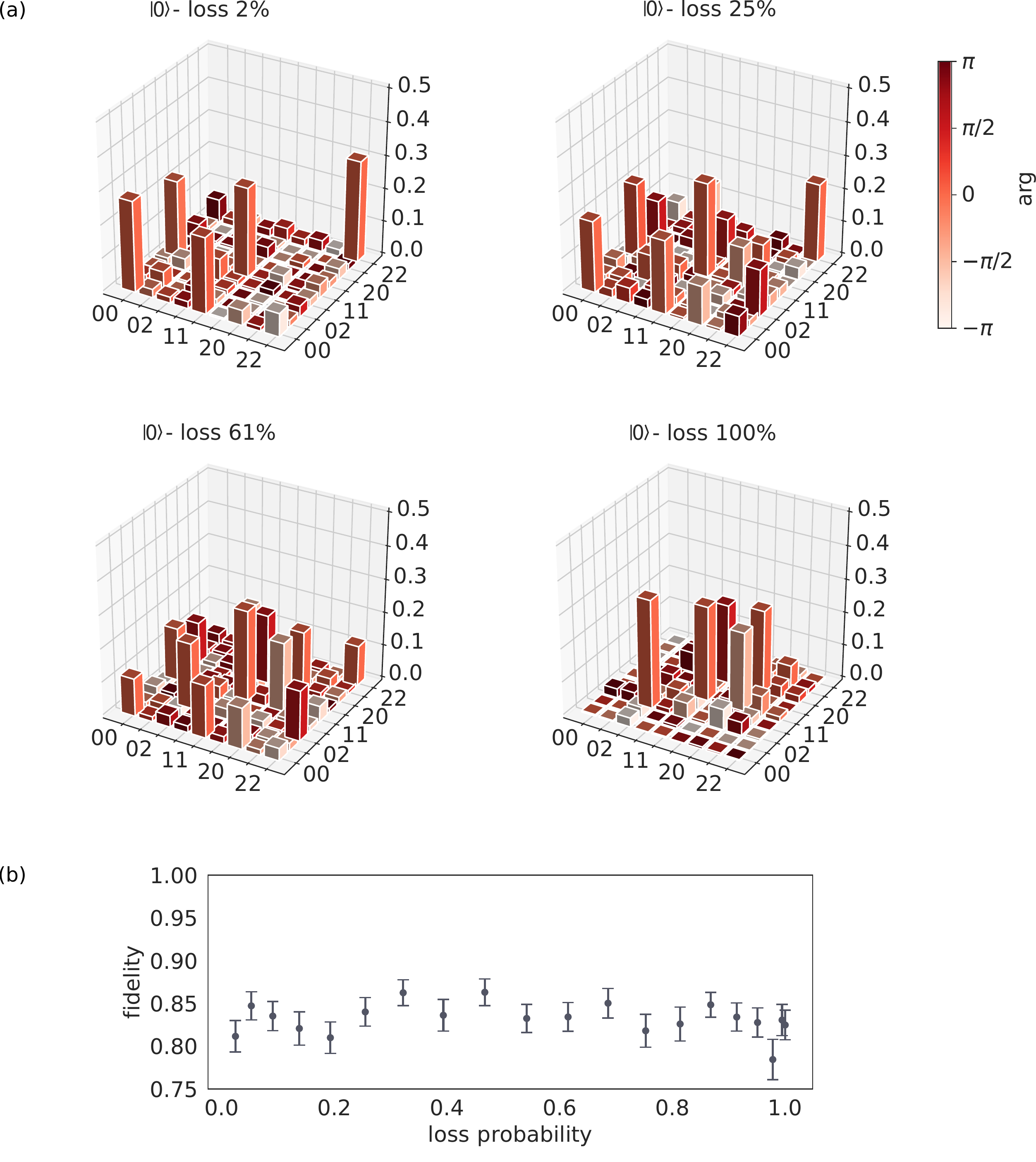}
\caption{\textbf{Qutrit process tomography on two repeated QND-detections for loss from $\ket{0}_q$ according to Fig.~\ref{SIFig:Tomography_Schematics}a}. \textbf{(a)} System qutrit Choi-operators mapping loss repeatedly onto ancilla $a_1$ and $a_2$ under several different loss probabilities. The processes for which we traced over both ancillas prior to reconstruction denote the loss-transition transferring population from $\ket{0}_q$ to $\ket{2}_q$. \textbf{(b)} Fidelities compared to the ideal operators remain approximately constant along all measured loss probabilities and show slightly decrased values compared to the results on the single QND detection from Fig.~\ref{SIFig:SingleQNDQutrit}.}
\label{SIFig:RepeatedQNDQutrit}
\end{figure}

For the sake of completeness we present similar Choi-operators on the repeated loss detection in Fig.~\ref{SIFig:RepeatedQNDQutrit} consecutively mapping the same loss event to two different ancilla qubits; shown for loss from $\ket{0}_q$. As the reconstructed Choi-operators follow the expected behaviour previously observed, their fidelities turn out slightly lower compared to Fig.~\ref{SIFig:SingleQNDQutrit} as expected due to the more complex experiment. 

\section{Noise model on QND loss detection}
\label{SISec:NoiseModel}

\begin{figure}[ht]
\centering
\includegraphics[width=0.85\textwidth]{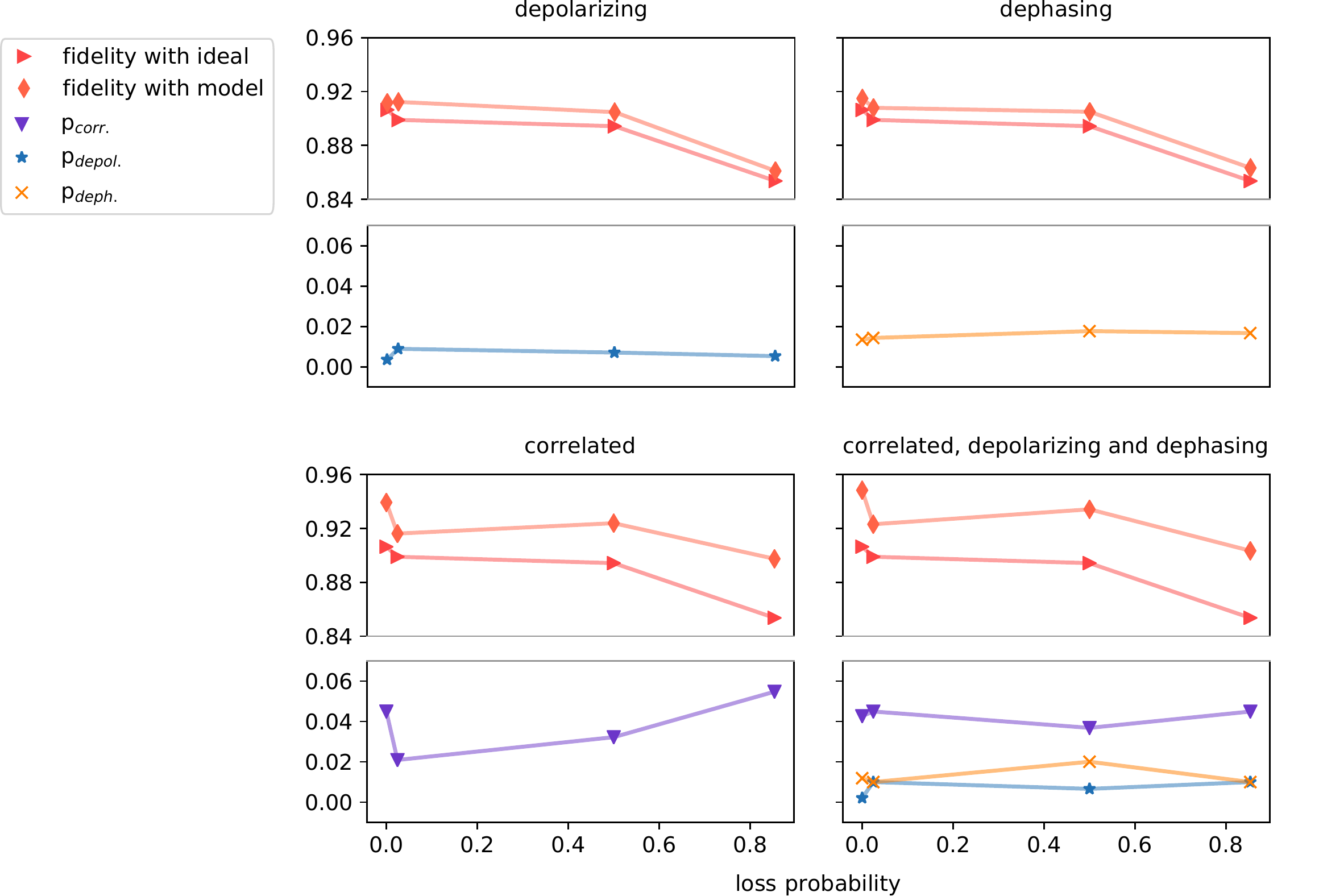}
\caption{\textbf{Noise-models QND-detection}. Various noise-models describing the experimental limitations on the ancilla-qutrit Choi-operator depicted in Fig.~\ref{SIFig:QubitQutrit}. The limitations are best described when combining correlated coherent rotations together with depolarizing and dephasing noise. Correlated errors clearly dominate as depolarizing and dephasing errors only lead to minor improvements. The error parameters on the bottom of each plot refer to depolarizing error $p_\text{depol.}$, dephasing error $p_\text{deph.}$ and correlated error $p_\text{corr.}$, the latter according to Eq.~\eqref{SIEq:CorrelatedNoiseModel}. Lines are connecting the points for clarity.}
\label{SIFig:NoiseModel}
\end{figure}

\begin{figure}
\centering
\includegraphics[width=0.95\textwidth]{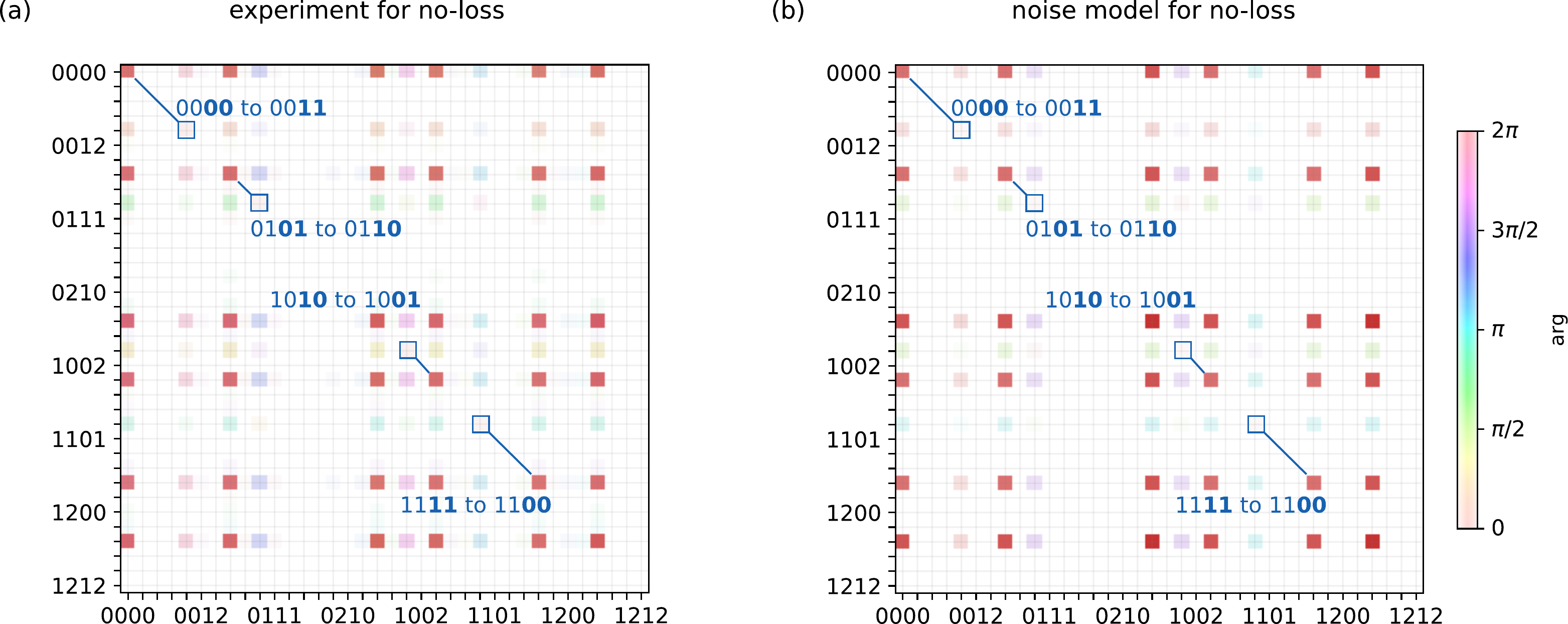}
\caption{\textbf{Comparison between experimental and noisy modeled Choi-operators}. \textbf{(a)} Experimentally estimated map according to Fig.~\ref{SIFig:QubitQutrit}a with additionally marked transitions denoting correlated errors describing our leading error mechanism, see Eq.~\eqref{SIEq:CorrelatedNoiseModel}. \textbf{(b)} Most suitable noisy modelled Choi-operator combining correlated, depolarizing and dephasing errors.}
\label{SIFig:NoiseModelChoi}
\end{figure}

\begin{table}[ht]
\centering
\begin{tabular}{c c c c c c c c c c c c}
    \toprule
    \midrule
    \multirow{1}[4]{*}{} & \multirow{1}[4]{*}{} & \multicolumn{2}{c}{depolarizing} & \multicolumn{2}{c}{dephasing} & \multicolumn{2}{c}{correlated} & \multicolumn{4}{c}{correlated, depol. and dephas.} \\ 
    \cmidrule(rl){1-1}\cmidrule(rl){2-2}\cmidrule(rl){3-4}\cmidrule(rl){5-6}\cmidrule(rl){7-8}\cmidrule(rl){9-12}
   loss & $\mathcal{F}_\text{ideal}$ & $\mathcal{F}_\text{model}$ & $p_\text{depol.}$ &   $\mathcal{F}_\text{model}$ & $p_\text{deph.}$ & $\mathcal{F}_\text{model}$ & $p_\text{corr.}$ & $\mathcal{F}_\text{model}$ & $p_\text{corr.}$ & $p_\text{depol.}$ & $p_\text{deph.}$ \\ 
        \cmidrule(rl){1-1}\cmidrule(rl){2-2}\cmidrule(rl){3-4}\cmidrule(rl){5-6}\cmidrule(rl){7-8}\cmidrule(rl){9-12}
    \multicolumn{1}{r}{\SI{0}{\%}}  & 0.906 & 0.912 & 0.004 & 0.915 & 0.013 & 0.939 & 0.045 & 0.948 &  0.042 & 0.012 & 0.002\\
    \multicolumn{1}{r}{\SI{2}{\%}}  & 0.899 & 0.912 & 0.009 & 0.908 & 0.014 & 0.916 & 0.021 & 0.923 & 0.045 & 0.010 & 0.010\\ 
    \multicolumn{1}{r}{\SI{50}{\%}}  & 0.894 & 0.905 & 0.007 & 0.905 & 0.018 & 0.924 & 0.032 & 0.934 & 0.037 & 0.020 & 0.007\\
    \multicolumn{1}{r}{\SI{85}{\%}}  & 0.854 & 0.861 & 0.005 & 0.863 & 0.017 & 0.897 & 0.054 & 0.903 & 0.045 & 0.010 & 0.010\\
    \midrule
    \bottomrule
\end{tabular}
\caption{\textbf{Summary on noise model parameters and results}: The parameters and fidelities refer to the best suitable model values describing the experimental noise from Fig.~\ref{SIFig:NoiseModel}: depolarizing error $p_\text{depol.}$, dephasing error $p_\text{deph.}$ and correlated error $p_\text{corr.}$ according to Eq.~\eqref{SIEq:CorrelatedNoiseModel}.}
\label{SITab:Params}
\end{table}

Here, we study various noise models in order to find the best suitable description of the experimental limitations underlying our QND-detection. Although very small contributions will be precluded by state preparation and measurement errors (SPAM-errors) the resulting models give us a rough estimate as a guide for where to look at upon which a microscopic noise model for the numerical simulations can be developed. Getting more insights on these error mechanisms is essential when observing implications of the quantum instrument in context of QEC protocols and is further an essential building block towards fault-tolerant quantum computation; see implication on QEC section in the main text. 

We refer to Eq.~\eqref{Eqn:SingleQutritQND} from above and express the action of the ideal QND map $U$ under a given loss rate $\phi$ acting on the combined ancilla-qutrit system in terms of the Choi operator $\rho^\text{CJ} = \mathds{1} \otimes U \cdot\ket{\Phi_+}\bra{\Phi_+}\cdot\mathds{1} \otimes  U^\dagger$ where $\ket{\Phi_+}\bra{\Phi_+}$ is the maximally entangled state of two copies of the ancilla-qutrit system.  An erroneous channel $\mathcal{E}_\text{noise}$ transforms the Choi operator $\rho^{\text{CJ}}$ to  $\rho^\text{CJ}_\text{noise} = (\mathds{1} \otimes \mathcal{E}_\text{noise})( \rho^\text{CJ} )$. Noise rates entering $\mathcal{E}_\text{noise}$ for given model parameters are then extracted by minimizing the distance between modelled noisy Choi-operators $\rho^\text{CJ}_\text{noise}$ and the experimentally determined ones $\rho^\text{CJ}_\text{exp}$ from Fig.~\ref{SIFig:QubitQutrit}. As a measure for the distance in the cost-function we minimize the infidelity:
\begin{equation}
\vert\vert 1-\mathcal{F}(\rho^\text{CJ}_\text{exp}, \rho^\text{CJ}_\text{noise})\vert\vert.
\end{equation}
Our initial considerations covered the study of the QND detection's failure modes, i.e.~false-positive and false-negative rates both quantified in the main part of the paper. Measuring process tomography however comes along with overhead in the form of preparation and measurement gates followed by two consecutive detections at the end of each experiment required for reading out the qutrit's state. Therefore SPAM errors are not to be neglected and lead to a significant bias on false-positive and false-negative rates. With this in mind we put the failure modes aside and focus on experimental limitations instead. In the following, we consider as model for $\mathcal{E}_\text{noise}$ a depolarizing, a dephasing channel, and coherent two- and single-qubit overrotations.

\textbf{Depolarizing and dephasing channels}.--- We start off by testing the agnostic models, namely depolarizing and dephasing channels as those represent error mechanisms typically considered in the field of quantum computation~\cite{NielsenChuang}. The effect of the latter can be understood by losing phase information between the quantum states involved. Coherences get lost and an arbitrary single qubit state in the Bloch sphere picture would finally shrink onto the $Z$-axis as no phase information is left. Depolarizing noise can be considered as simultaneous dephasing in $X$, $Y$ and $Z$ basis eventually leading to a complete mixed state which for a single qubit can be illustrated by shrinking the Bloch sphere towards its center. Note that, we implement those models such that they act both on the ancilla and the qutrit using only a single noise parameter~\cite{Wei2013}. The upper row of Fig.~\ref{SIFig:NoiseModel} depicts the fidelities (top part) for the individual models at the optimized parameters (bottom part). Both results indicate similar improvements compared to the fidelity with the ideal QND map from Eq.~\eqref{Eqn:SingleQutritQND}. Numbers on fidelities and optimized parameters for depolarizing noise $p_\text{depol.}$ and dephasing noise $p_\text{deph.}$ are further summarized in Tab.~\ref{SITab:Params}. The parameters typically lie around 1\% or below, yet the small increase in fidelity indicates other error mechanisms to be more dominant. 

\textbf{Correlated two-qubit overrotations}.---The erroneous peaks in the experimentally estimated Choi-operators from Fig.~\ref{SIFig:QubitQutrit} imply that additional rotations should be taken into account by the agnostic models. Those dominant error peaks are found originating from correlated rotations between ancilla and system qubit as illustratively labeled in Fig.~\ref{SIFig:NoiseModelChoi}a. Note that the error terms are restricted to qubit level and partial coherences are still present. Hence, if the system qutrit's state is~$\ket{2}_q$ no correlated error is induced on the ancilla qubit. Therefore, correlated errors are due to faulty entangling MS-gates. A potential noise model covering correlated rotations in such a way reads:
\begin{equation}
\rho \mapsto \mathcal{E}_\text{noise}(\rho) =  U_\text{corr} \rho U_\text{corr}^{\dag}
\end{equation}
with
\begin{align}
\begin{split}
 U_\text{corr} &= \cos\frac{\alpha}{2}\ \mathds{1}_a\otimes\mathds{1}_q +i \sin\frac{\alpha}{2} \ (X_a\otimes X_{q}+\mathds{1}_a\otimes\ket{2}\bra{2}_q) 
\label{SIEq:CorrelatedNoiseModel}
\end{split}
\end{align}
where $\alpha$ describes the correlated under-/ overrotations and relates to the corresponding error probability via $p_\text{corr.}=\sin(\alpha/2)^2$. For comparison a value  $p_\text{corr.}$ of $0.5$ would induce a maximally entangling two-qubit operation on ancilla and system qubit. We first test the model alone followed by combining it with depolarizing and dephasing noise. The resulting fidelities (top part) at the optimized model parameters (bottom part) are shown in the second row of Fig.~\ref{SIFig:NoiseModel} and clearly overcome the ones on the agnostic models denoting correlated rotations to be our leading noise mechanism. The effect of the additional depolarizing and dephasing noise (bottom right) leads to slight improvements. The modeled Choi-operator on this combined noise model is plotted for the no-loss case in Fig.~\ref{SIFig:NoiseModelChoi}b showing strong similarities to the experimental one from part a underlining a good agreement between model and experiment. Numbers on fidelities and optimized parameters for all models are summarized in Tab.~\ref{SITab:Params}.

\textbf{Correlated and single overrotations}.---Finally, we combine the action of correlated rotations with single-qubit rotations on the ancilla and the qutrit and we consider the coherent error model given by
\begin{equation}\label{SIEqn:CorrelatedSingleRotationChannel}
\rho \mapsto \mathcal{E}_\text{noise}(\rho) =  {R}\ U_\text{corr} \rho U_\text{corr}^{\dag} \ {R}^{\dag}
\end{equation}
where $\rho$ is the state obtained after the application of the loss operation $U$ of Eq.~\eqref{Eqn:SingleQutritQND} (see also Fig.~\ref{Fig:7QubitCode}a of the main text), $U_\text{corr}$ is a correlated two-qubit overrotation defined in Eq.~\eqref{SIEq:CorrelatedNoiseModel} and $R = \mathrm{R}^X_a(\beta) \mathrm{R}^X_q(\beta)$ with
\begin{gather}\label{eqn_rotation_x_a}
R^X_a(\beta) = \cos (\beta / 2) - i\sin (\beta / 2) X_a \\ 
R^X_q(\beta) = \cos (\beta / 2)(1 - \ket{2}\bra{2}_q)  - i\sin (\beta / 2) X_q + \ket{2}\bra{2}_q  \label{eqn_rotation_x_q} 
\end{gather}
are overrotations with angle $\beta$ of the ancilla and the qutrit system which corresponds to the single qubit flip error rate $p_\text{single} = \sin(\beta / 2)^2$. After the measurement of the ancilla, the quantum process arising from the erroneous channel in Eq.~\eqref{SIEqn:CorrelatedSingleRotationChannel} can be written as
\begin{equation}\label{eqn_supp_coherent_channel}
\rho \mapsto \ket{0}\bra{0}_a \otimes \mathcal{R}_0(\rho_q)  +  \ket{1}\bra{1}_a \otimes \mathcal{R}_1(\rho_q)
\end{equation}
where $\rho_q$ is the state related to the qutrit only and the processes $\mathcal{R}_0$ and $\mathcal{R}_1$ describe the maps that transform the qutrit state in the case of no-loss detected (ancilla qubit in $\ket{0}_a$) and of loss detected (ancilla qubit in $\ket{1}_a$).  The Choi operators $\Lambda_0$ and $\Lambda_1$ of the maps $\mathcal{R}_0$ and $\mathcal{R}_1$ can be computed for all values of the over-rotated angles $\alpha$ and $\beta$. In particular, if we consider small deviations for  $\alpha$ and $\beta$, $\Lambda_0$ and $\Lambda_1$ read at second order

\colorlet{gray_1}{green!10}
\colorlet{gray_2}{blue!10}
\colorlet{gray_3}{red!30}
\colorlet{gray_4}{magenta!30}
\colorlet{gray_5}{orange!30}
\colorlet{gray_6}{teal!30}

\newcommand{\grayone}[1]{%
  \begingroup\setlength{\fboxsep}{0pt}
  \colorbox{gray_1}{#1}%
  \endgroup
}
\newcommand{\graytwo}[1]{%
  \begingroup\setlength{\fboxsep}{0pt}
  \colorbox{gray_2}{#1}%
  \endgroup
}
\newcommand{\graythree}[1]{%
  \begingroup\setlength{\fboxsep}{0pt}
  \colorbox{gray_3}{#1}%
  \endgroup
}
\newcommand{\grayfour}[1]{%
  \begingroup\setlength{\fboxsep}{0pt}
  \colorbox{gray_4}{#1}%
  \endgroup
}

\newcommand{\grayfive}[1]{%
  \begingroup\setlength{\fboxsep}{0pt}
  \colorbox{gray_5}{#1}%
  \endgroup
}

\newcommand{\graysix}[1]{%
  \begingroup\setlength{\fboxsep}{0pt}
  \colorbox{gray_6}{#1}%
  \endgroup
}
{\small
\begin{equation}
\Lambda_0 = 
\left(
\begin{array}{ccccccccc}
\cellcolor{gray_1} 1 -\frac{\alpha ^2}{4} - \frac{\beta^2}{2}& \left(-\frac{\alpha }{4}-\frac{i}{2}\right) \beta  & 0 & \left( -\frac{\alpha }{4}-\frac{i}{2}\right) \beta  & 1 -\frac{\alpha ^2}{4} - \frac{\beta^2}{2}& 0 & 0 & 0 & \frac{i}{2}  \beta  \\
 \left(-\frac{\alpha }{4}+\frac{i}{2}\right) \beta  & \cellcolor{gray_2} \frac{\beta ^2}{4} & 0 & \frac{\beta ^2}{4} & \left(-\frac{\alpha }{4}+\frac{i}{2}\right) \beta  & 0 & 0 & 0 & -\frac{1}{4} \beta ^2 \\
 0 & 0 & 0 & 0 & 0 & 0 & 0 & 0 & 0 \\
 \left( -\frac{\alpha }{4}+\frac{i}{2}\right) \beta  & \frac{\beta ^2}{4} & 0 & \cellcolor{gray_2}\frac{\beta ^2}{4} & \left(-\frac{\alpha }{4}+\frac{i}{2}\right) \beta  & 0 & 0 & 0 & -\frac{1}{4}\beta^2 \\
1 -\frac{\alpha ^2}{4} - \frac{\beta^2}{2} & \left(-\frac{\alpha }{4}-\frac{i}{2}\right) \beta  & 0 & \left(-\frac{\alpha }{4}-\frac{i}{2}\right) \beta  & \cellcolor{gray_1} 1 -\frac{\alpha ^2}{4} - \frac{\beta^2}{2} & 0 & 0 & 0 & \frac{i}{2} \beta  \\
 0 & 0 & 0 & 0 & 0 & 0 & 0 & 0 & 0 \\
 0 & 0 & 0 & 0 & 0 & 0 & 0 & 0 & 0 \\
 0 & 0 & 0 & 0 & 0 & 0 & 0 & 0 & 0 \\
  -\frac{i}{2} \beta  & -\frac{1}{4} \beta ^2 & 0 & -\frac{1}{4} \beta ^2 &  -\frac{i}{2} \beta  & 0 & 0 & 0 & \cellcolor{gray_3} \frac{\beta ^2}{4} 
\end{array}
\right)
\end{equation}

\begin{equation}
\Lambda_1 = 
\left(
\begin{array}{ccccccccc}
\cellcolor{gray_5} \frac{\beta ^2}{4} & \frac{\alpha  \beta }{4} & 0 & \frac{\alpha  \beta }{4} & \frac{\beta ^2}{4} & 0 & 0 & 0 & \left(\frac{\alpha }{4}-\frac{i}{2}\right) \beta  \\
 \frac{\alpha  \beta }{4} & \cellcolor{gray_4}\frac{\alpha ^2}{4} & 0 & \frac{\alpha ^2}{4} & \frac{\alpha  \beta }{4} & 0 & 0 & 0 & \frac{\beta ^2}{4}-\frac{i \alpha }{2} \\
 0 & 0 & 0 & 0 & 0 & 0 & 0 & 0 & 0 \\
 \frac{\alpha  \beta }{4} & \frac{\alpha ^2}{4} & 0 & \cellcolor{gray_4}\frac{\alpha ^2}{4} & \frac{\alpha  \beta }{4} & 0 & 0 & 0 & \frac{\beta ^2}{4}-\frac{i \alpha }{2} \\
 \frac{\beta ^2}{4} & \frac{\alpha  \beta }{4} & 0 & \frac{\alpha  \beta }{4} & \cellcolor{gray_5}\frac{\beta ^2}{4} & 0 & 0 & 0 & \left(\frac{\alpha }{4}-\frac{i}{2}\right) \beta  \\
 0 & 0 & 0 & 0 & 0 & 0 & 0 & 0 & 0 \\
 0 & 0 & 0 & 0 & 0 & 0 & 0 & 0 & 0 \\
 0 & 0 & 0 & 0 & 0 & 0 & 0 & 0 & 0 \\
 \left(\frac{\alpha }{4}+\frac{i}{2}\right) \beta  & \frac{\beta ^2}{4}+\frac{i \alpha }{2} & 0 & \frac{\beta ^2}{4}+\frac{i \alpha }{2} & \left(\frac{\alpha }{4}+\frac{i}{2}\right) \beta  & 0 & 0 & 0 & \cellcolor{gray_6} 1-\frac{\beta ^2}{4} \\
\end{array}
\right)
\end{equation}
}
where we have labeled the qutrit basis states in the order $\ket{00}, \ket{01}, \ket{02}, \dots \ket{22}$. In the next section we discuss how to approximate the channel in Eq.~\eqref{eqn_supp_coherent_channel} with Clifford gates.

\textbf{Effective Clifford channel}.---Before deriving the analytical expression for the Clifford channel, the form of Choi operators $\Lambda_0$ and $\Lambda_1$ allows us to make a qualitative discussion on the events that will form the Clifford channel approximating Eq.~\eqref{eqn_supp_coherent_channel}. In $\Lambda_0$ and $\Lambda_1$ we can easily identify the following events happening to the ancilla-qutrit system: If the ancilla is in $\ket{0}_a$, the qutrit state is left unchanged with probability
\grayone{$1  -  {\alpha^2}/{4} -  {\beta^2}/{2} $} or it undergoes an $X_q$ bit-flip error with probability 
\graytwo{${\beta^2}/{4} $}. When the ancilla is instead in $\ket{1}_a$, the qutrit state is left unchanged in the loss state $\ket{2}\bra{2}_q$ with probability \graysix{$1 - {\beta^2}/{4}$}. 

We can also identify the origin of the false negative and false positive events: From $\Lambda_0$ we see that the qutrit will be projected on the loss state $\ket{2}\bra{2}_q$ with probability \graythree{${\beta^2}/{4} $} while the ancilla will be in the no-loss detected state $\ket{0}_a$. This corresponds to a false negative event whose origin can be traced back to the single qubit overrotation $R$ of Eq.~\eqref{SIEqn:CorrelatedSingleRotationChannel}.

From $\Lambda_1$ we see that when the qutrit is generated in the computational space by $\ket{0}_q$ and $\ket{1}_q$, the ancilla will be found in the loss detected state $\ket{1}_a$. In particular, the qutrit will be left unchanged with probability \grayfive{${\beta^2}/{4} $} and it will undergo an $X_q$ bit-flip error with probability \grayfour{${\alpha^2}/{4}$}. These events correspond to false positive events whose origin can be traced back to the single qubit overrotation $R$ and to the correlated overrotation $U_\text{corr}$ of Eq.~\eqref{SIEqn:CorrelatedSingleRotationChannel}. 

The previous considerations on the Choi operators $\Lambda_0$ and $\Lambda_1$ can be made more precise by computing explicitly the process in Eq.~\eqref{eqn_supp_coherent_channel}  with the help of Eqs.~\eqref{SIEq:CorrelatedNoiseModel}, \eqref{eqn_rotation_x_a} and \eqref{eqn_rotation_x_q} and by retaining only the terms that can be written in the Kraus form $P\rho P^\dag$ where $P$ is a Pauli operator. In this way, we can approximate the channel in Eq.~\eqref{eqn_supp_coherent_channel} as 
\begin{equation}\label{supp_eqn_inchoerent_channel}
\begin{split}\rho \mapsto &  p_a\ P_{01}\rho P_{01}^\dag + p_b\ X_q\rho X_q  +  p_c\ X_a X_q\rho X_q X_a + p_d\ X_a P_{01}\rho  P_{01}^\dag X_a \\ &+  q_a \  X_a \ket{2}\bra{2} \rho \ket{2}\bra{2} X_a + q_b \ \ket{2}\bra{2} \rho \ket{2}\bra{2} 
\end{split}
\end{equation}
where $\rho$ is the density matrix of the whole ancilla and qutrit system, $P_{01} = 1 - \ket{2}\bra{2}_q$ is the projector on the computational space $\{\ket{0}_q, \ket{1}_q\}$ of the qutrit and the probabilities take the form
\begin{gather}
p_a=\sin^2\alpha \sin^4\beta + \cos^2\alpha \cos^4\beta \sim 1 - {\alpha^2}/{4} - {\beta^2}/{2}   \\
p_b={\sin^2\beta}/{4} \sim {\beta^2}/{4} \\ 
p_c=\sin^2\alpha \cos^4\beta + \cos^2\alpha \sin^4\beta \sim {\alpha^2}/{4} \\
p_d = {\sin^2\beta}/{4} \sim {\beta^2}/{4} \\ 
q_a= \cos^2\left({\beta}/{2}\right) \sim 1 - {\beta^2}/{4} \\ 
q_b= \sin^2\left({\beta}/{2}\right)  \sim  {\beta^2}/{4}.
\end{gather}

\begin{figure*}
    \centering
    \includegraphics[width=0.9\textwidth]{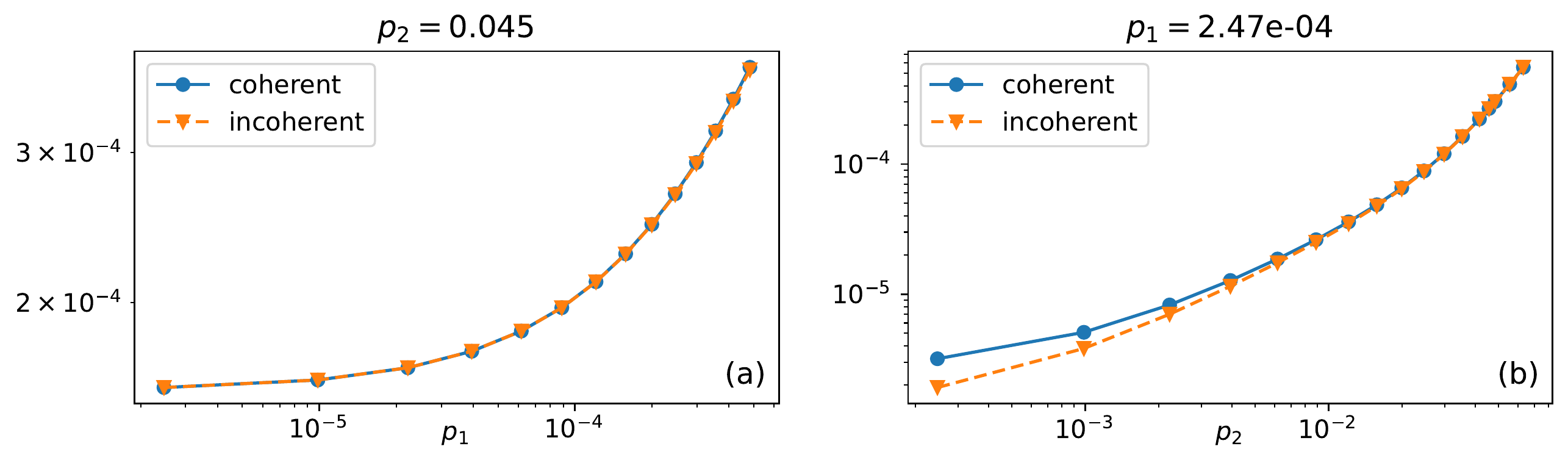}
    \caption{\textbf{Comparison between the coherent and incoherent implementations of the faulty QND loss-detection unit in the case of no losses.} \textbf{(a)} Logical error rate as a function of the correlated overrotation rate $p_1$ for the parameter $p_2 = 0.045$ obtained from the experimental data. \textbf{(b)} Logical error rate as a function of the correlated overrotation rate $p_2$ for the parameter $p_1 =$ 2.47e-4 obtained from the experimental model.}
        \label{supp_fig_coherent_incoherent}
\end{figure*}

The channel in Eq.~\eqref{supp_eqn_inchoerent_channel} can be then implemented in the following way:
\begin{enumerate}
\item If the ancilla is in $\ket{0}_a$ we
\begin{enumerate}
\item leave the qutrit state in the computational space with probability \grayone{$1  - {\beta^2}/{2}  - {\alpha^2}/{4} $};
\item apply an $X_q$ bit-flip error to the qutrit with probability \graytwo{${\beta^2}/{4} $};
\item apply an $X_q$ bit-flip error to the qutrit and an $X_a$ bit-flip error to the ancilla with probability \grayfour{${\alpha^2}/{4} $} (corresponding to a false positive event from the correlated overrotation);
\item leave the qutrit state as it is and flip the ancilla with probability \grayfive{${\beta^2}/{4} $} (corresponding to a false positive event from the single rotations).
\end{enumerate}

\item If the ancilla is in $\ket{1}_a$ we
\begin{enumerate}
\item leave the qutrit state in the loss state $\ket{2}_q$ with probability \graysix{$1 - {\beta^2}/{4}$};
\item flip the ancilla to the no-loss detection state  $\ket{0}_a$ with probability \graythree{${\beta^2}/{4} $} (corresponding to a false negative from the single rotations).
\end{enumerate}
\end{enumerate}
The comparison between the coherent channel in Eq.~\eqref{SIEqn:CorrelatedSingleRotationChannel} and the effective Clifford channel previously described is shown in Fig.~\ref{supp_fig_coherent_incoherent} and in Fig.~\ref{fig_coherent_incoherent} of the main text.

\section{Losses in the 7-qubit code}
\label{SISec:ColorCode}
In this section, we discuss the correction from losses for the 7-qubit color code, in the ideal scenario of perfect QND loss detection and stabilizer measurements. We also assume that losses occur on each qubit independently with loss probability~$p$. 

A loss event is correctable if the density matrix of the losses is fully mixed or, more generally, it does not contain any information on the encoded logical state. With this criterion, we can then check the loss events that can be corrected. Obviously, the event (happening with probability $P_0 = (1-p)^7$) where no loss occurs is  correctable. The events where one loss occurs are also correctable. For showing this, let us consider for instance the encoded $\ket{0_L}$ state 
\begin{equation}\label{supp_eqn_logical_zero_loss}
\ket{0_L} \sim (\mathds{1} + S_{x}^{(1)})(\mathds{1} + S_{x}^{(2)})(\mathds{1} + S_{x}^{(3)})\ket{0}^{\otimes 7}
\end{equation}
where $S_{x}^{(j)}$ are the stabilizer generators and let us suppose that the loss affects the qubit $q_1$ (see Fig.~\ref{Fig:7QubitCode}a of the main text). By introducing the two orthogonal states $\ket{\chi_0} = P_{x}^{(2)} P_{x}^{(3)}\ket{0}^{\otimes 6}$ and  $\ket{\chi_1} = X_2 X_3 X_4 P_{x}^{(2)} P_{x}^{(3)}\ket{0}^{\otimes 6}$ (where $P_{x}^{(j)}  = \mathds{1} + S_{x}^{(j)}$ with $j=2,3$ are chosen because the loss does not belong to $S_{x}^{(j)}$), the state $\ket{0_L}$ can be written  explicitly as
\begin{equation} 
\ket{0_L} \sim \ket{0_1} \ket{\chi_0} + \ket{1_1} \ket{\chi_1}.
\end{equation}
As $\ket{\chi_0}$ and  $\ket{\chi_1}$ are orthogonal, the reduced density matrix of the loss $q_1$ obtained by tracing out the 6 other qubits will be $\rho_1 \sim \ket{0_1}\bra{0_1}+ \ket{1_1}\bra{1_1} $, i.e. it will be fully mixed. Therefore the events with one loss (happening with probability $P_1 = 7p(1-p)^6$) can be correctable. A similar reasoning applies to all the events where two losses happen  ($P_2 = 21 p^2(1-p)^5$) and to the events where three losses that do not form a logical operator happen as well.  The events with three losses that form a logical operator are instead not correctable. There are precisely 7 of such events (corresponding to the logical operators $\mathcal{L} = \{[1,2,5], [1,3,6], [1,4,7],[2,3,7],[4,3,5],[5,6,7], [2,4,6]\}$ in Fig.~\ref{Fig:7QubitCode}a of the main text). The last one ($ [2,4,6] $) is given by the product of the logical operator acting on all the 7 qubits multiplied by all the 3 stabilizer generators. This implies that the probability to successfully recover the logical state is $P_3 = [\binom{7}{3} - 7] p^3(1-p)^4 = 28 p^3(1-p)^4$. In the case of four losses, we have that in 7 cases out of $\binom{7}{4} = 35$ the reduced density matrix of the losses does not depend on the encoded logical state. These cases correspond to the losses happening on the qubits of the stabilizer generators and their products and are given by 
\begin{equation}
\mathcal{S} = \{[1, 2, 3, 4], [2, 3, 5, 6], [3, 4, 6, 7], [1, 4, 5, 6], \linebreak[0] [1, 2, 6, 7], [2, 4, 5, 7], [1, 3, 5, 7]\}.
\end{equation}

This can be shown by considering for instance four losses happening on the stabilizer $[1, 2, 3, 4]$. A bit of algebra shows that the logical states $\ket{0_L}$ and $\ket{1_L}$ can be written as
\begin{gather}
\ket{0_L} =  \ket{G}\ket{000} + X_2 X_3\ket{G}\ket{110} + X_3 X_4 \ket{G}\ket{011} + X_2 X_4\ket{G}\ket{101} \\
\ket{1_L} =  \ket{G}\ket{111} + X_2 X_3\ket{G}\ket{001} + X_3 X_4 \ket{G}\ket{100} + X_2 X_4\ket{G}\ket{010}.
\end{gather}
where $\ket{G} = \ket{0000} + \ket{1111}$ is a GHZ state of the qubits 1, 2, 3, 4 where the losses happen. Tracing on the qubits 5, 6, 7 transforms any logical state $\ket{\psi_L} = c_0 \ket{0_L} + c_1 \ket{1_L}$ into a mixture with equal probabilities of the four states $\{\ket{G}, X_2 X_3\ket{G},  X_3 X_4 \ket{G}, X_2 X_4\ket{G}\}$ that is independent on the coefficients $c_0$ and $c_1$. Finally, no event with five, six or seven losses can be corrected. The total probability of a successful correction is given by the sum of all the probabilities $P_j$ and reads
\begin{equation}\label{prob_success}
\begin{split}
p_\text{success} & = (1-p)^7 + 7p(1-p)^6 + 21 p^2(1-p)^5 + 28 p^3(1-p)^4 + 7p^4(1-p)^3 \\
& = 1 -7 p^3+21 p^5 -21 p^6 + 6 p^7.
\end{split}
\end{equation}
\end{document}